\documentclass[12pt]{article}

\newcommand{\blind}{1}

\usepackage[margin=1in]{geometry}
\RequirePackage{amsthm,amsmath,amsfonts,amssymb}
\usepackage{natbib}
\setcitestyle{aysep={}}
\RequirePackage[colorlinks,citecolor=blue,urlcolor=blue]{hyperref}
\RequirePackage{graphicx}
\usepackage[utf8]{inputenc}
\usepackage{hyperref}

\usepackage{subcaption}
\graphicspath{ {./graphs/} }
\usepackage{multirow}
\usepackage{setspace}

\usepackage[noend]{algpseudocode} 
\usepackage{algorithm}
\usepackage{color}
\graphicspath{{fig/}}
\usepackage{xcolor}
\usepackage{mathrsfs}
\usepackage{bbm}

\theoremstyle{plain}

\newtheorem{theorem}{Theorem}[section]
\newtheorem{lemma}[theorem]{Lemma}
\newtheorem{proposition}[theorem]{Proposition}
\newtheorem{corollary}[theorem]{Corollary}

\theoremstyle{remark}

\newtheorem{remark}{Remark}

\algnewcommand{\IIf}[1]{\State\algorithmicif\ #1\ \algorithmicthen}
\algnewcommand{\EndIIf}{\unskip\ \algorithmicend\ \algorithmicif}
\algnewcommand\algorithmicinitialize{\textbf{Initialize:}}
\algnewcommand\Initialize{\item[\algorithmicinitialize]}

\date{}

\begin{document}

\def\spacingset#1{\renewcommand{\baselinestretch}%
{#1}\small\normalsize} \spacingset{1}

\if1\blind
{
  \title{\bf A Subsequence Approach to Topological Data Analysis for Irregularly-Spaced Time Series}
  \author{Sixtus Dakurah\thanks{
    The authors were supported by NSF Grant DMS-2038556.  Support for this research was also provided by the University of Wisconsin-Madison Office of the Vice Chancellor for Research and Graduate Education with funding from the Wisconsin Alumni Research Foundation. 
    The authors thank Chris Geoga from the University of Wisconsin-Madison for useful discussion related to this work.
    }\hspace{.2cm}\\
    Department of Statistics, University of Wisconsin-Madison\\
    and \\
    Jessi Cisewski-Kehe\footnotemark[1] \\
    Department of Statistics, University of Wisconsin-Madison}
  \maketitle
} \fi

\if0\blind
{
  \bigskip
  \bigskip
  \bigskip
  \begin{center}
    {\LARGE\bf A Subsequence Approach to Topological Data Analysis for Irregularly-Spaced Time Series}
\end{center}
  \medskip
} \fi

\bigskip
\begin{abstract}
A time-delay embedding (TDE), grounded in the framework of Takens's Theorem, provides a mechanism to represent and analyze the inherent dynamics of time-series data. Recently, topological data analysis (TDA) methods have been applied to study this time series representation mainly through the lens of persistent homology. Current literature on the fusion of TDE and TDA are adept at analyzing uniformly-spaced time series observations. This work introduces a novel {\em subsequence} embedding method for irregularly-spaced time-series data. We show that this method preserves the original state space topology while reducing spurious homological features. Theoretical stability results and convergence properties of the proposed method in the presence of noise and varying levels of irregularity in the spacing of the time series are established. Numerical studies and an application to real data illustrates the performance of the proposed method.
\end{abstract}

\noindent%
\vfill

\newpage
\section{Introduction}
\label{sec:intro}
A time series measurement $x(t) \in \mathbb{R}$ at time $t$ can be considered as the outcome of a data-generating space of some dynamical system (i.e., mathematical models that describes the evolution of variables over time) with state vector $\mathbf{s}(t) \in \mathbb{R}^N$. Constructing a meaningful approximation of this underlying data-generating space when only the scalar time series is observed can uncover latent patterns and structures not readily apparent in the raw time series. Time-delay embeddings (TDEs) are often employed for this state space reconstruction. The TDE transforms the time-series data from the time-domain to an estimate of the state space, which can reveal properties of the system such as periodicity and other structures not apparent in the time domain. The principle underlying TDEs is Takens's Theorem, which asserts that even if the actual dynamics (i.e., the system's behavior over time) are not known, a single time series can be treated as a one-dimensional projection of the path traced by the system's state vector in a multi-dimensional space. An approximation to the actual dynamics can be constructed from this projection \citep{takens2006detecting}. Takens proved that assuming uniformly-spaced and noise-free measurements of unlimited length, there exists a diffeomorphism (i.e., a smooth and invertible function) between the true high-dimensional system  and the (TDE) reconstructed system. This theorem forms the foundation for much of the discussions on reconstructing the multi-dimensional state of a system from a single time series \citep{ali2007chaotic}.

More recently, there is renewed interest in coupling the TDE method with tools from topological data analysis (TDA) to study the underlying dynamics of time-series data using various geometric and topological features of the TDE such as clusters, loops, voids, and their higher dimensional analogs \citep{gholizadeh2018short}. In particular, the reconstructed state space obtained from the TDE reveals consistent patterns and structures that reflect the shape and characteristics of the underlying time-series dynamics \citep{takens2006detecting}. TDA is a computational method for studying the shape of data, which can be applied to characterize the topological features of these reconstructed state spaces. The characterization is often carried out using {\em persistent homology}, a tool of TDA, which employs a multi-scale approach to quantify certain topological features \citep{edelsbrunner2000topological,edelsbrunner2022computational} (see Section \ref{subsec:vr-pers-homology} for more details). TDA and TDE have been successfully applied to quantify periodicity in time-series data \citep{perea2015sliding}, analyze human speech \citep{brown2009nonlinear}, detect motion patterns in video \citep{tralie2018quasi}, and in wheeze detection \citep{emrani2014persistent}.

In the applications noted above, the observed time series is uniformly-spaced. However, time series is often not uniformly-spaced due to measurement lapses \citep{stark1997takens}, process errors \citep{casdagli1991state}, or inherent features of the data generating process \citep{stark1997takens,lekscha2018phase}, etc. The standard Takens's theorem does not handle  irregularly-spaced time series, but several options exist in the literature to address issues related to irregularly-spaced time series observations to make it amenable to TDE. Broadly, these can be classified into {\em imputation} or {\em exclusion} methods. Imputation methods involve predicting the missing observations, and then the analysis is carried out assuming a uniformly-spaced time series has been observed \citep{harvey1984estimating,casdagli1991state,lekscha2018phase}. Exclusion methods initially ignore the presence of missing values and assume a uniformly-spaced set. The TDE maps are then constructed and any map with a missing value is excluded \citep{boker2018robustness,johnson2022empirical}.  If the imputation model is misspecified, it can produce structures in the TDE that do not reflect true properties of the data, and the exclusion method can significantly alter the shape of the TDE space \citep{huke2007embedding,boker2018robustness}. Since TDA can provide quantification of qualitative properties of the reconstructed state space, the drawbacks of the imputation and exclusion methods may distort topological features constructed from the TDE spaces.

In this work, we propose a {\em subsequence} method for carrying out a TDE of irregularly-spaced time-series data that preserves certain properties of the reconstructed state space. The level of irregularity of the time-series data is controlled by the {\em regularity score} (defined in Section \ref{sec:subsequence}). We show that the proposed method preserves the topological features of the original underlying state space of the time series while reducing spurious shape features. Theoretically, we prove stability and convergence results of the proposed subsequence method in the presence of noise and for varying levels of irregularity in the observed time series. 
Further, we demonstrate the competitiveness of the proposed subsequence method through simulation studies and an application to real data.  We view this work as laying a necessary foundation for new statistical TDA methodology that has the potential for characterizing properties of time-series signals beyond period estimation; this point is expanded upon in the Discussion Section of the supplementary material (Section~\ref{sec:discussion}) included at the end of this article after the references.

\section{Background on Topological Data Analysis}
\label{sec:prelim-topology}
One of the main constructs for studying the topological features of data is persistent homology \citep{edelsbrunner2000topological,edelsbrunner2022computational}, which detects different dimensional holes in data. Persistent homology provides a multi-scale characterization of the homological features (holes) of a topological space by keeping track of when these topological features first appear (birth) and when they disappear (death) in a filtration. The main topological object in this work is a point cloud resulting from the $(M+1)$-dimensional TDE space of time-series data. The following provide an introduction to persistent homology that are used in the proposed method.

\subsection{Vietoris-Rips Filtration and Persistent Homology}
\label{subsec:vr-pers-homology}
Homology is an area of mathematics that looks for holes in a topological space, and persistent homology looks for holes in data.  Topological holes are identified using ideas from algebraic topology and are characterized by different dimensional homology groups \citep{hatcher2002algebraic,edelsbrunner2022computational}. 
The zero-dimensional homology group (${H}_0$) contains connected components (clusters), the one-dimensional homology group (${H}_1$) contains loops, the two-dimensional homology group ($H_2$) contains voids like the interior of a balloon, and more generally, the k-dimensional homology group (${H}_k$) can be thought of as representing k-dimensional holes.

The data for which we use persistent homology in the proposed method are point clouds derived from a timeseries.  If we only computed the homology of the point-cloud data, it would not be useful because the data points are not connected resulting in only $H_0$ features. Instead, an intermediate structure is defined with simplicial complexes. A zero-simplex is a point, a one-simplex is a segment, a two-simplex is a triangle, and, more generally, a $k$-simplex $\sigma = (v_0, \cdots, v_k)$ is a $k$-dimensional polytope of $k+1$ affinely independent points (i.e., zero-simplexes) $v_0, \cdots, v_k$.
A simplicial complex ${K}$ is a finite set of simplices such that for any $\sigma_1, \sigma_2 \in {K}$, $\sigma_1 \cap \sigma_2$ is a face (i.e., the convex hull of any non-empty subset of points that define the simplex) of both simplices, or the empty set; and a face of any simplex $\sigma \in {K}$ is also a simplex in ${K}$.
To compute the homology, simplicial complexes are built along a sequence of filtration values, which is discussed next.

\begin{figure}[ht]
     \centering
     \begin{subfigure}[b]{0.3\textwidth}
         \centering
         \includegraphics[width=\textwidth]{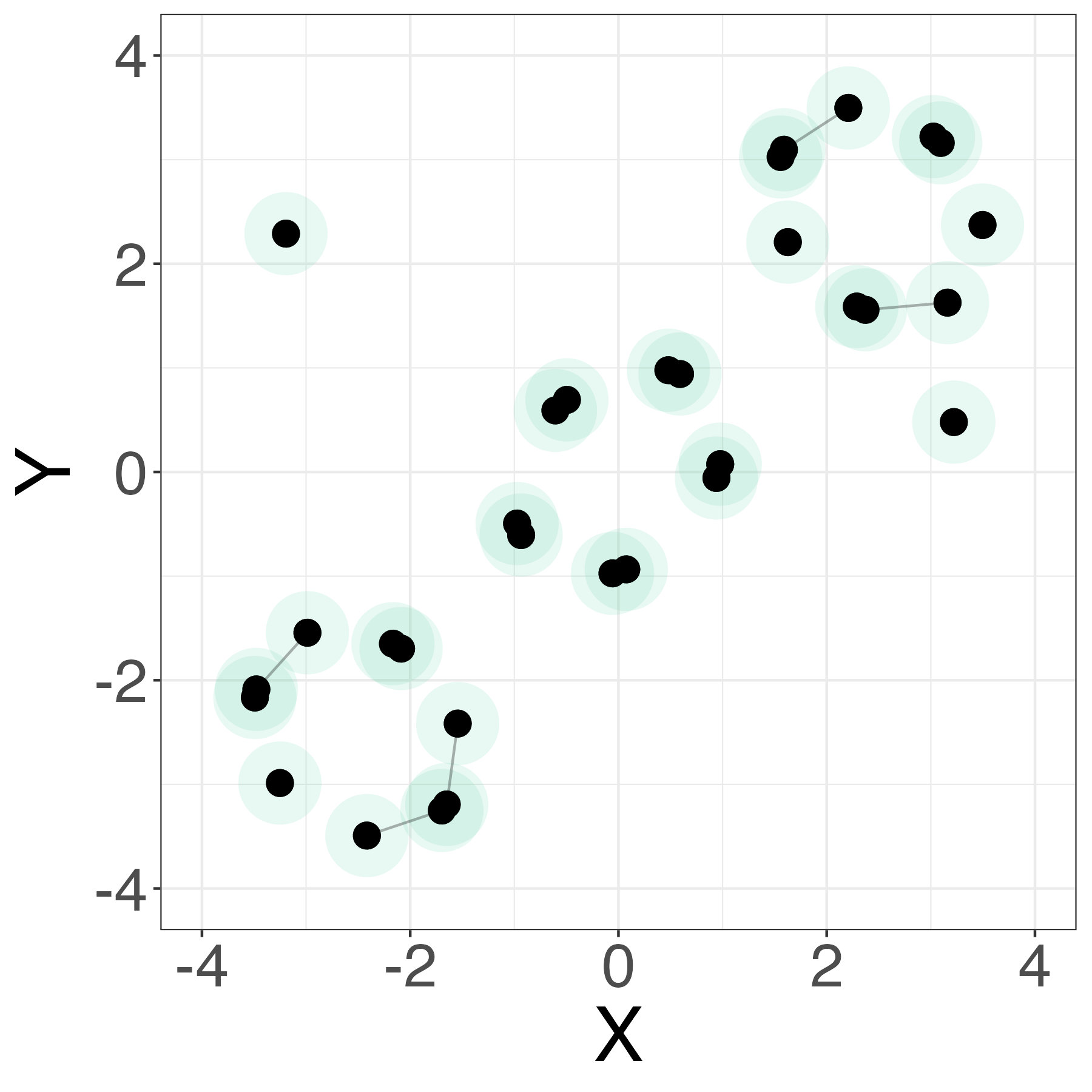}
         \caption{Point set with $\delta=0.8$}
         \label{fig:point-cloud-delta08}
     \end{subfigure}
     \hfill
     \begin{subfigure}[b]{0.3\textwidth}
         \centering
         \includegraphics[width=\textwidth]{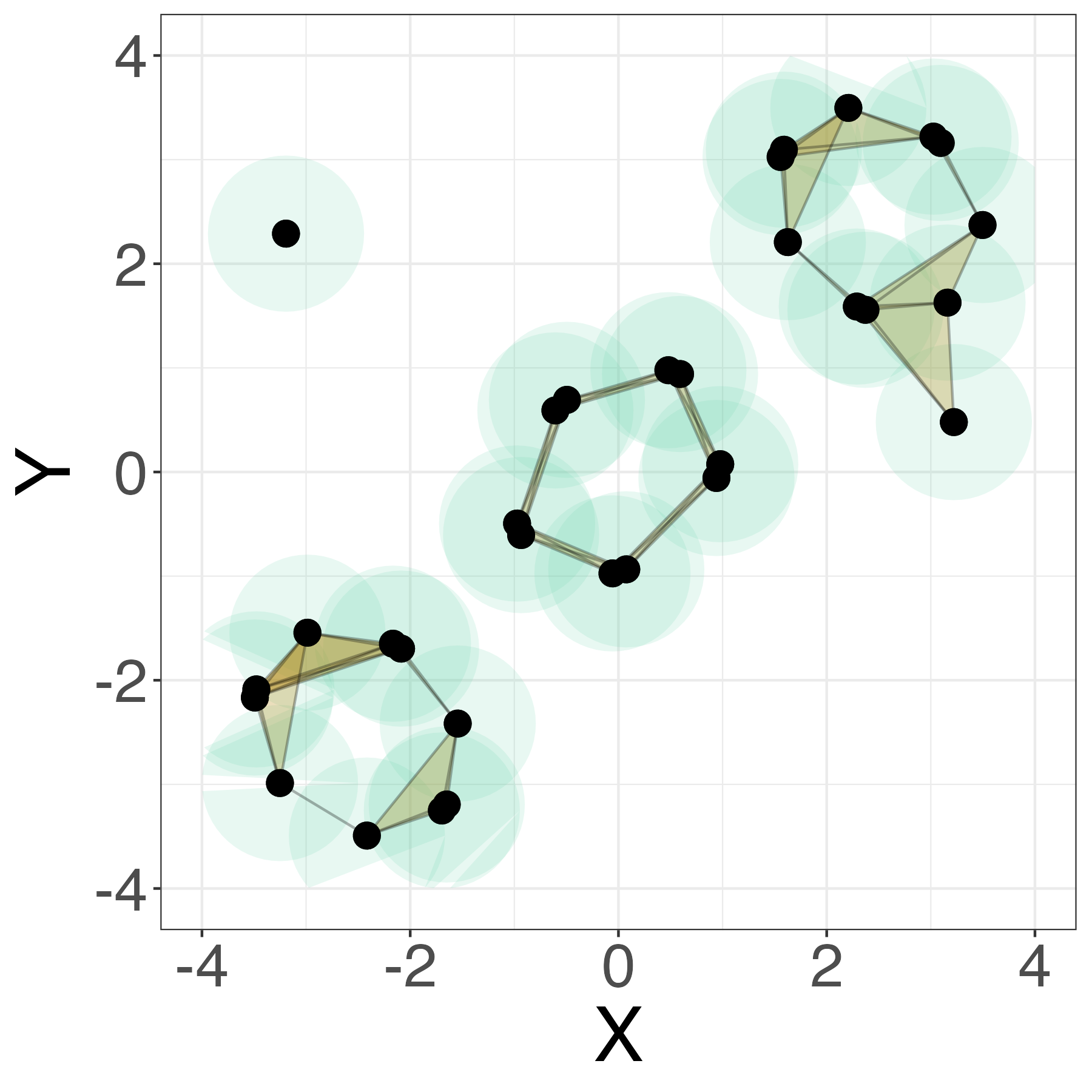}
         \caption{Point set with $\delta=1.5$}
         \label{fig:point-cloud-delta15}
     \end{subfigure}
     \hfill
     \begin{subfigure}[b]{0.3\textwidth}
         \centering
         \includegraphics[width=\textwidth]{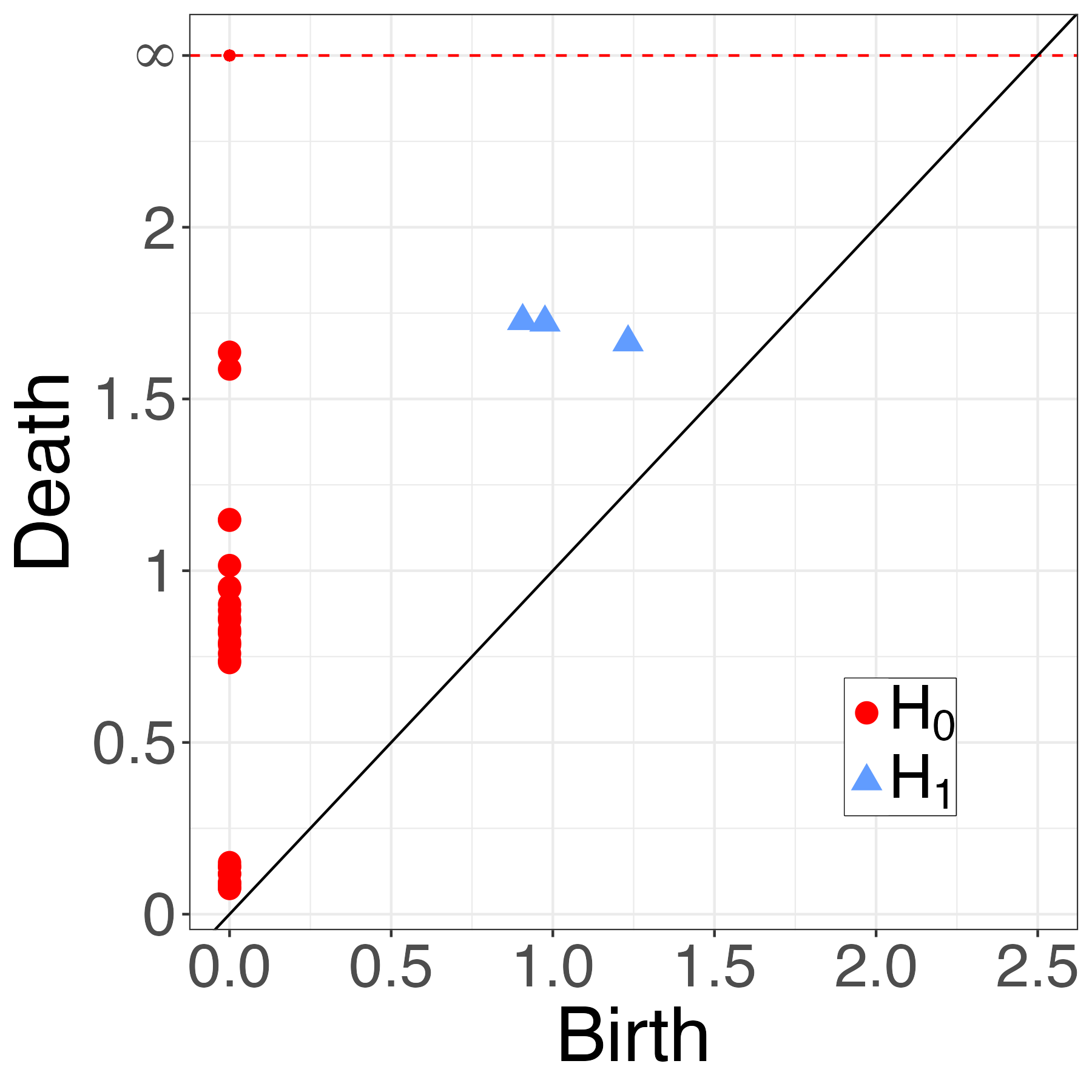}
         \caption{Persistence diagram}
         \label{fig:persistence-diagram}
     \end{subfigure}
        \caption{Illustration of VR complexes and persistence diagram. The zero-simplices (black points) were sampled around three circles. Balls of diameter $0.8$ (a) and $1.5$ (b) are drawn around the points, resulting in one-simplices (black lines) and two-simplices (yellow triangles). The associated persistence diagram (c) has $H_0$ features (red points) and $H_1$ features (blue triangles) that represent the three circles.}
        \label{fig:vietoris-rips-filtration}
\end{figure}

A common approach to constructing simplicial complexes in TDA  is the Vietoris-Rips (VR) complex \citep{vietoris1927hoheren,edelsbrunner2022computational}. A VR complex is constructed over a finite set of points $S = \{v_1, v_2, \cdots, v_n\}$ using a distance parameter $\delta\geq0$. For any subset of k-points $\{ v_{i_1}, \cdots, v_{i_k} \}$, a $(k-1)$-dimensional simplex is formed when the pairwise Euclidean distance between all points is at most $\delta$. A collection of all such simplices forms the VR complex denoted as $VR(S, \delta)$. The composition of the VR complex progresses hierarchically as $\delta$ increases. This leads to the concept of filtration, which defines an inclusion relation between the simplicial complexes for a set of $\delta$ values. More formally, for an ordered sequence of $\delta$ values: $0 < \delta_1 < \delta_2 < \cdots < \delta_q < \infty$, the VR complexes admits a nested structure,
$VR(S, 0) \subset VR(S, \delta_1) \subset \cdots \subset VR(S, \delta_q) \subset VR(S, \infty).$
Figure \ref{fig:vietoris-rips-filtration} provides an illustration of a VR complex. The black points (zero-simplices) in Figure \ref{fig:point-cloud-delta08} and  Figure \ref{fig:point-cloud-delta15} denotes the data with cyan balls of diameter $\delta=0.8$ and $\delta=1.5$, respectively, along with the resulting one-simplices and two-simplices. The simplicial complex in Figure \ref{fig:point-cloud-delta08} is a subset of the simplicial complex displayed in  Figure \ref{fig:point-cloud-delta15}.

The inclusion relation between the VR complexes induces a map between the homology groups,
${H}_k(VR(S, 0)) \xrightarrow{} {H}_k(VR(S, \delta_1)) \xrightarrow{} \cdots \xrightarrow{} {H}_k(VR(S, \infty)).$
The notion of persistent homology is developed through these homology maps by tracking the changes in the generators (i.e., the different dimensional holes) of these nested homology groups. 
The birth time and death time of homology group generators along this sequence encodes topological changes in the groups. For a homology group ${H}_k$, we denote the birth time and death time of the $j$-th homology group generator (i.e., the $j$-th hole) by $b_j$ and $d_j$, respectively. The persistence, or lifetime, of the generator is given by $d_j - b_j$, and a longer persistence is often considered to be topological signal while shorter persistence often represents topological noise. If we let $k_j$ be the homology group dimension of the $j$-th generator, and ${J}$ the index set of the generators of the homology groups, then the set $\text{Dgm}(S) = \{(b_j, d_j, k_j): \forall j \in J\} \cup \Delta$ fully characterizes the persistence (life span) of the homology generators, and is used to construct a graphical summary referred to as the {\it persistence diagram}; see example in Figure \ref{fig:persistence-diagram}. The $\Delta$ represents points where the birth time is equal to the death time, and it included in the definition of persistence diagrams for technical reasons \citep{mileyko2011probability}.

A persistence diagram is a summary statistic that characterizes the birth and death times of different dimensional holes detected in a dataset.  These diagrams can be useful on their own for visualizing topological information of complex data, but are often used for inference or prediction tasks (e.g., \citealt{fasy2014confidence, bubenik2015statistical, adams2017persistence, robinson2017hypothesis, xu2019finding, berry2020functional, pun2022persistent, glenn2024confidence}).  In this work, persistence diagrams are used to characterize the TDE from time series and to quantify the performance of the proposed and comparison methods for reconstructing the state space of irregularly-spaced time-series data. 
 
\section{Method}
The persistence of topological features forms the basis for most of our analysis on the TDE reconstructed state spaces. A TDE transforms a time series over a sliding window of size $M\tau$ to a point cloud in $\mathbb{R}^{M+1}$. While Takens's theorem guarantees exact reconstruction for uniformly-spaced time-series data with appropriate $M$ and $\tau$, no such guarantees exist for irregularly-spaced time series. Here, we propose a new reconstruction method for irregularly-spaced time series that better preserves the topological features of the reconstructed state space compared to existing methods. 

\subsection{A Note on Terminology}
\label{sec:note-terminologies}
For this work, the term {\em uniformly-spaced} time series is used to describe time series that have equally-spaced time intervals between successive observations, and is considered the ``true'' times series for purposes of evaluating the proposed method. The term {\em irregularly-spaced} time series refers to observations with unequally-spaced time interval between successive observations. To characterize how the two forms of time series are related, it is assumed throughout this work that the irregularly-spaced time series is a subset of the uniformly-spaced time series. Hence the irregularly-spaced time series always have fewer time measurements than the corresponding uniformly-spaced time series. 

The concept of TDE is also referred to in the literature as a {\em delayed coordinate embedding}, a {\em sliding window embedding}, or simply a {\em Takens embedding}. For this work, only the term TDE is used. For any irregularly-spaced time series, it is assumed there is a ``true'' underlying uniformly-spaced time series. The use of ``TDE'' exclusively refers to an embedding constructed from this true underlying uniformly-spaced time series. After applying the proposed subsequence method, the resulting embedding is referred as the {\em subsequence embedding} (SSE). In instances where an exposition applies to both the TDE and SSE, the term {\em embedding map} is used as a collective reference to the two concepts.

For a given time interval, it is assumed that missing or unobserved values occurs with a given probability. That is, for a given time point in a time interval, a measurement is not observed at that point with some probability. Such probabilistic mechanism governing the observations of time series values is not uncommon in the literature (e.g., \citealt{dunsmuir1981estimation}). This probability can be fixed for all time points or it can vary for each time point. This characterization is referred to as the {\em missingness structure} of the time series in context. 

\subsection{Basics of Time-Delay Embeddings}
\label{sec:basic-tde}
For this work, the discussion on TDEs is restricted to univariate time series. Assume this univariate time series is generated by a system with a state vector $\mathbf{s}(t)$ on a manifold which is a subset of some $N$-dimensional space $\mathbb{R}^N$. The state vector $\mathbf{s}(t)$ is not directly observable, however some measurement of it, denoted $x(t) = h(\mathbf{s}(t))$, is observed through the measurement function $h(\cdot)$. The measurement function $h(\cdot)$ can be thought of as rule that transforms the high-dimensional state vector into the observed univariate time series $x(t)$. For instance, in astronomy, $\mathbf{s}(t)$ might include variables such as positions, velocities, and luminosities of various celestial bodies, such as exoplanets, stars, or galaxies. However, the measurement function is specifically designed to extract a single scalar value from this vector. The specific form of $h(\cdot)$ is influenced by many considerations, for example, the limitation of observational tools. For a star, the measurement function could be designed to extract a key observable from the state vector, such as its luminosity. Thus the measurement $h(\mathbf{s}(t))$ reflects the observed brightness of the star at any given time $t$.

\begin{figure}[ht]
 \centering
 \includegraphics[width=0.75\linewidth,clip=true]{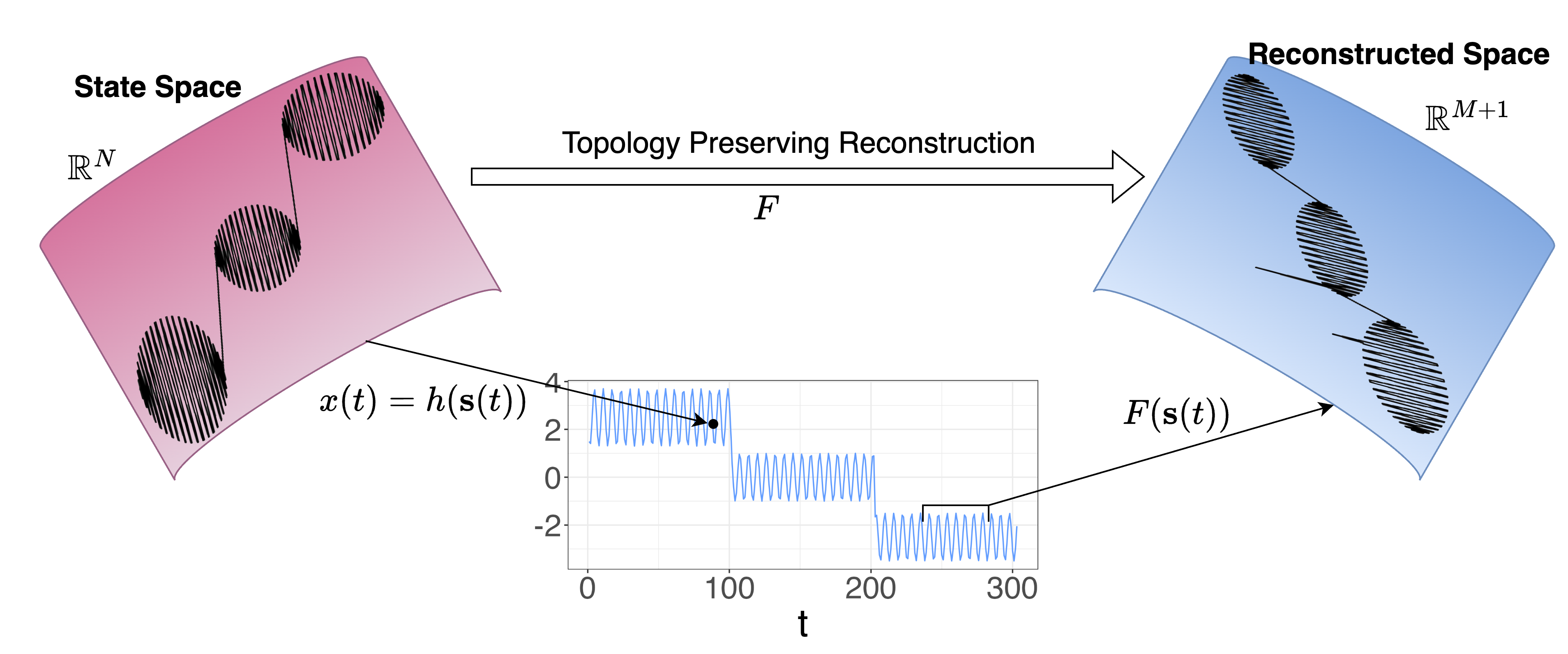}
 \caption{Illustration of the embedding process. Top-left: the state space, typically not observed. Middle-bottom: the time series obtained via the measurement function $h(\cdot)$. Top-right: the reconstructed space from the TDE matrix $F$, which preserves the topology of the original state space.}
 \label{fig:tde-illustration}
\end{figure}

The scalar value $x(t)$ is the observed time series measurement. Define the function $F: \mathbb{R}^N \xrightarrow{} \mathbb{R}^{M+1}$ as the embedding map with the form $F(\mathbf{s}(t)) = \left[x(t), x(t+\tau), \cdots, x(t + M\tau)\right]$. If the measurement function $h(\cdot)$ is noise-free, and the embedding dimension $M+1$ is chosen to be more than twice the dimension of the attractor (i.e., the $N$-dimensional region toward which the system evolves) of the system's state space, Takens's theorem guarantees that the embedding map $F(\mathbf{s}(t))$ has a one-to-one correspondence between the original state space of the system (from which the time series is derived) and the reconstructed state space formed by $F(\mathbf{s}(t))$ \citep{takens2006detecting}. This ensures that the dynamics of the system can be studied in the reconstructed space as if it were being studied in the original space. Figure~\ref{fig:tde-illustration} demonstrates this reconstruction process by mapping the scalar time series to the TDE matrix $F$ to reconstruct the state space.

The choice of embedding dimension $M+1$ and step size $\tau$ is a subject of considerable research in the literature (e.g., \citealt{cao1997practical,kim1999nonlinear}). In this work, the embedding dimension is chosen manually. However, the method of false nearest neighbors is one common method for determining this dimension, which identifies points in a low-dimensional space that appear to be near each other but are not actually neighbors when the data is viewed in a high-dimensional space. By systematically increasing the embedding dimension, and evaluating the percentage of false nearest neighbors, the dimension can be set where this percentage drops significantly, indicating a suitable dimension. More details about this and other procedures for determining $M$ and $\tau$ can be found in \cite{cao1997practical,kim1999nonlinear}. A large value of $M$ is often preferred as it enables the embedding to capture more details inherent in the time series. If $M$ is too large, there may be an insufficient number of points in the embedding space. Furthermore, if $M\tau$ is too small due to a small $\tau$, relatively fewer points fall in each embedding window. This results in points repeatedly appearing in windows, which can lead to redundant information. If $M\tau$ is too large due to a large value of $\tau$, the reconstructed state space can be distorted because relevant periodic behavior of the time series may not be captured \citep{casdagli1991state}. Hence the choice of $\tau$ and $M$ is such that $M\tau$ is not too large or too small, but is application-dependent and requires empirical testing.

\subsection{Subsequence Method}
\label{sec:subsequence}
The TDE construction in the previous section assumes the observed time series is uniformly-spaced, but a time series is often irregularly-spaced in real data.  We propose a method to extract uniformly-spaced subsequences from the observed irregularly-spaced time series and prove its topology-preserving properties, along with consistency and convergence results.

\subsubsection{Subsequence construction}
\label{sssec:subseq-const}
Let $\mathbf{x} = \left(x(t) : t \in \mathcal{T}\right)$ be a time series of length $n$ where $\mathcal{T} = \{ t_1, \cdots, t_n \} \subset \mathbb{N}$. Further assume that this time series is not uniformly spaced, that is, $t_{i+1} - t_i \ne t_{i+2} - t_{i+1}$, for some $t_i \in \mathcal{T}$ such that $t_i < t_{i+1}$. In this work, a subsequence of the set $\mathbf{x}$ is defined as any subset that omits elements of $\mathbf{x}$ without changing the order of the remaining elements. This definition does not guarantee that $t_{i+1} - t_i = t_{i+2} - t_{i+1},$ $\forall t_i \in \mathcal{T}$, which is a condition we want to achieve with the proposed subsequence construction. 
Let $\mathbf{x}_{p, r} \subseteq \mathbf{x}$ be a subset of the original time series with time index $\mathcal{T}_{p,r} \subseteq \mathcal{T}$ with the condition that $t_{p, i+1} - t_{p, i} = r,$ $\forall t_{p,i} \in \mathcal{T}_{p, r}$. The set $\mathbf{x}_{p, r}$ is the $p$-th subsequence of {\em regularity} $r$, and it is a uniformly-spaced subsequence. For any non-uniformly spaced time series, we can build a collection of such subsequence for various values of $r$. The goal is to first obtain the longest subsequence for a small $r$.  As the subsequence length reduces for a given $r$, the regularity value $r$ can increase to obtain more uniformly-spaced subsequences. An algorithm for computing this collection of subsequences is displayed in Algorithm~\ref{alg:subsequence}, which is adapted from an algorithm that finds the longest arithmetic progression in a sequence developed in \citep{erickson1999finding}. In the statement of the algorithm, the following notation is used: (i) the union symbol $\cup$ denotes the addition of a set (element) to a set (vector), (ii) the number of elements in a set or a vector $A$ is denoted by $|A|$, and (iii) the notation $A\backslash B$ represents subset of elements in set $A$ obtained by excluding all elements from set $B$. 
\begin{algorithm}[ht!]
\caption{Uniform subsequence construction}
\label{alg:subsequence}
\begin{algorithmic}[1]
\Require Time points $\mathcal{T} = \{ t_1, \cdots, t_n \}$, regularity score $r$, minimum sequence length $m$.
\Ensure $r \le\ t_n - t_1$,\quad $m \le n$.
\Initialize $\mathcal{T}_p \gets \{\dots\}$, temporary time index, $\mathbf{T}_{\text{reg}} \gets \{\}$, uniformly-spaced subsequences.
\While{number of elements in $\mathcal{T}$ is greater than m}
\For{$i = 1:(|\mathcal{T}| - 1)$}
\State $\mathcal{T}_{\text{sub}} \gets \mathcal{T}[i]$ \Comment{Initialize a subsequence.}
\For{$j = (i+1):|\mathcal{T}|$}
    \If{$\mathcal{T}[j] - \mathcal{T}_{\text{sub}}[j-i] = r$} \Comment{Check the regularity condition.}
    \State $\mathcal{T}_{\text{sub}} \gets \mathcal{T}_{\text{sub}} \cup \mathcal{T}[j]$; \quad
    \textbf{if} {$|\mathcal{T}_{\text{sub}}| > |\mathcal{T}_p|$} \textbf{then} 
    $\mathcal{T}_p \gets \mathcal{T}_{\text{sub}}$ 
    \Else \quad \textit{break} \Comment{Initialize with the next point in the sequence.}
    \EndIf
\EndFor
\EndFor 
\If{$|\mathcal{T}_p| \ge m$ and $\mathcal{T}_{p}$ is not identical to any other subsequence in $\mathbf{T}_{reg}$}
\State $\mathbf{T}_{reg} \gets \mathbf{T}_{reg} \cup \mathcal{T}_p$;
 \quad $\mathcal{T} \gets \mathcal{T}\backslash\mathcal{T}_p$ \Comment{Remove the subset from the sequence.}
\Else\quad \textit{break} \Comment{No uniformly-spaced subsequence of the required length exist.}
\EndIf
\EndWhile\\
\Return $\mathbf{T}_{\text{reg}}$ \Comment{Set of all regularly spaced subsequences each of regularity $r$.}
\end{algorithmic}
\end{algorithm}
Algorithm \ref{alg:subsequence} returns all possible uniformly-spaced time points from the time index set $\mathcal{T}$ with regularity score $r$.  Note that for uniformly-spaced $\mathcal{T}$, it returns the full sequence. The uniformly-spaced observations can now be obtained by simply matching these observations to the time points in each subsequence.
Not all the subsequences returned by Algorithm \ref{alg:subsequence} are required in the SSE (see Remark \ref{rmk:number-ps}). 

\subsubsection{Subsequence embedding method} \label{sec:sse}
Takens's theorem guiding the construction of the TDE in Section \ref{sec:basic-tde} involves a single measurement function $h(\cdot)$, which generates each time series measurement \citep{takens2006detecting}. A generalization considers each coordinate in the embedding maps as a measurement function (see Remark $2.9$ in \cite{sauer1991embedology} and Theorem $2$ in \cite{deyle2011generalized}). Such generalizations allow for the extension of Takens's theorem to multiple measurement functions involving multiple time series. This motivates the proposed SSE method where each subsequence is viewed as distinct time series.

To construct the proposed SSE, a single distinct measurement function is defined on each subsequence. 
Let $h_p(\cdot)$ be the measurement function associated with the $p$-th subsequence. Then the $p$-th embedding mapping has the form 
\begin{equation}
F_p(\mathbf{s}(t_{p, i})) = \left[x_{p, r}(t_{p, i}), x_{p, r}(t_{p, i}+\tau_p), \cdots, x_{p,r}(t_{p, i} + M\tau_p)\right], \quad x_{p, r}(t_{p, i}) = h_p(\mathbf{s}(t_{p, i})).
\label{eqn:p-map-embedding}
\end{equation}
The delay step $\tau_p$ is fixed for each subsequence map.  The map is also constructed under the assumption that the length of each subsequence $n_p > \max(M+1, M*\tau_p)$. This ensures that there are sufficient observations within each subsequence to construct a point in the embedding space.
The embedding matrix from the $p$-th subsequence has the form
\begin{equation}
\mathbf{F}_p = 
    \begin{bmatrix}
        F(\mathbf{s}(t_{p, 1}))^\top & F(\mathbf{s}(t_{p, 2}))^\top & \cdots & F(\mathbf{s}(t_{p, n_p - M}))^\top 
    \end{bmatrix}^\top.
    \label{eqn:embedding-matrix-sub}
\end{equation}
The embedding matrix of the irregularly-spaced time series is then obtained by vertically stacking the embedding matrices from each of the subsequences. This is denoted as $\mathbf{F} = \begin{bmatrix} \mathbf{F}_1; & \cdots & ; \mathbf{F}_P\end{bmatrix}$, where $P$ is the total number of uniformly-spaced subsequences. The matrix $\mathbf{F}$ is of dimension $\left(\sum_{p = 1}^P (n_p - M)\right) \times (M+1)$.
Note that when the original time series is uniformly-spaced, the SSE method is identical to the TDE. To see this, observe that the longest subsequence in the uniformly-spaced time series is the original sequence.

\begin{figure}[ht!]
     \centering
     \begin{subfigure}[b]{0.6\textwidth}
         \centering
         \includegraphics[width=\textwidth, height=0.6\textwidth]{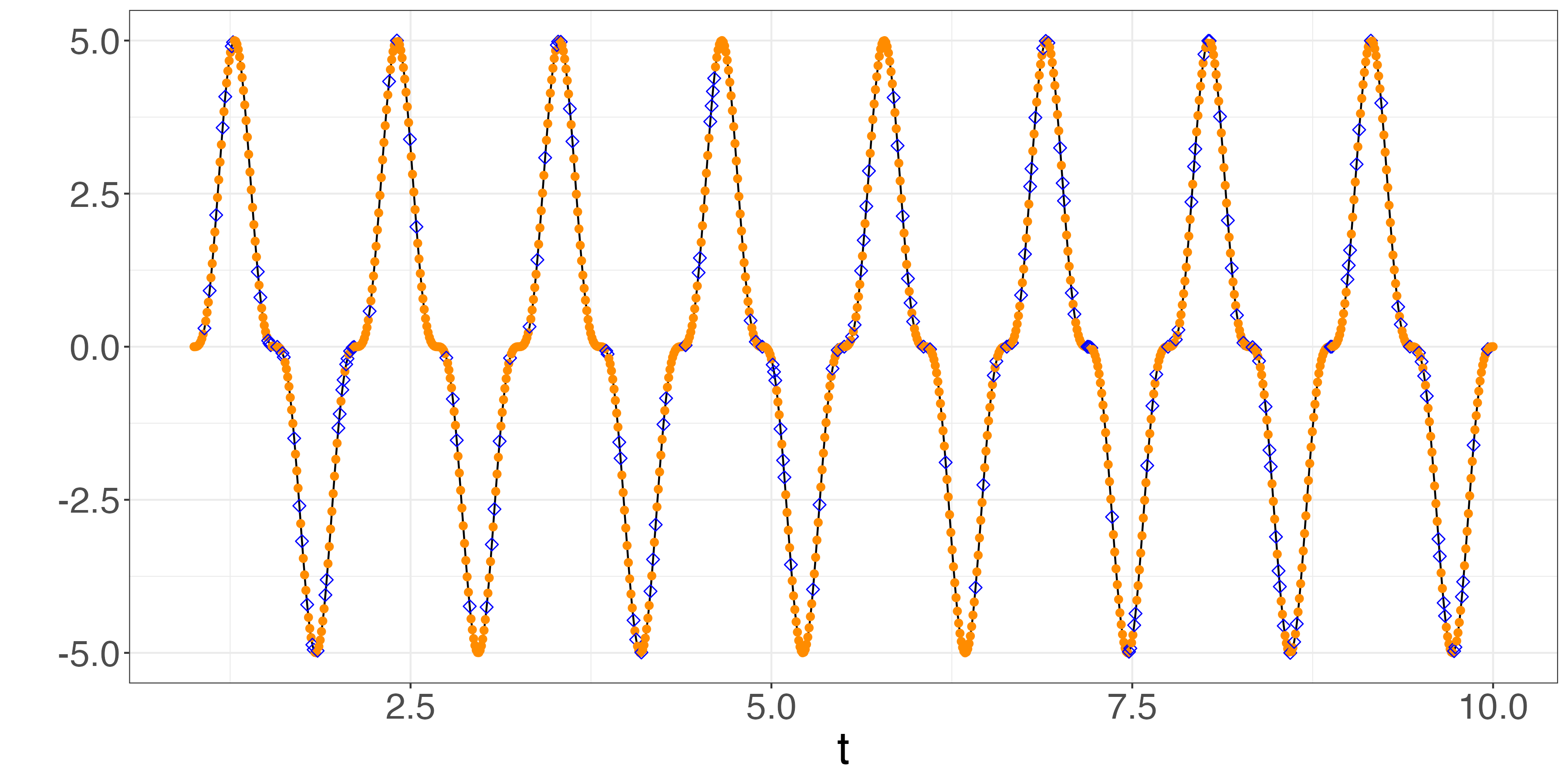}
         \caption{Time series measurements at 1000 time points}
         \label{fig:time-series-subseq}
     \end{subfigure}
     \hfill
     \begin{subfigure}[b]{0.32\textwidth}
        \centering
         \begin{subfigure}[b]{1\textwidth}
             \centering
             \includegraphics[width=\textwidth, height=0.56\textwidth]{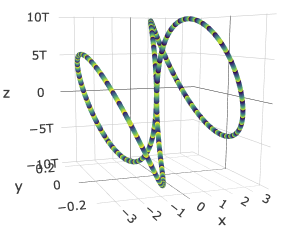}
             \label{fig:embedding-regular}
         \end{subfigure}
        \vfill \vspace{-0.5cm}
         \begin{subfigure}[b]{1\textwidth}
             \centering
             \includegraphics[width=\textwidth, height=0.56\textwidth]{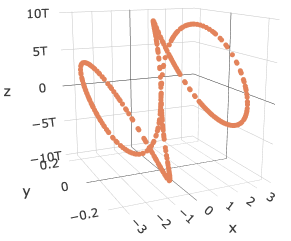}
             \caption{Reconstructed spaces}
             \label{fig:embedding-irregular}
         \end{subfigure}
     \end{subfigure}
     \vfill \vspace{0.5cm}
     \begin{subfigure}[b]{1\textwidth}
         \centering
         \begin{subfigure}[b]{0.49\textwidth}
             \centering
             \includegraphics[width=\textwidth]{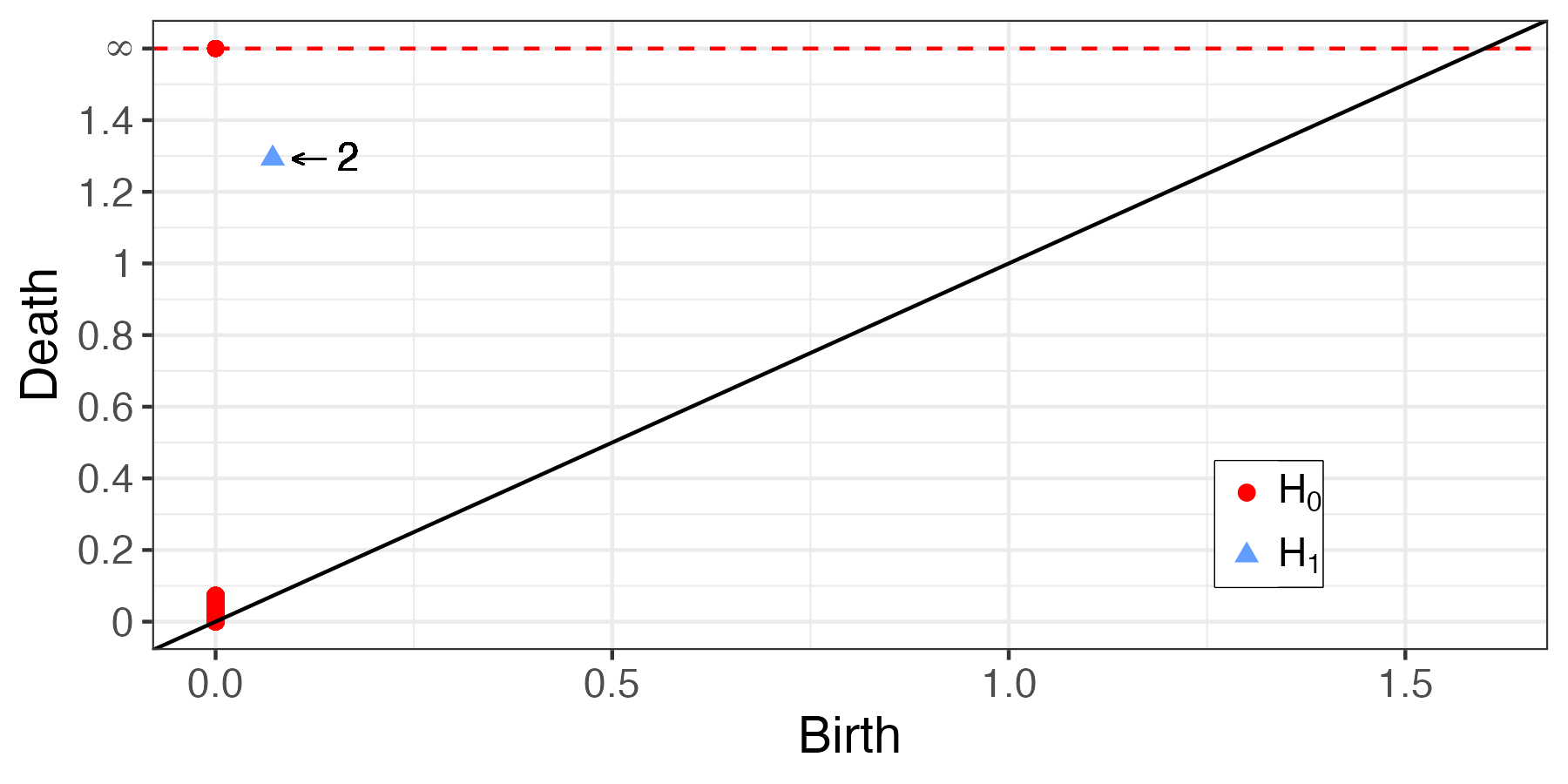}
             \caption{TDE persistence diagram}
             \label{fig:persistence-diagram-regular}
         \end{subfigure}
        \hfill
         \begin{subfigure}[b]{0.49\textwidth}
             \centering
             \includegraphics[width=\textwidth]{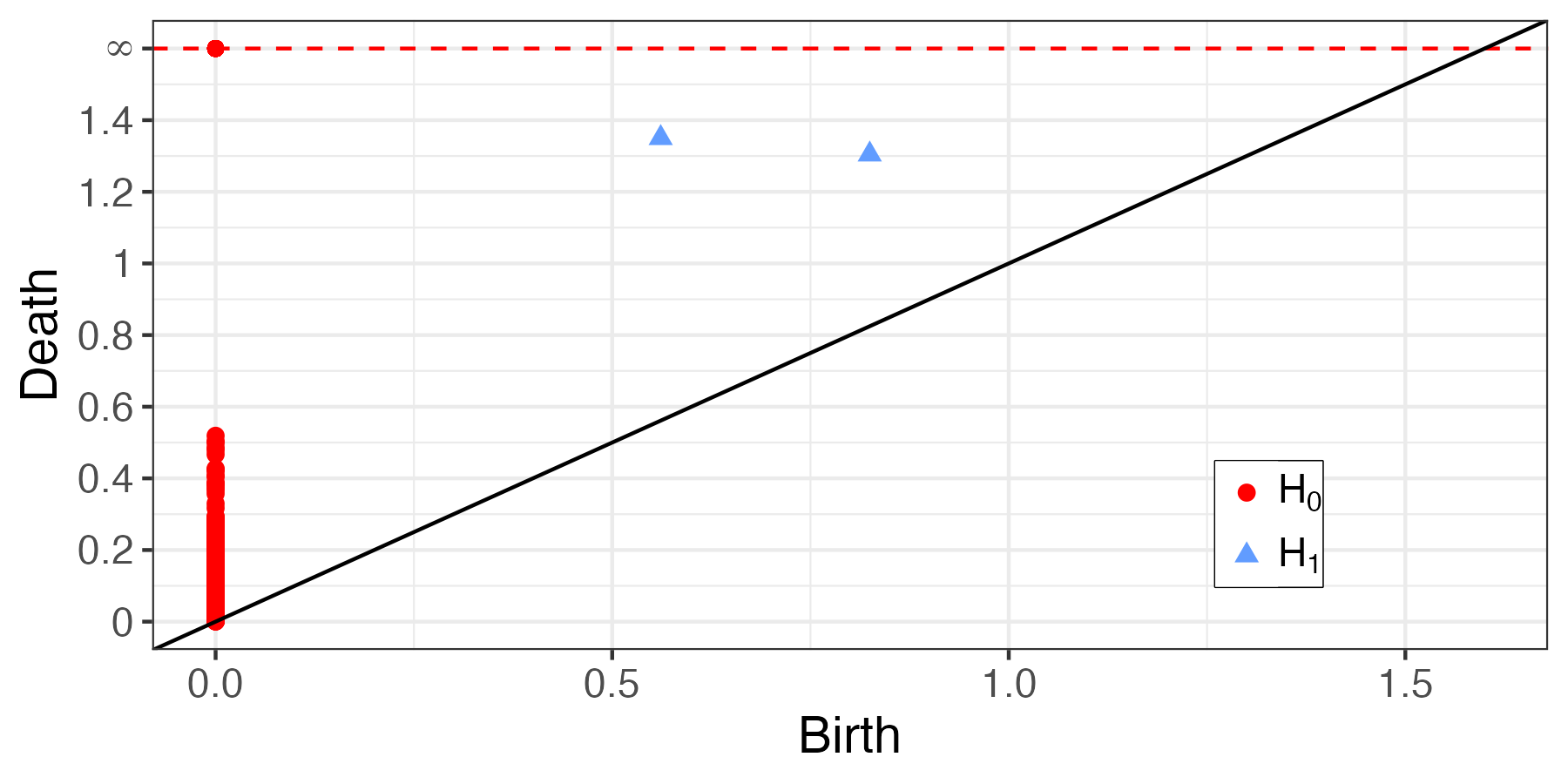}
             \caption{SSE persistence diagram}
             \label{fig:persistence-diagram-irregular}
         \end{subfigure}
     \end{subfigure}
    \caption{SSE method illustration. (a) One thousand time series measurements (blue and orange points). About $20\%$ were designated as missing (hollow blue diamonds) to obtain irregularly-spaced observations (orange points). The TDE of the full time series ((b)-top) and the SSE of the irregularly-spaced time series ((b)-bottom); both time series were embedded in $\mathbb{R}^4$ and their first three principal components are plotted. The persistence diagram of the TDE (c) and SSE (d).}
    \label{fig:subsequence-embedding-iluustration}
\end{figure}

To illustrate the SSE framework, Figure \ref{fig:time-series-subseq} shows measurements at 1000 uniformly-spaced time points (orange points and blue diamonds combined) of which about $20\%$ are designated as missing values (blue diamonds), which creates an irregularly-spaced time series (orange points). Both the uniformly-spaced and irregularly-space time series were embedded into $\mathbb{R}^4$ using the TDE and SSE methods, respectively.
Figure \ref{fig:embedding-irregular}-top gives the TDE of the uniformly-spaced 1000 measurements and contains two identical elliptical shapes. Figure \ref{fig:embedding-irregular}-bottom shows the proposed SSE of the irregularly-spaced time series and also contains two similar elliptical shapes, however, there is visible non-uniform spacing of the points compared to the TDE space. This is primarily due to the SSE using a subset of the original time series (i.e., it constructs a uniform subsample from the irregularly-spaced time series based on Algorithm \ref{alg:subsequence}); the SSE space may be considered as a sparse representation of the TDE space.
The persistence diagram for the TDE is shown in Figure \ref{fig:persistence-diagram-regular}. Since Figure \ref{fig:embedding-irregular}-top  has two identical elliptical shapes, the $H_1$ features have overlapping birth and death time, hence the appearance of a single blue triangle. Figure \ref{fig:persistence-diagram-irregular} shows the persistence diagram for the SSE, and correctly identifies the two loops but the birth and death time are non-overlapping due to the non-identical spacing of the points in the two elliptical shapes. 
In general, the SSE converges to the TDE in terms of the topological similarity of the reconstructed spaces and in the closeness of the persistence diagrams as the time sampling becomes more uniform. A more formal theoretical justification of this assertion, and other technical considerations are discussed in the next section.

\begin{remark}
    The choice of the number of subsequences $P$ and the number of subsequences of different regularity scores $r$ depend on the context and goals of the analysis. To reconstruct an $(M+1)$-dimensional state space, subsequences must satisfy $n_p > \max(M+1, M*\tau_p)$. A set of subsequences with the same regularity score can lead to better reconstruction accuracy as it captures the dominant patterns of the underlying data-generating space of the time series more coherently. Combining subsequences with different regularity scores can improve topological approximations as it captures a wider range of structures of the underlying data-generating space. Thus, there is a trade-off between a better topology approximation and improved reconstruction accuracy. If subsequences with the same regularity score capture most of the time series, combining sequences with different regularity scores may offer limited benefits. While the simulations and real data analysis in this work utilize subsequences with the same regularity score, the methodology and theoretical results apply to subsequences with the same or different regularity scores.
    \label{rmk:number-ps}
\end{remark}

\section{Stability and Convergence Results}
\label{sec:stability-convergence-results}
The reconstructed state space using the proposed SSE is an estimate of the ``true'' state space based on a uniformly-sampled time series (i.e., the TDE space). Persistence diagrams are used to quantify the stability of the estimate by measuring its closeness to the true state space.  The results are stated below, and the proofs are included in the supplementary materials Section~\ref{sec:supp_proofs}.

\subsection{Stability Theory in Persistence Homology}

The set of persistence diagrams from the true state space (TDE) and the reconstructed state space (SSE) can be endowed with a distance measure, such as the popular bottleneck distance. The bottleneck distance gives the minimal bijection between any two diagrams as a measure of the distance between the diagrams. Let $\mathbf{F}^1$ and $\mathbf{F}^2$ be two  finite compact subsets of $\mathbb{R}^{M+1}$, with $\text{Dgm}(\mathbf{F}^1)$ and $\text{Dgm}(\mathbf{F}^2)$ as their corresponding persistence diagrams. The bottleneck distance, $\text{d}_\text{B}$, between the two persistence diagrams is defined as 
\begin{equation}
    \text{d}_\text{B}\left(\text{Dgm}(\mathbf{F}^1), \text{Dgm}(\mathbf{F}^2)\right) = \inf_{\gamma} \sup_{\mu \in \text{Dgm}(\mathbf{F}^1)} || \mu - \gamma(\mu) ||_\infty,
    \label{eqn:bottleneck-dist}
\end{equation}
where the infimum is taken over all bijections $\gamma: \text{Dgm}(\mathbf{F}^1) \xrightarrow{} \text{Dgm}(\mathbf{F}^2)$. For the metric space $(\mathbb{R}^{M+1},  \lVert \cdot \rVert_2)$, where $\lVert \cdot \rVert_2$ denotes the $l^2$-norm, define the distance function as: $d(F, \mathbf{F}^1) = \inf_{F^1 \in \mathbf{F}^1} \lVert F - F^1 \rVert_2,$ $\forall F \in \mathbb{R}^{M+1}.$
Then the Hausdorff distance, $d_H$, is given by
\begin{equation}
    d_H(\mathbf{F}^1, \mathbf{F}^2) = \max \left\{ \sup_{F^1 \in \mathbf{F}^1 }d(F^1, \mathbf{F}^2), \sup_{F^2 \in \mathbf{F}^2 }d(F^2, \mathbf{F}^1)\right\}.
    \label{eqn:dist-hausdorff}
\end{equation}
The Hausdorff distance quantifies the level to which two metric spaces represent the same object. A modification of the Hausdorff distance that is blind to isometric transformations is the Gromov-Hausdorff distance. Let $\nu_1: \mathbf{F}^1 \hookrightarrow \mathbb{R}^{M+1}$ and $\nu_2: \mathbf{F}^2 \hookrightarrow \mathbb{R}^{M+1}$ be isometric (distance preserving) embeddings into the metric space $(\mathbb{R}^{M+1}, \lVert \cdot \rVert_2)$. Then the Gromov-Hausdorff distance between $\mathbf{F}^1$ and $\mathbf{F}^2$ is given by: $ d_{GH}(\mathbf{F}^1, \mathbf{F}^2) = \inf_{\nu_1, \nu_2}d_H(\nu_1(\mathbf{F}^1), \nu_2(\mathbf{F}^2))$.
A fundamental result on persistence diagrams is that they are stable summaries in many settings (i.e., a small change in a point cloud result in a small change in the corresponding persistence diagram) \citep{chazal2021introduction}. The stability relation of the persistence diagrams is stated as 
\begin{equation}
    \text{d}_\text{B}\left(\text{Dgm}(\mathbf{F}^1), \text{Dgm}(\mathbf{F}^2)\right) \le 2 \text{d}_{\text{GH}}(\mathbf{F}^1, \mathbf{F}^2) \le 2\text{d}_{\text{H}}(\mathbf{F}^1, \mathbf{F}^2).
    \label{eqn:stability-bn}
\end{equation}
In what follows, these stability results are established for the proposed SSE method and a denoising procedure to reduce the noise present in the observed time series.

\subsection{Stability of Denoising Procedure} \label{sec:denoise}

Time-series data is typically observed with noise. The level of noise in the reconstructed space influences the presence and the persistence of the topological features. A Fourier denoising procedure is proposed to filter out noise in the observed time series.

Let $\mathbf{x} = \begin{bmatrix}x(t_1), & x(t_2), & \cdots &, x(t_n) \end{bmatrix}^\top$ be an observed time series vector. The first step in the denoising procedure is to transform this observed signal to the frequency domain. The discrete Fourier transform (DFT)  of $x(t_{k})$, denoted as $\Tilde{x}(t_{k})$ is given by
\begin{equation}
    \Tilde{x}(t_{k}) = \sum_{r = 1}^{n} x(t_{r})e^{-j2\pi w_rf_{k}} =  \sum_{r = 1}^{n} x(t_{r}){\phi}_{kr}, \quad 1 \le k \le n,
    \label{eqn:dft-formal}
\end{equation}
where $j$ is the imaginary unit ($j^2 = -1$), $\phi_{kr} = e^{-j2\pi w_r f_{k}}$, $0 \le w_r \le 1$ are sample points, $0 \le f_{k} \le n$ are frequencies, and  $\Tilde{x}(t_{k})$ is the $k$-th sample of the power spectrum at $f_{k}$. 

To filter out noise, the power spectral density is computed for each observation $\tilde{x}(t_k)$. Then a threshold is chosen, and any $\tilde{x}(t_k)$ with power spectral density less than the threshold is set to zero. In selecting the threshold, the goal is to choose a value that does not smooth out the peaks in the true signal. The derivation that follows assumes the selected threshold preserves the peaks in the true signal.
To simplify notations, the thresholded observations are also denoted as $\tilde{x}(t_k)$. The thresholded $\tilde{x}(t_k)$ are transformed back to the time domain to get the noise-reduced signal. which typically involves multiplying $\tilde{x}(t_k)$ by the inverse of a Fourier transform matrix.

Let $\Tilde{\mathbf{x}}$ be the DFT of the time series vector with corresponding Fourier basis $\boldsymbol{\phi}_k$,  where
\begin{equation}
\Tilde{\mathbf{x}} = \begin{bmatrix} \Tilde{x}(t_{1}), & \Tilde{x}(t_{2}), & \cdots, & \Tilde{x}(t_{n}) \end{bmatrix}^\top,  
\boldsymbol{\phi}_k = \begin{bmatrix}\phi_{k1}, & \phi_{k2}, & \cdots, & \phi_{kn} \end{bmatrix}^\top, k=1, \ldots, n.
\label{eqn:inv-dft-coef}
\end{equation}
The forward transform in Equation~\eqref{eqn:dft-formal} can be vectorized: 
\begin{equation}
    \mathbf{\Tilde{x}} = \boldsymbol{\Phi} \mathbf{x}, \quad
    \boldsymbol{\Phi} = \begin{bmatrix} \boldsymbol{\phi}_1  &  \boldsymbol{\phi}_2 & \cdots & \boldsymbol{\phi}_{n}  \end{bmatrix}^\top.
\end{equation}
The backward transform can then be determined by inverting the matrix $\boldsymbol{\Phi}$. However, due to the non-uniformity in the spacing of the time series $\mathbf{x}$, the columns of $\boldsymbol{\Phi}$ are not orthogonal, and it is not directly invertible, so the pseudo-inverse is used instead. The backward transform then has the form: $\mathbf{x} = \frac{1}{n} (\boldsymbol{\Phi}^H\boldsymbol{\Phi})^\dagger \boldsymbol{\Phi}^H \Tilde{\mathbf{x}}$,
%
%
where $\mathbf{A}^H$ and $\mathbf{A}^\dagger$ denote the complex conjugate transpose and the Moore-Penrose inverse of the matrix $\mathbf{A}$, respectively. 
The matrix $(\boldsymbol{\Phi}^H\boldsymbol{\Phi})^\dagger\boldsymbol{\Phi}^H$ projects the frequency vector $\mathbf{\Tilde{x}}$ onto the column space of $\boldsymbol{\Phi}$.  
Let $\Pi_{\boldsymbol{\Phi}}(\mathbf{x})$ denote this projection operation. 
The modulus of off-diagonal elements of $\boldsymbol{\Phi}^H\boldsymbol{\Phi}$ is bounded as: $\left|\sum_{k = 1}^n e^{j 2\pi (w_{l_1} - w_{l_2})f_k}\right|
    \le \sum_{k = 1}^n \left| e^{j 2\pi (w_{l_1} - w_{l_2})f_k}\right| = n$,
where the equality follows from the definition of the complex modulus $|{z}| = \sqrt{{z}\bar{z}}$, with $\bar{z}$ as the conjugate of the complex value $z$. The matrix $\boldsymbol{\Phi}\boldsymbol{\Phi}^H$ has the structure
\begin{equation}
    (\boldsymbol{\Phi}\boldsymbol{\Phi}^H)_{l_1, l_2} = \sum_{k = 1}^n e^{ (-1)^\delta j2\pi (f_{l_2} - f_{l_1})w_k}, \quad \delta = \mathbbm{1}(l_1 \le l_2),
    \label{eqn:toeplitz-form}
\end{equation}
for indicator function $\mathbbm{1}(l_1 \le l_2)$ , and $(\boldsymbol{\Phi}\boldsymbol{\Phi})^H_{l_1,l_2}$ denotes the value in the $l_1$-th row and $l_2$-th column of $\boldsymbol{\Phi} \boldsymbol{\Phi}^H$.
Then $\boldsymbol{\Phi}\boldsymbol{\Phi}^H$ is Toeplitz when the frequency components are uniformly-spaced such that $f_k = k$, and $\boldsymbol{\Phi}\boldsymbol{\Phi}^H$ is fully specified by its first row elements \citep{huoliu1998iterative}. 

Using these results, the following proposition asserts that the DFT preserves topological features and is stable with respect to the bottleneck distance. In particular, the bottleneck distance between the persistence diagrams of the embeddings of the noise-free and smoothed (i.e., noise-reduced) time series is bounded above by the embeddings of the observed noisy and noise-free time series.
\begin{proposition}
Given $\mathbf{x}^* \in \mathbb{R}^n$ as a possibly irregularly-spaced scalar time series with additive noise of the form $\mathbf{x}^* = \mathbf{x} + \mathbf{\varepsilon}$, where $\mathbf{x}$ is a noise-free scalar time series, and $\varepsilon$ is a zero-mean noise term, then let $\mathbf{x}^\prime$ be the time series vector after applying the proposed Fourier denoising to $\mathbf{x}^*$, and $\mathbf{F}$, $\mathbf{F}^*$, and  $\mathbf{F}^\prime$ be the embedding matrices associated with $\mathbf{x}$, $\mathbf{x}^*$, and $\mathbf{x}^\prime$, respectively. Also, let $\text{Dgm}(\mathbf{F})$ and $\text{Dgm}({\mathbf{F}^\prime})$ denote the persistence diagrams associated with the Vietoris-Rips complex constructed from $\mathbf{F}$ and $\mathbf{F}^\prime$, respectively. Then the bottleneck distance between these two persistence diagrams is bounded as
\begin{equation}
    \text{d}_\text{B}(\text{Dgm}({\mathbf{F}}), \text{Dgm}({\mathbf{F}^\prime})) \le \frac{4n - 2}{\gamma}\left(\sup\limits_{i, p}||F^*(\mathbf{s}(t_{p, i})) - F(\mathbf{s}(t_{p, i}))||_2\right),
\label{eqn:stable-denoising}
\end{equation}
where $0 < \gamma \le 1, \quad 1 \le i \le n_p - M\tau_p, \quad 1 \le p \le P.$
\label{prop:stable-denoising}
\end{proposition}
\begin{remark}
 When the samples are uniformly-spaced in both the time and frequency domain, the matrix $\boldsymbol{\Phi}$ is Hermitian with orthogonal columns, hence the factor $({2n - 1})/{\gamma}$ $(\text{not } ({4n - 2})/{\gamma}$, as the scaling by $2$ is still required$)$ is not required for the bound to hold. The factor $({2n - 1})/{\gamma}$ makes the bound conservative. However, if the denoising is well constructed, this has minimal impact on the scale of the bound. Numerical experiments in the supplementary materials Section~\ref{sec:denoising-simulation} suggest this is the case for the settings considered.
\end{remark}

\subsection{Stability of the Subsequence Embedding Method}
In this section, an error bound and stability results for the SSE approximation of the TDE are established. To simplify the notation, the embedding matrices are represented as sets where the elements of the set are the row vectors of the corresponding TDE or SSE. Also, for an embedding from a single uniformly-spaced time series, the step-size is assumed to be $\tau$. When constructing from a set of $P$ subsequences, a step-size of $\tau_p$ is assumed for the $p$-th subsequence, where $1 \le p \le P$.

Because the SSE can have fewer elements than the TDE, the following lemma addresses how to expand the SSE without affecting its topology by repeating already existing points in the SSE, so distances can be computed between the TDE and the expanded SSE.
\begin{lemma}[Topology-preserving transform]
    Let $\mathbf{F}^1$ be an embedding matrix from a uniformly-spaced time sequence of length $n$ with the form: 
    \begin{equation}
        \mathbf{F}^1= \left\{F^1(\mathbf{s}(t_{1})), F^1(\mathbf{s}(t_{2})), \cdots, F^1(\mathbf{s}(t_{n - M\tau}))  \right\} \subset \mathbb{R}^{M+1}.
    \end{equation}
    Also, let $\mathbf{F}^2$ be an embedding matrix from a set of $P$ subsequences with the form:
    \begin{equation}
        \mathbf{F}^2 = \left\{F^2(\mathbf{s}(t_{1, 1})), \cdots, F^2(\mathbf{s}(t_{1, n_1-M\tau_1})), \cdots, F^2(\mathbf{s}(t_{P, 1}), \cdots, F^2(\mathbf{s}(t_{P, n_P - M\tau_P}))  \right\} \subset \mathbb{R}^{M+1},
    \end{equation}
    where $\sum_{p = 1}^P (n_p - M\tau_p) \le n - M\tau$. Consider the set extension $\widehat{\mathbf{F}}^2 = \left\{ \mathbf{F}^2, {\mathbf{F}}^2_k \right\}$, where ${\mathbf{F}}^2_k$ is a subset of $k$ elements from $\mathbf{F}^2$. Then the persistence diagrams associated with $\mathbf{F}^2$ and $\widehat{\mathbf{F}}^2$ are identical, that is,  $\text{Dgm}(\mathbf{F}^2) \equiv \text{Dgm}(\widehat{\mathbf{F}}^2)$, and the three embedded spaces are related through the bottleneck distance as follows: $\text{d}_\text{B}(\text{Dgm}(\mathbf{F}^1), \text{Dgm}(\mathbf{F}^2)) = \text{d}_\text{B}(\text{Dgm}(\mathbf{F}^1), \text{Dgm}(\widehat{\mathbf{F}}^2)).$
    \label{lem:equivalence}
\end{lemma}

Lemma \ref{lem:equivalence} asserts that a particular set of points can be added to an embedding matrix without changing the SSE's persistence diagram. This is used to establish a bound on the SSE as an approximation to the TDE in the following proposition. In Lemma \ref{lem:equivalence}, when $k = (n - M\tau) - \sum_{p = 1}^P (n_p - M\tau_p)$, the row dimension of $\widehat{\mathbf{F}}^2$ is the same as that of $\mathbf{F}^1$. In such instances, when the interest is in a row-wise comparison of $\widehat{\mathbf{F}}^2$ and $\mathbf{F}^1$ the subsequence indexing in the time variable for any $\widehat{F}^2(\mathbf{s}(t_{p, i})) \in \widehat{\mathbf{F}}^2$ is ignored and a row is simply written as $\widehat{F}^2(\mathbf{s}(t_{i}))$, where $1 \le i \le n - M \tau$. 
\begin{proposition}
    Let $\mathbf{x}^1 $ be a uniformly-spaced time series vector of length $n$ with TDE matrix $\mathbf{F}^1$. Let $\mathbf{x}^2 \subset \mathbf{x}^1$ be a time series vector where some of the elements are missing or unobserved. Denote the SSE matrix constructed from $\mathbf{x}^2$ as $\mathbf{F}^2$. Define the extension $\widehat{\mathbf{F}}^2 = \left\{ \mathbf{F}^2, {\mathbf{F}}^2_k \right\}$, where ${\mathbf{F}}^2_k$ is a subset of $k = (n - M\tau) - \sum_{p = 1}^P (n_p - M\tau_p)$ elements from $\mathbf{F}^2$. Then the bottleneck distance between $\mathbf{F}^1$ and $\mathbf{F}^2$ is bounded as: $\text{d}_{\text{B}}( \text{Dgm}(\mathbf{F}^1), \text{Dgm}(\mathbf{F}^2) ) \le 2\sup\limits_{1 \le i \le n - M\tau}\lVert F^1(\mathbf{s}(t_i)) - \widehat{F}^2(\mathbf{s}(t_i))\rVert_2.$
    \label{prop:stability-embed}
\end{proposition}
\noindent
The choice of the $k$ subsets of embedding vectors $\mathbf{F}^2_k$ in Proposition \ref{prop:stability-embed} is arbitrary as any subset satisfies the bound. However, since they are chosen to match the subset $\{ \mathbf{F}^\prime (\mathbf{s}_{t_l}): \sum_{p = 1}^P (n_p - M\tau_p) + 1 \le l \le  n- M\tau \}$ of $\mathbf{F}^1$, the bound can be improved. The minimum bound can be attained by choosing a subset in $\mathbf{F}^2$ that has the smallest Euclidean distance to the subset $\{ \mathbf{F}^\prime (\mathbf{s}_{t_l}): \sum_{p = 1}^P (n_p - M\tau_p) +1 \le l \le  n- M\tau \}$. This is summarized as a corollary below. 
\begin{corollary}
    Let $\mathbf{x}^1 $ be a uniformly spaced time series vector of length $n$ with TDE matrix $\mathbf{F}^1$.  Let $\mathbf{x}^2 \subset \mathbf{x}^1$ be a time series vector where some of the elements are unobserved, and $\mathbf{F}^2$ its SSE matrix. Define the extension $\widehat{\mathbf{F}}^2 = \left\{ \mathbf{F}^2, {\mathbf{F}}^2_k \right\}$, where ${\mathbf{F}}^2_k$ is a random subset of $k = (n - M\tau) - \sum_{p = 1}^P (n_p - M\tau_p)$ elements from $\mathbf{F}^2$. For some $F^1 \in \mathbf{F}^1$, define the set ${\mathbf{F}}^2_{k, \text{min}}$ as follows:
   \begin{equation}
    {\mathbf{F}}^2_{k, \text{min}} = \left\{ F^2_\text{min} \in \mathbf{F}^2: \lVert F^2_\text{min} - F^1 \rVert_2 \le \lVert F^2- F^1 \rVert_2, \forall F^2 \in \mathbf{F}^2, \text{ s.t. } {F}^2_\text{min} \ne {F}^2\right\}.
  \end{equation}
    That is, ${\mathbf{F}}^2_{k, \text{min}}$ is a subset of $k$ embedding vectors in $\mathbf{F}^2$ with minimum distance to some points in $\mathbf{F}^1$. Let $\widehat{\mathbf{F}}^2_{\text{min}} = \left\{ \mathbf{F}^2, {\mathbf{F}}^2_{k, \text{min}} \right\}$, then it follows that
    \begin{equation}
        \sup\limits_{1 \le i \le n}\lVert F^1(\mathbf{s}(t_i)) - \widehat{F}^2_\text{min}(\mathbf{s}(t_i))\rVert_2 \le \sup\limits_{1 \le i \le n}\lVert F^1(\mathbf{s}(t_i)) - \widehat{F}^2(\mathbf{s}(t_i))\rVert_2,
    \end{equation}
    where $F^1(\mathbf{s}(t_i)) \in \mathbf{F}^1$, $\widehat{F}^2(\mathbf{s}(t_i)) \in \widehat{\mathbf{F}}^2$, and $\widehat{F}^2_\text{min}(\mathbf{s}(t_i)) \in \widehat{\mathbf{F}}^2_\text{min}$.
    \label{corr:stability-embed-corr}
\end{corollary}

\noindent
An immediate consequence of Proposition \ref{prop:stability-embed} and Corollary \ref{corr:stability-embed-corr} is that the SSE matrix approximates the TDE for varying level of regularity. In particular, for a time series $\mathbf{x} = \{ x(t): t \in \mathcal{T}\}$, and $\mathcal{T} = \{ t_1, \cdots, t_n \} \subset \mathbb{N}$, the sequence of embedding matrices for each $r$ where $1 \le r \le t_n - t_1$ is finite. Hence for a fixed $n$, the limiting persistence diagram as $r \xrightarrow[]{} 1$ is close to the TDE's persistence diagram. If $r = 1$, the SSE is exactly the same as the TDE, and the persistence diagrams would be identical. This reinforces the fact that the proposed reconstruction preserves the topological structures more accurately as the level of irregularity in the observed time series decreases. A more formal treatment of these observations is presented next.

\subsection{Convergence Results}
Throughout this section, it is assumed that the number of missing values increases at a slower rate than the sample size of the time series. Specifically, the missing values grows at a rate of $o(\log m)$, for sample size $m$. This assumption and others are formalized as follows.

Let $\mathbf{x}_1, \mathbf{x}_2, \cdots,$ be irregularly-spaced time series vectors where  $|\mathbf{x}_i| < |\mathbf{x}_j|$, $i < j$. Denote by $\mathbf{F}_m$, the SSE associated with $\mathbf{x} \in \{ \mathbf{x}_1, \mathbf{x}_2, \cdots, \}$, i.e., 
\begin{equation}
\mathbf{F}_m = \left\{F(\mathbf{s}(t_{1})), F(\mathbf{s}(t_{2})), \cdots, F(\mathbf{s}(t_{m}))  \right\}.
\end{equation}
Note the correspondence between the subscript $m$ and the number of points in $\mathbf{F}_m$. $\mathbf{F}_m$ depends on which time series is selected; however, indexing over this selection is not needed for the following results. 
Recall that $\mathbf{F}_m$ is a compact subset of $(\mathbb{R}^{M+1}, \lVert \cdot \rVert_2)$. Let the space $(\mathbb{R}^{M+1}, \lVert \cdot \rVert_2)$ be endowed with the unknown probability measure $\vartheta$ such that the set of $m$ time points are randomly sampled according to $\vartheta$. Let $\vartheta$ be supported on an embedding $\mathbf{F}_\vartheta$, and let $\varphi$ be the associated density function. 
Consider the following set of assumptions.
\begin{itemize}
    \item[\textbf{A1.}] The sample size increases such that $\mathbf{x}_i \subset \mathbf{x}_j$ whenever $i < j$.
    \item[\textbf{A2.}] Let $\varepsilon_m(r)$ be a function of $m$ and the regularity score $r$ such that $\varepsilon_m(1) \xrightarrow{} 0$ as $m \xrightarrow{} \infty$. 
    \item[\textbf{A3.}] For any point $F_\vartheta \in \mathbf{F}_\vartheta$, $\vartheta(B(F_\vartheta, \delta)) \ge \min(\kappa\delta^{M+1}, 1)$, where $B(F_\vartheta, \delta)$ is a closed ball of radius $\delta > 0$ around $F_\vartheta$, with constant $\kappa>0$.
    \item[\textbf{A4.}] It is possible to create joint distributions based on the marginals of $\mathbf{F}_m$ that satisfy
    $$
    \sup_{\mathbf{F}_m} \left| \frac{\varphi\left(F(\mathbf{s}(t_{1})), \cdots, F(\mathbf{s}(t_{m})) \right) - \varphi\left(F(\mathbf{s}(t_{1}))\right)\times \cdots\times \varphi\left(F(\mathbf{s}(t_{m}))\right) }{\varphi\left(F(\mathbf{s}(t_{1}))\right) \times  \cdots \times \varphi\left(F(\mathbf{s}(t_{m}))\right)} \right| \le \eta_m,
    $$
    where $\eta_m$ is such that $\sum_{m = 1}^\infty \frac{\eta_m}{m^\beta\log(m)} < \infty$ for any $\beta > 1$. 
\end{itemize}
Assumption \textbf{A4} is to address the possible lack of independence of the vectors in $\mathbf{F}_m$. Under this assumption, the dependence can be controlled and the vectors in $\mathbf{F}_m$ are regarded as the so-called $\eta_m$-almost independent samples, which allows for $\mathbf{F}_m$ to converge in Hausdorff distance to $\mathbf{F}_\vartheta$ \citep{aaron2017detection,picado2020denoising}.
The SSE matrix $\mathbf{F}_m$ can be regarded as an estimator of $\mathbf{F}_\vartheta$ and convergence results can be established in the context of assumptions \textbf{A1}-\textbf{A4}. These results are analogous to convergence results established on support estimation of $d$-dimensional sets \citep{cuevas2004boundary}, its generalization to metric spaces, and on the space of persistence diagrams \citep{mileyko2011probability,chazal2014convergence}. The following result gives the rate of convergence in estimating $\mathbf{F}_\vartheta$. 
\begin{theorem}
    Let $\mathbf{x}_1, \mathbf{x}_2, \cdots,$ be a sequence of irregularly-spaced time series vectors satisfying assumption \textbf{A1}, and  $\mathbf{F}_m = \left\{F(\mathbf{s}(t_{1})), F(\mathbf{s}(t_{2})), \cdots, F(\mathbf{s}(t_{m}))  \right\} \subset \mathbb{R}^{M+1}$ be the SSE associated with some $\mathbf{x} \in \{ \mathbf{x}_1, \mathbf{x}_2, \cdots,\}$, satisfying assumption \textbf{A2}. If the probability measure $\vartheta$ satisfies assumption \textbf{A3} and \textbf{A4}, then with probability one,
    \begin{equation}
        \lim_{m \xrightarrow{}\infty} \sup \left(\varepsilon_m(r)\right)^{-\frac{1}{M+1}} \text{d}_{H}(\mathbf{F}_m, \mathbf{F}_\vartheta) \le K,
    \end{equation}
    where $K$ is a constant  depending on $\kappa$ and the embedding dimension $M+1$.
    \label{prop:hd-conv}
\end{theorem}

From the stability relation in Equation \eqref{eqn:stability-bn}, the Hausdorff metric can be replaced with the Gromov-Hausdorff metric and the results still holds. This also gives a similar convergence results on the space of persistence diagrams with respect to the bottleneck distance and is summarized as:
\begin{corollary}
     Let $\mathbf{x}_1, \mathbf{x}_2, \cdots,$ be a sequence of irregularly-spaced time series vectors satisfying assumption \textbf{A1}, and let $\mathbf{F}_m = \left\{F(\mathbf{s}(t_{1})), F(\mathbf{s}(t_{2})), \cdots, F(\mathbf{s}(t_{m}))  \right\} \subset \mathbb{R}^{M+1}$ be the SSE associated with some $\mathbf{x} \in \{ \mathbf{x}_1, \mathbf{x}_2, \cdots,\}$, satisfying assumption \textbf{A2}. If the probability measure $\vartheta$ satisfies assumption \textbf{A3} and \textbf{A4}, then with probability one,
    \begin{equation}
        \lim_{m \xrightarrow{}\infty} \sup \left(\varepsilon_m(r)\right)^{-\frac{1}{M+1}} \text{d}_{B}(\text{Dgm}(\mathbf{F}_m), \text{Dgm}(\mathbf{F}_\vartheta)) \le K,
    \end{equation}
    where $K$ is a a constant  depending on $\kappa$ and the embedding dimension $M+1$.
    \label{eqn:bn-conv}
\end{corollary}

\section{Numerical Studies}
\label{sec:simulation-application}
This section presents numerical studies that shows the performance of the proposed SSE method.

\subsection{Reconstruction Accuracy} \label{sec:reconstruction_accuracy}
\begin{figure}[ht]
     \centering
     \begin{subfigure}[b]{0.49\textwidth}
         \centering
         \includegraphics[width=\textwidth, height=0.55\textwidth]{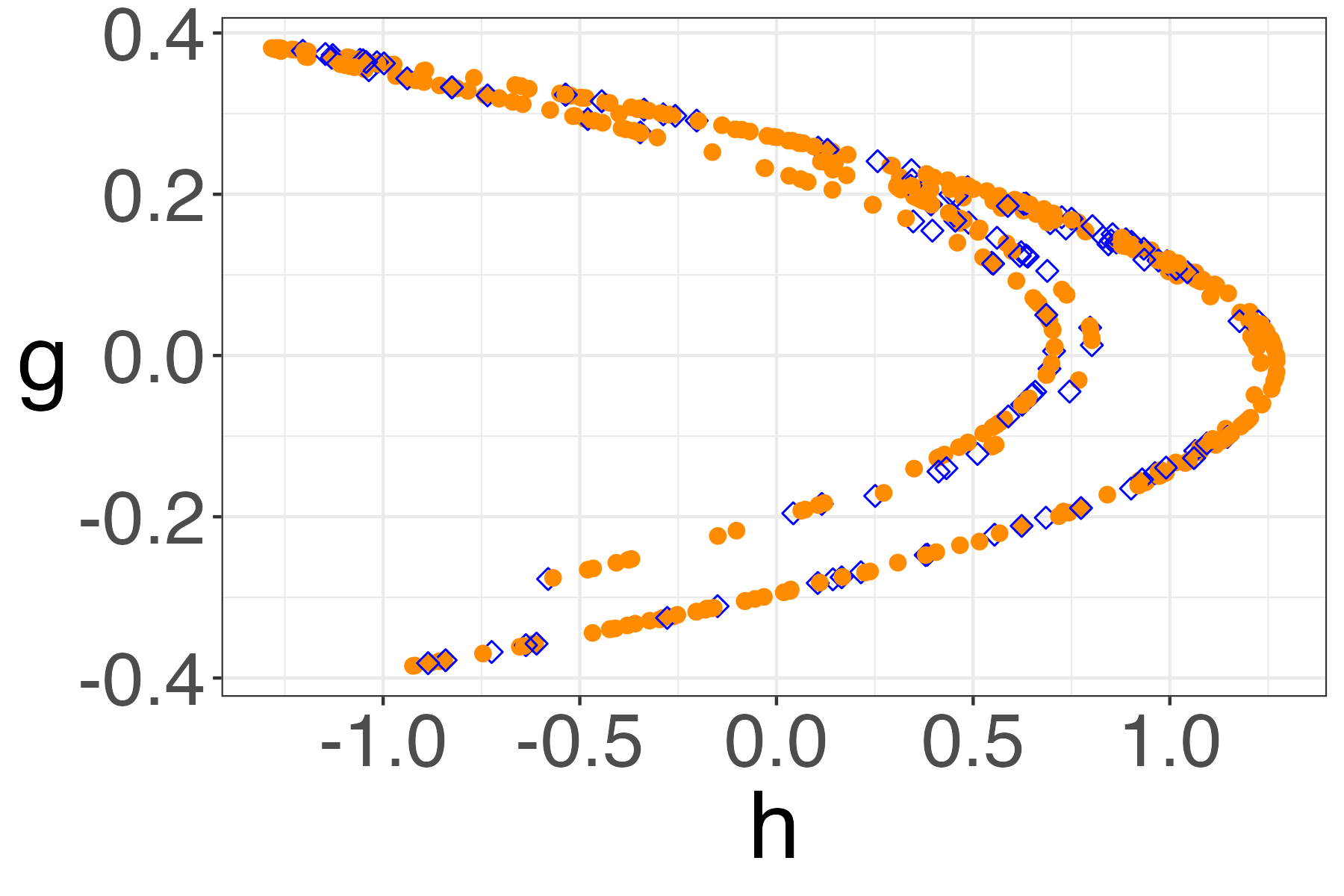}
         \caption{H{\'e}non map}
         \label{fig:henon-map-f}
     \end{subfigure}
     \hfill
     \begin{subfigure}[b]{0.49\textwidth}
        \centering
         \includegraphics[width=\textwidth, height=0.55\textwidth]{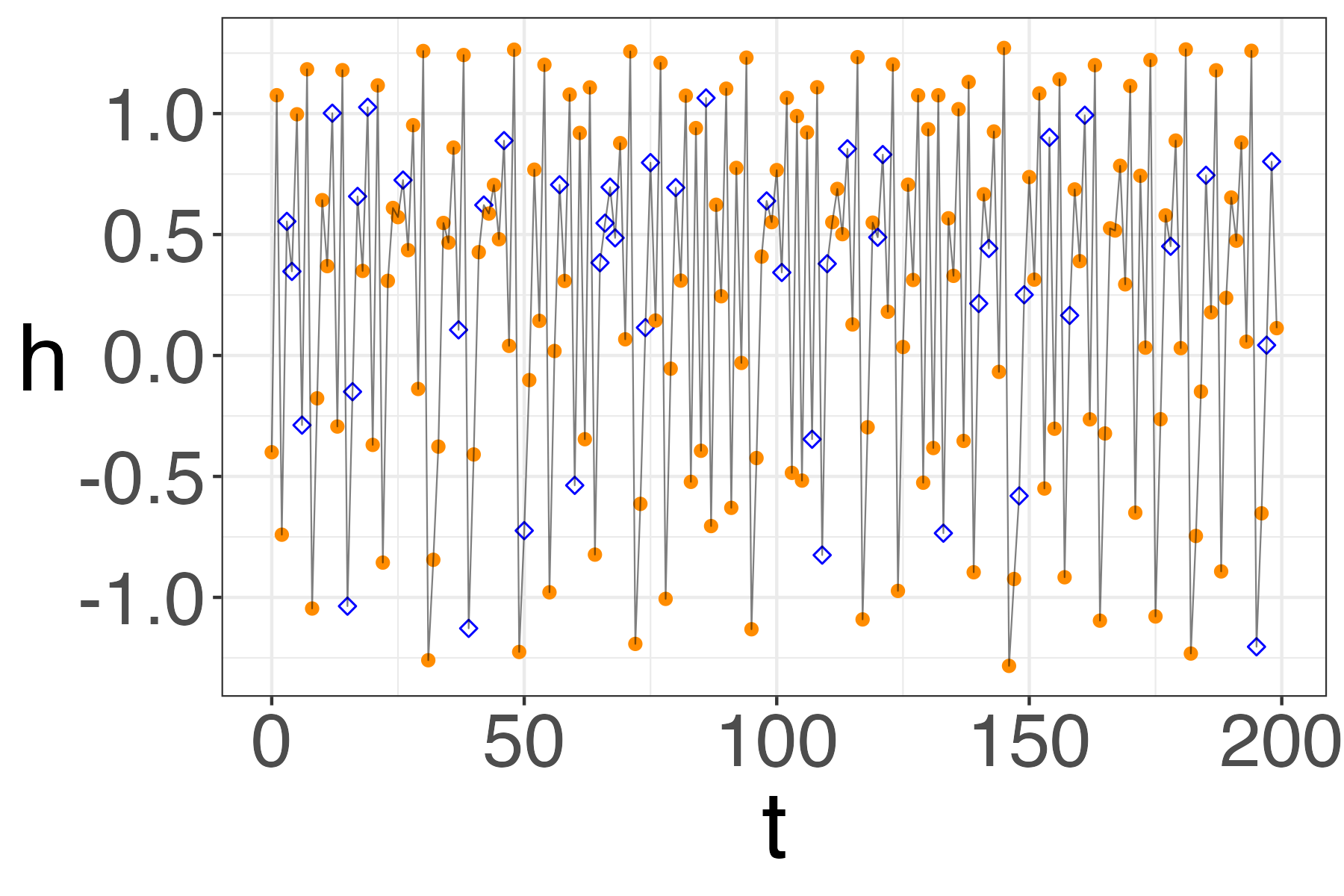}
         \caption{The $h$ dimension}
         \label{fig:henon-map-dimension-h}
     \end{subfigure}
    \caption{The H{\'e}non map used in assessing reconstruction accuracy. (a) The H{\'e}non map with 500 points (blue and orange) where the blue diamonds are designated as missing. (b) The $h$-dimension of the H{\'e}non map; only 200 points are displayed for visual clarity.}
    \label{fig:henon-map}
\end{figure}

%
This simulation assesses the SSE method's effectiveness in preserving the original state space geometry using the H{\'e}non map as an illustrative example \citep{henon2004two}. The H{\'e}non map recursively maps a point $(h_t, g_t) \in \mathbb{R}^2$ as follows: $h_{t+1} = 1 - ah_t^2 + g_t$,  $g_{t+1} = bh_t$, with $a = 1.4$ and $b = 0.3$, their classical values. The map is initialized at $(h_0, g_0) = (0, 0)$, and simulated with $500$ points with observations designated as missing with a given probability. Figure \ref{fig:henon-map} shows the 2D H{\'e}non map and the corresponding time series for one dimensions. The measurement function (see Section \ref{sec:basic-tde}) extracts observations along the  $h$-dimension, hence $\{h_t\}$ are used to reconstruct the space. Observations along the $g$-dimension could be used instead.

The {\em correlation dimension} is used to assess how well the geometry of the original state space is preserved in the reconstruction. Specifically, for a given $\varepsilon > 0$, it measures the probability that two random points in a space are within $\varepsilon$-distance of each other. To compute the correlation dimension, the correlation sum is computed using the following:
\begin{equation}
    \text{Corr}(\varepsilon) = \lim_{m \xrightarrow{} \infty} \frac{2}{m(m-1)}\sum_{i = 1}^m \sum_{j = i+1}^m \mathbbm{1}\left(\lVert F(\mathbf{s}(t_i)) - F(\mathbf{s}(t_j))\rVert_2 \le \varepsilon \right),
\end{equation}
for some embedding map $\mathbf{F} = \{ F(\mathbf{s}({t_1})), \cdots, F(\mathbf{s}({t_m})) \}$. Then the correlation dimension is estimated as: $\lim_{\varepsilon \xrightarrow{} 0 } {\log(\text{Corr}(\varepsilon))}/{\log(\varepsilon)}.$
%
%
If the reconstructed space preserves relevant geometrical invariants, its correlation dimension should match that of the true state space. Other accuracy measures include the box-counting dimension, Hausdorff dimension, and information dimension. However, the correlation dimension is more robust to sample size, making it less noisy with fewer samples \citep{grassberger1983measuring}.

The SSE method is compared to common statistical interpolation methods used to impute missing data. A range of methods were considered\footnote{The comparison methods considered are available in the the R package, {\em imputeTS} \citep{moritz2017imputets}:  {\em linear}, {\em spline}, and {\em Stineman} interpolation methods, {\em Kalman Smoothing} with a structural model and autoregressive integrated moving average model, a {\em moving average} method with exponential and linear weighting, {\em seasonal decomposition} (imputation by interpolation is done on the deseasonalized component), {\em seasonal split} (imputation by interpolation is done on each split), imputing with the {\em previous observation} (LOCF) or {\em next observation} (NOCB), imputing with the {\em mean}, {\em median}, {\em mode}, and by a {\em random point} in the dataset. 
}, but only the best three methods are presented, which were implemented using the R package, {\em imputeTS} \citep{moritz2017imputets}: \\
\noindent \textbf{(1) Kalman Smoothing (KS)}: This fits a structural time series model via maximum likelihood, using the linear local trend as the structural class (see referenced package for more details). \\
\noindent \textbf{(2) Last Observation Carried Forward (LOCF)}: This methods replaces each missing value with the most immediate prior observed value.  \\
\noindent \textbf{(3) Next Observation Carried Backward (NOCB)}: This is similar to the LOCF, but instead replaces each missing value with the most immediate \emph{next} observed value. 
%

The results are presented in terms of the correlation dimension, with standard errors generated by applying each method to 100 independently generated instances of the H{\'e}non map. A noise model (with no missing values) served as a baseline, with observations from a normal distribution (mean zero) and standard deviation equal to the probability of observing a missing value.

\begin{figure}[ht!]
     \centering
     \begin{subfigure}[b]{0.325\textwidth}
         \centering
         \includegraphics[width=\textwidth]{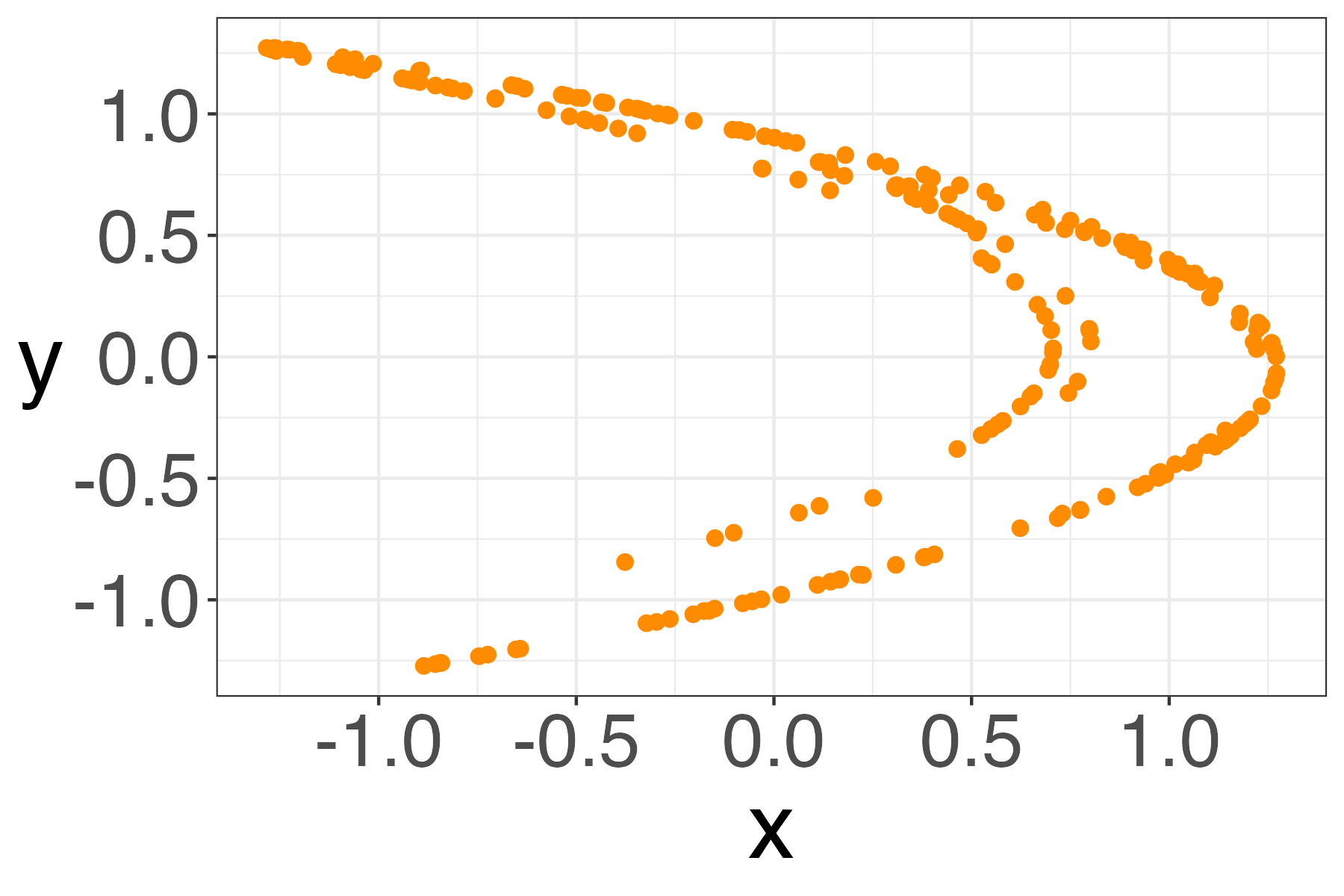}
         \caption{SSE}
         \label{fig:henon-map-reconstructed}
     \end{subfigure}
     \begin{subfigure}[b]{0.325\textwidth}
        \centering
         \includegraphics[width=\textwidth]{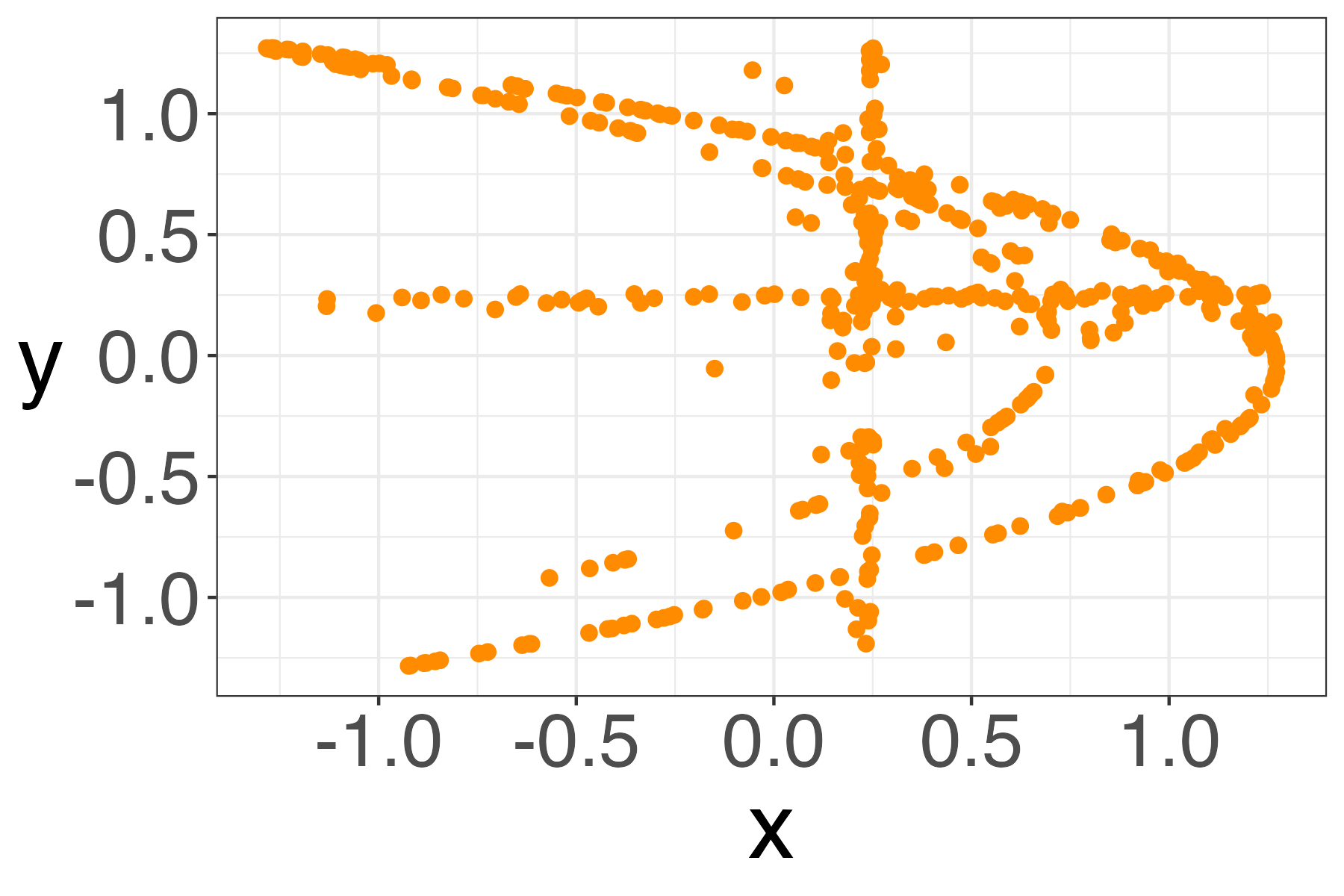}
         \caption{KS imputation}
         \label{fig:henon-map-reconstructed-kalman}
     \end{subfigure}
     \begin{subfigure}[b]{0.325\textwidth}
         \centering
         \includegraphics[width=\textwidth]{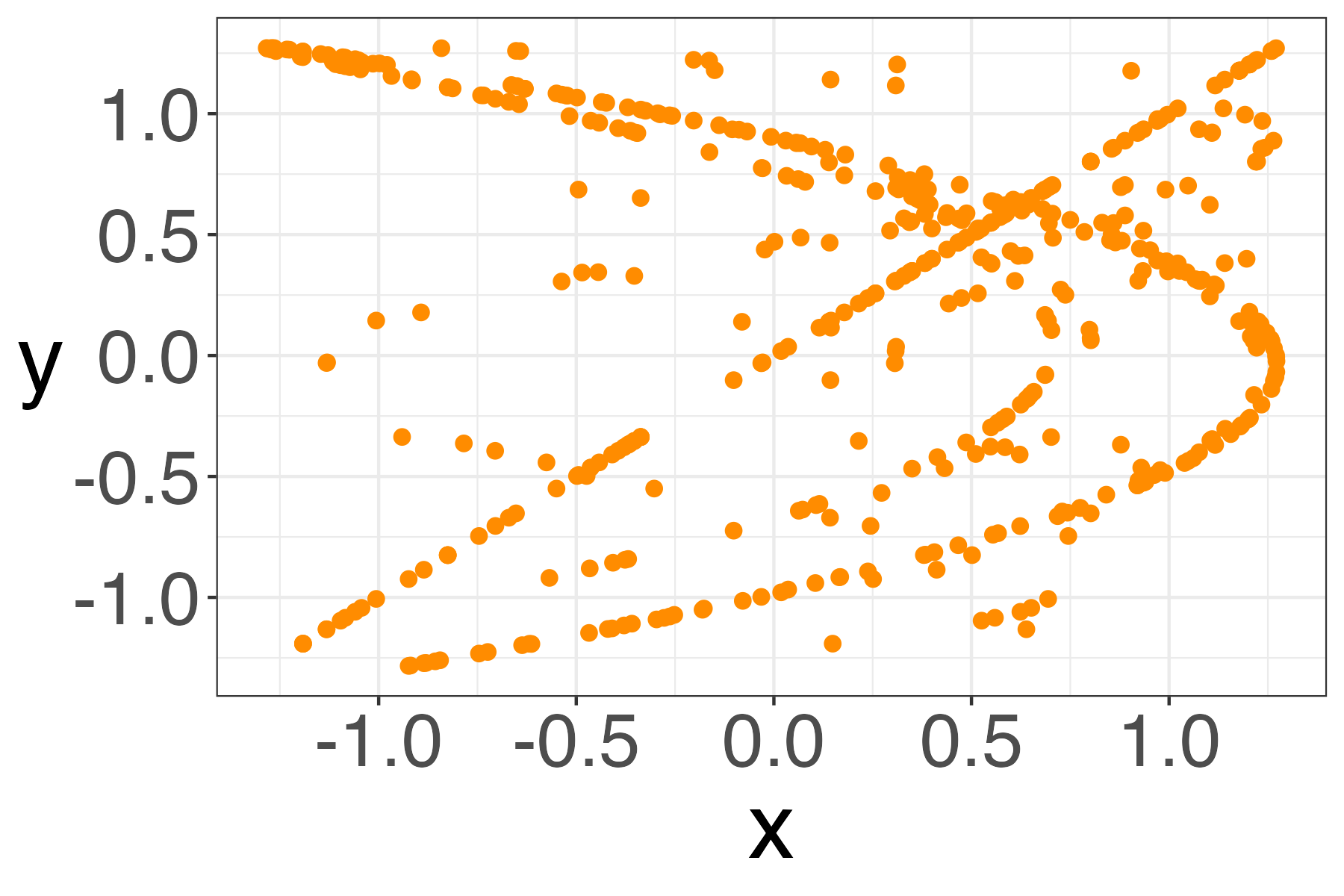}
         \caption{LOCF imputation}
         \label{fig:henon-map-reconstructed-locf}
     \end{subfigure}
    \caption{Reconstructed state spaces of the Hénon map for: (a) proposed SSE method, (b) KS imputation, and (c) LOCF imputation.}
    \label{fig:henon-map-reconstructions}
\end{figure}

%
Figure \ref{fig:henon-map-reconstructions} shows example reconstructed spaces using the proposed method and two imputation methods (the NOCB result is nearly identical to the LOCF and is not shown) with $500$ samples and a $0.25$ missingness probability. Note that for the comparison methods, after imputation the TDE method is used to estimate the state space. Only the SSE method preserves the original geometry, while the imputation methods introduce extraneous features.

\begin{figure}[t!]
 \centering
 \includegraphics[width=0.5\linewidth,clip=true]{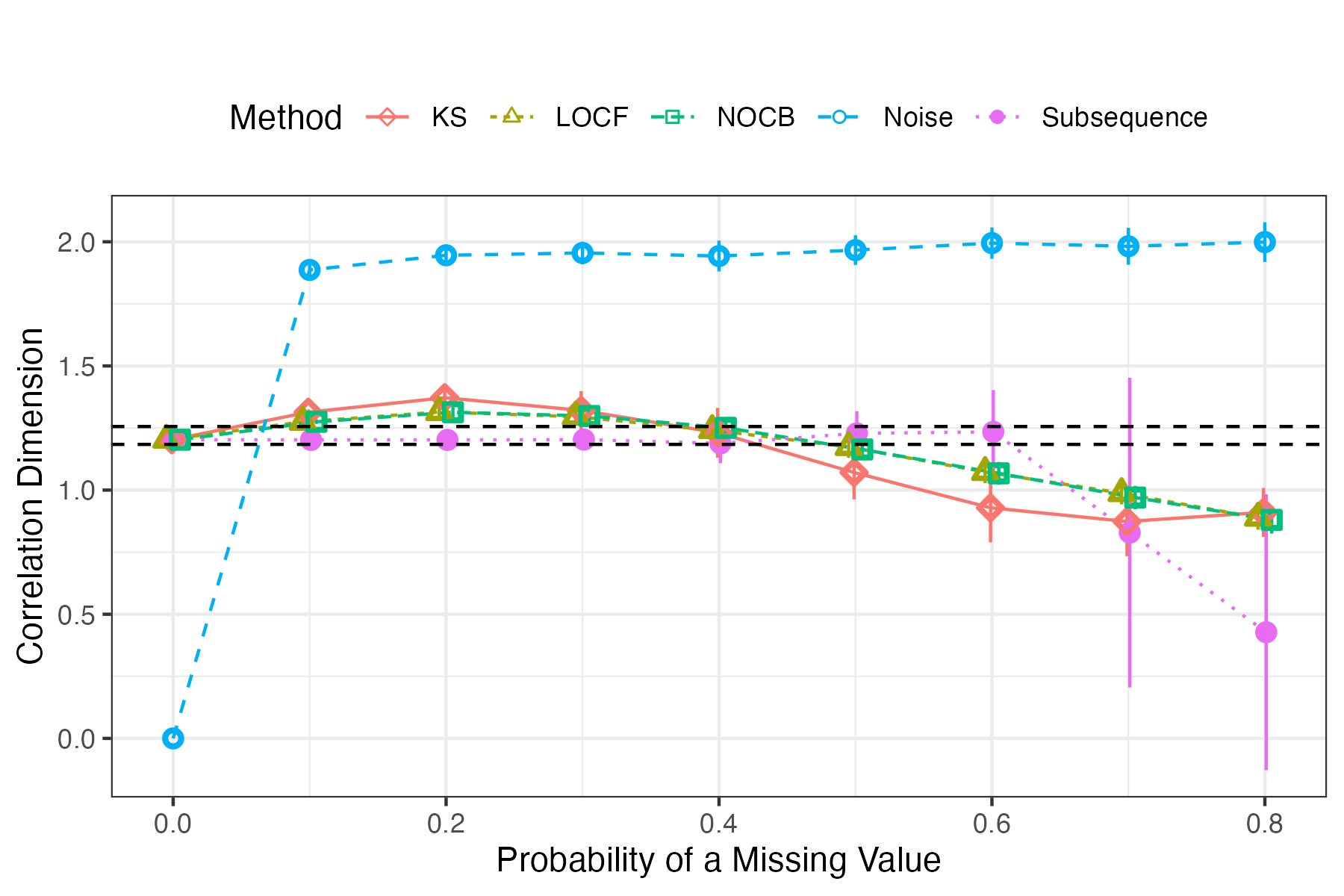}
 \caption{Reconstruction accuracy results of the H{\'e}non map based on the correlation dimension.  The points in different shapes are the mean correlation dimension after 100 repetitions using the proposed SSE method (pink dotted), the three imputation methods, and a baseline noise model (blue dashed), and the vertical bars represent the corresponding standard errors. The black dashed lines indicate the established empirical bounds of the H{\'e}non map.  } 
 \label{fig:correlation-dimension-results}
\end{figure}
Figure \ref{fig:correlation-dimension-results} shows the correlation dimension versus missingness probability for the SSE method and the three imputation methods. The black dashed lines indicate the established empirical estimate for the H{\'e}non map's correlation dimension $(1.22 \pm 0.04)$ \citep{grassberger1983characterization,sprott2001improved}. The SSE method performs well up to a missingness probability of $0.6$, staying within or close to the empirical bounds. Beyond $0.6$, its comparable to the three imputation methods. However, the SSE method is more variable due to the fewer points used to compute the correlation dimension compared to the other methods (which always have 500 points).

\subsection{Periodicity Quantification} \label{sec:period}
The periodicity of a time series can be quantified based on the $H_1$ features in the persistence diagram. This relies on the idea that periodic patterns yields elliptic curves in the reconstructed state space, and the roundness of the curves is an indicator of the periodicity in the time series \citep{perea2015sliding}. The roundness of these ellipses can be quantified by examining the maximum persistence of their associated $H_1$ features. For a time series vector $\mathbf{x}$ with its embedding map $\mathbf{F}$, its periodicity score $\text{ps}(\mathbf{x})$ can be defined as \citep{perea2015sliding}:
\begin{equation}
    \text{ps}(\mathbf{x}) = \max_{ {(b, d)} \in \text{Dgm}(\mathbf{F})} (d - b)/{\sqrt{3}},
    \label{eqn:periodicity-score}
\end{equation}
where $\text{Dgm}(\mathbf{F})$ is the persistence diagram, and $\max_{ {(b, d)} \in \text{Dgm}(\mathbf{F})} (d - b)$ is restricted to the $H_1$ features.
%
%
For this calculation, the embedding map $\mathbf{F}$ is pointwise-centered and scaled. The motivation for the periodicity score is that during the VR filtration for a dataset with a large enough sample size, a unit circle ($H_1$ feature) dies when an inscribed equilateral triangle appears in the VR complex at filtration value $\sqrt{3}$, hence the maximum periodicity score of one is realized when either the TDE or SSE spaces has a well-sampled circle; a $\text{ps}(\mathbf{x})$ closer to one  indicates a stronger periodic signal in $\mathbf{x}$.

 %
 
 \subsubsection{Simulation}
\begin{figure}[ht!]
     \centering
     \begin{subfigure}[b]{0.4\textwidth}
         \centering
         \includegraphics[width=\textwidth, height=0.55\textwidth]{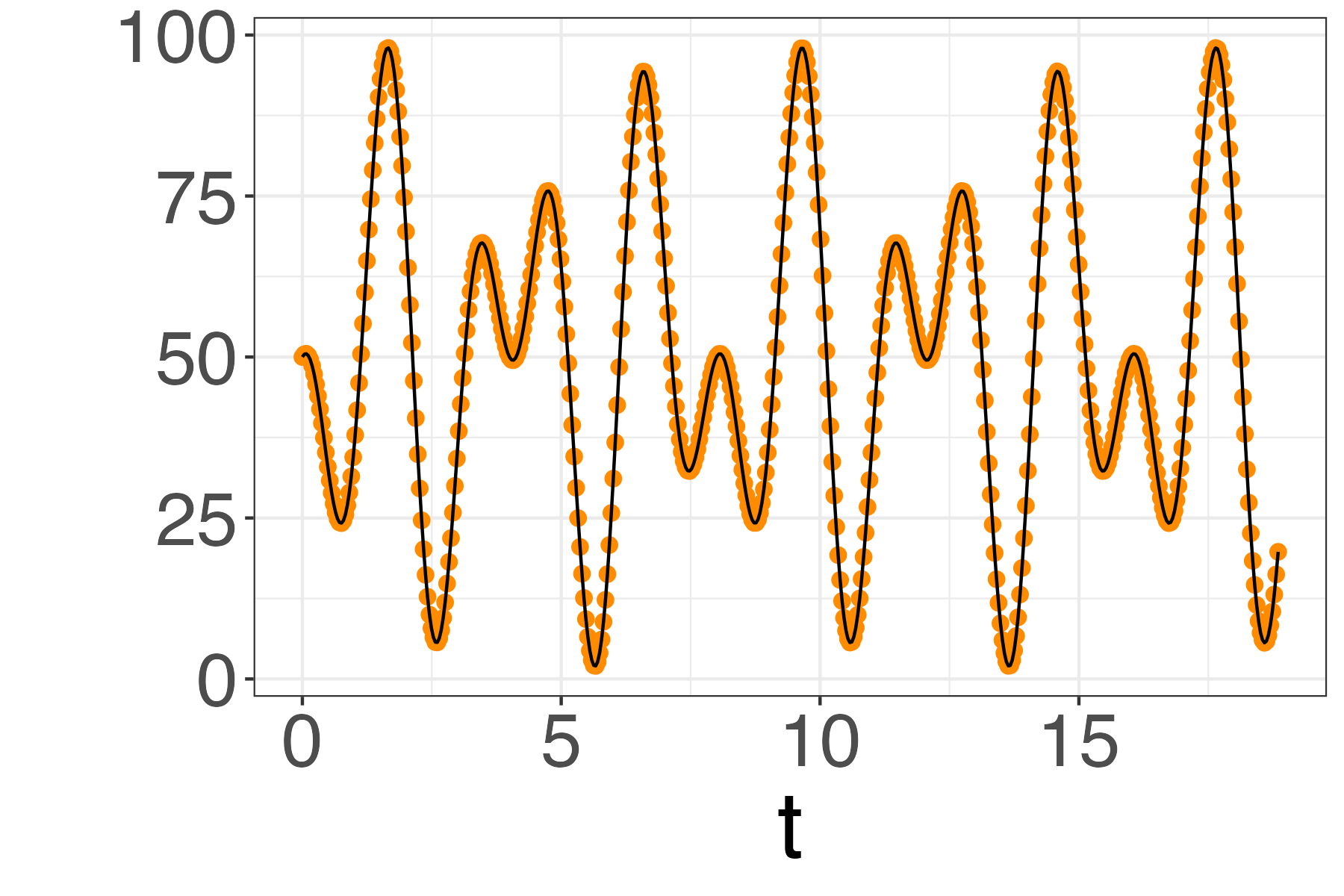}
         \caption{Periodic signal}
         \label{fig:periodic-signal-perioidicity}
     \end{subfigure}
     \begin{subfigure}[b]{0.4\textwidth}
        \centering
         \includegraphics[width=\textwidth, height=0.55\textwidth]{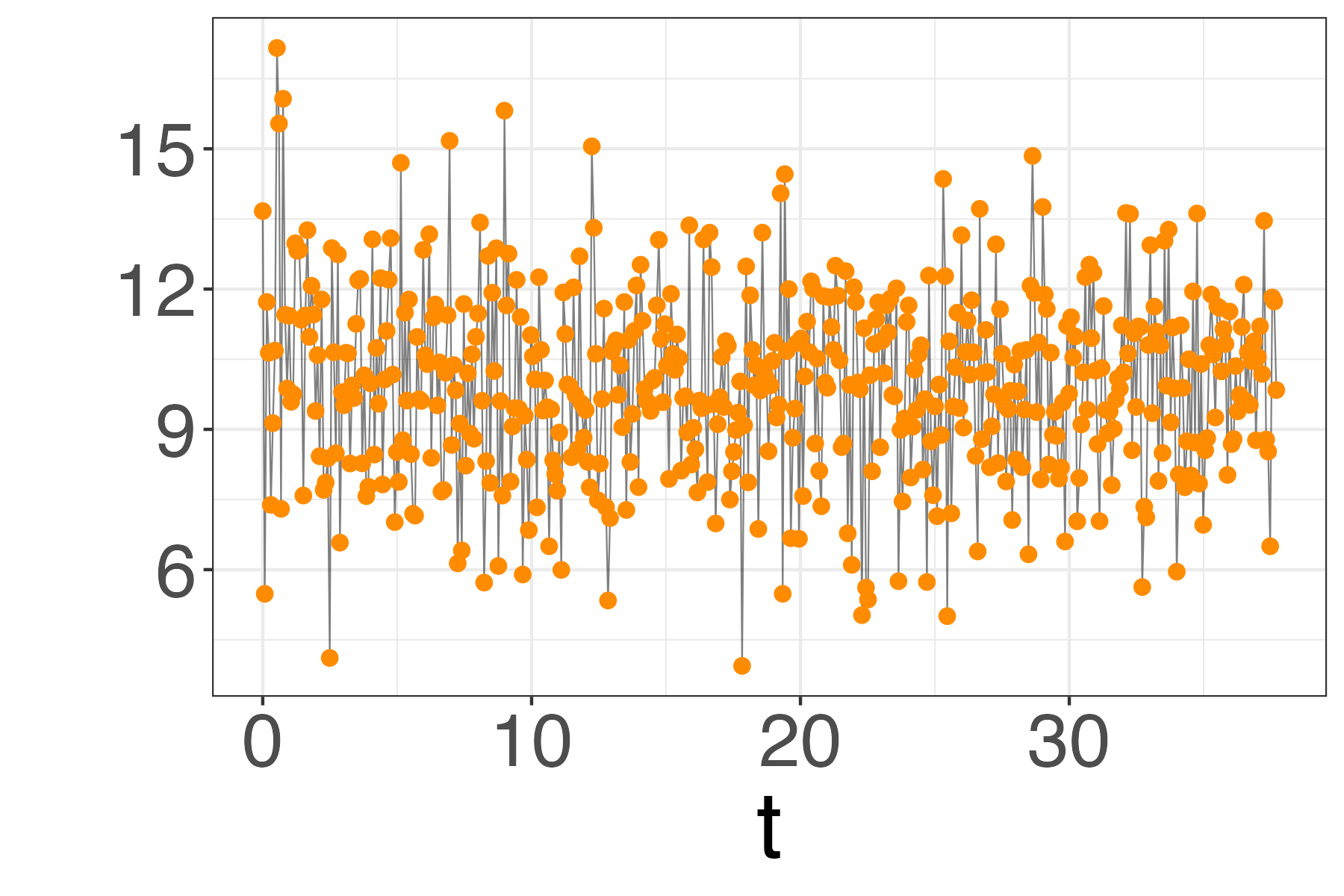}
         \caption{Non-periodic signal}
         \label{fig:noisy-signal-perioidicity}
     \end{subfigure}
    \caption{Sample periodic (a) and non-periodic (b)  signals used in the periodicity quantification simulation of Section~\ref{sec:period}. Each time series include 500 time points.}
    \label{fig:periodicity-illustration}
\end{figure}

To evaluate this framework, two different set of signals were generated with sample sizes $n \in \{50, 100, 500, 1000\}$ and missingness probabilities $\{0, 0.1, 0.2, 0.3, 0.4\}$. The first set follows $f(t) = 50\times\cos(\pi t/4 - \lambda \pi)\times\sin({\pi t}/{2}) + 50$ with $\lambda \in (0, 1)$ and $t \in [0, 12\pi]$, having a longest period of $8$. The second set is a non-periodic signals drawn from a $N(10,2)$. Figure \ref{fig:periodicity-illustration} shows samples of both signals.
To construct the embedding from the time series, the time points are rescaled to integers and the step-size of $\tau = 1$. 
The periodicity score $\text{ps}(\mathbf{x})$ is then compared to those obtained from the Lomb-Scargle periodogram method for both uniformly-spaced and irregularly-spaced observations \citep{lomb1976least,scargle1982studies,ruf1999lomb}, the sliding windows method \citep{perea2015sliding}, and the $\text{JTK}\_\text{Cycle}$ algorithm for uniformly-spaced samples \citep{hughes2010jtk_cycle}.

%
%
The results are summarized in Table \ref{tab:SimulationStudy-1} (periodic signal) and in Table \ref{tab:SimulationStudy-2} (non-periodic signal). Table \ref{tab:SimulationStudy-1} shows that all the methods rate the periodic signals as highly periodic with increasing sample size. The proposed SSE method consistently identifies a distinct $H_1$ across all sample sizes and missing observations despite noisy features in the persistence diagram. The JTK$\_$Cycle and the Lomb-Scargle method requires specifying a period search range. The proposed SSE method has the added advantage that its periodicity score has a geometric interpretation \citep{perea2015sliding}. 
\begin{table}[ht!]
\caption{Results for the periodic signal summarized as p-values for $\text{JTK}\_\text{Cycle}$ and Lomb-Scargle with estimated period in brackets, and as periodicity scores for Sliding Windows and SSE methods.}
\renewcommand{\arraystretch}{0.5}%
\centering
\begin{tabular}{ccccccc}
    \hline
  n & $\pi$ & $M$ &Sliding Windows & $\text{JTK}\_\text{Cycle}$ & Lomb-Scargle & Proposed SSE Method \\ \hline
  \multirow{4}{*}{ 50 }& $0.00$ & $2$ & $0.74$ & $0.00$ $(7.69)$ & $0.00$  $(8.02)$ & $0.74$\\
  & $0.10$ & $2$ & $-$ & $-$ & $0.00$ $(8.03)$ & $0.74$\\
  & $0.20$ & $2$ & $-$ & $-$ & $0.00$ $(8.03)$ & $0.70$\\
  & $0.30$ & $2$ & $-$ & $-$ & $0.00$ $(8.04)$ & $0.67$\\ 
  & $0.40$ & $2$ & $-$ & $-$ & $0.00$ $(8.04)$ & $0.44$\\ 
  \cline{1-7}
  \multirow{5}{*}{ 100 } & $0.00$ & $6$ & $0.53$ & $0.00$ $(8.00)$ & $0.00$ $(8.02)$ & $0.53$\\
  & $0.10$ & $6$ & $-$ & $-$ & $0.00$ $(8.02)$ & $0.53$\\
  & $0.20$ & $6$ & $-$ & $-$ & $0.00$ $(8.02)$ & $0.53$\\
  & $0.30$ & $4$ & $-$ & $-$ & $0.00$ $(8.20)$ & $0.49$\\
  & $0.40$ & $3$ & $-$ & $-$ & $0.00$ $(8.02)$ & $0.43$\\
  \cline{1-7}
  \multirow{6}{*}{ 500 } & $0.00$ & $8$ & $0.93$ & $0.00$ $(2.64)$ & $0.00$ $(8.02)$ & $0.93$\\
  & $0.10$ & $8$ & $-$ & $-$ & $0.00$ $(8.02)$ & $0.85$\\
  & $0.20$ & $6$ & $-$ & $-$ & $0.00$ $(8.02)$ & $0.71$\\
  & $0.30$ & $2$ & $-$ & $-$ & $0.00$ $(8.02)$ & $0.63$\\
  & $0.40$ & $2$ & $-$ & $-$ & $0.00$ $(8.02)$ & $0.60$\\
  \cline{1-7}
  \multirow{8}{*}{ 1000 } & $0.00$ & $26$ & $0.90$ & $0.00$ $(1.28)$ & $0.00$ $(8.02)$ & $0.90$\\
  & $0.10$ & $15$ & $-$ & $-$ & $0.00$ $(8.02)$ & $0.74$\\
  & $0.20$ & $12$ & $-$ & $-$ & $0.00$ $(8.02)$ & $0.69$\\
  & $0.30$ & $6$ & $-$ & $-$ & $0.00$ $(8.02)$ & $0.44$\\
  & $0.40$ & $4$ & $-$ & $-$ & $0.00$ $(8.01)$ & $0.43$\\ \hline
\end{tabular}
\label{tab:SimulationStudy-1}
\end{table}
\begin{table}[ht!]
\caption{Results for the non-periodic signal are given as p-values for $\text{JTK}\_\text{Cycle}$ and Lomb-Scargle with estimated period in brackets, and as periodicity scores for Sliding Windows and SSE methods.}
\renewcommand{\arraystretch}{0.5}%
\centering
\begin{tabular}{ccccccc}
    \hline
    n & $\pi$ & $M$ & Sliding Windows & $\text{JTK}\_\text{Cycle}$ & Lomb-Scargle & Proposed SSE Method \\ \hline
  \multirow{4}{*}{ 50 } & $0.00$ & $3$ & $0.30$ & $1.00$ $(13.29)$  & $0.15$ $(2.05)$ & $0.30$\\
  & $0.10$ & $3$  & $-$ & $-$ & $0.16$ $(2.04)$ & $0.32$\\
  & $0.20$ & $3$  & $-$ & $-$ & $0.28$ $(2.04)$ & $0.23$\\ 
  & $0.30$ & $3$  & $-$ & $-$ & $0.17$ $(2.04)$ & $0.25$\\ 
  & $0.40$ & $3$  & $-$ & $-$ & $0.41$ $(32.87)$ & $0.14$\\ 
  \cline{1-7}
  \multirow{5}{*}{ 100 } & $0.00$ & $3$ & $0.26$ & $1.00$ $(12.88)$ & $0.10$ $(1.00)$ & $0.26$\\
  & $0.10$ & $3$ & $-$ & $-$ & $0.08$ $(1.00)$ & $0.22$\\
  & $0.20$ & $3$ & $-$ & $-$ & $0.27$ $(1.00)$ & $0.25$\\
  & $0.30$ & $3$ & $-$ & $-$ & $0.55$ $(16.39)$ & $0.32$\\
  & $0.40$ & $3$ & $-$ & $-$ & $0.33$ $(16.39)$ & $0.29$\\
  \cline{1-7}
  \multirow{6}{*}{ 500 } & $0.00$ & $11$ & $0.14$ & $0.28$ $(0.45)$ & $0.11$ $(0.15)$ & $0.14$\\
  & $0.10$ & $9$ & $-$ & $-$ & $0.23$ $(0.80)$ & $0.16$\\
  & $0.20$ & $0$ & $-$ & $-$ & $0.13$ $(0.20)$ & $0.10$\\
  & $0.30$ & $7$ & $-$ & $-$ & $0.29$ $(0.79)$ & $0.18$\\
  & $0.40$ & $3$ & $-$ & $-$ & $0.19$ $(0.45)$ & $0.28$\\
  \cline{1-7}
  \multirow{8}{*}{ 1000 } & $0.00$ & $3$ & $0.13$ & $1.00$ $(1.28)$ & $0.87$ $(0.25)$ & $0.13$\\
  & $0.10$ & $3$ & $-$ & $-$ & $0.50$ $(0.40)$ & $0.14$\\
  & $0.20$ & $3$ & $-$ & $-$ & $0.55$ $(0.26)$ & $0.15$\\
  & $0.30$ & $3$ & $-$ & $-$ & $0.22$ $(0.26)$ & $0.16$\\
  & $0.40$ & $3$ & $-$ & $-$ & $0.66$ $(0.26)$ & $0.20$\\ \hline
\end{tabular}
\label{tab:SimulationStudy-2}
\end{table}

For the non-periodic signal, all the methods performed reasonably well across all samples and missingness mechanisms. The performance of the SSE method in the non-periodic setting is not surprising. This is because as more observations are missing, the sampled time points appear random, and the resulting time series looks more like random random noise than signal.


\subsubsection{Real Data Analysis}
Irregularly-spaced times series data are common in astronomy such as those discussed in \cite{vanderplas2018understanding} and in exoplanet detection methods \citep{zhao2020expres,zhao2022expres}. In this section, we examine an asteroid dataset from the Lincoln Near-Earth Asteroid Research (LINEAR) survey, which tracks near-Earth asteroids. The data includes $280$ magnitude measurements (brightness) of LINEAR object ID 11375941 over five and a half years. Magnitude measurements are unitless; lower values indicate brighter objects. Further details on the data and preprocessing are in \cite{sesar2011exploring,palaversa2013exploring,vanderplas2018understanding}. 

\begin{figure}[ht!]
     \begin{subfigure}{0.32\textwidth}
         \centering
         \includegraphics[width=\textwidth]{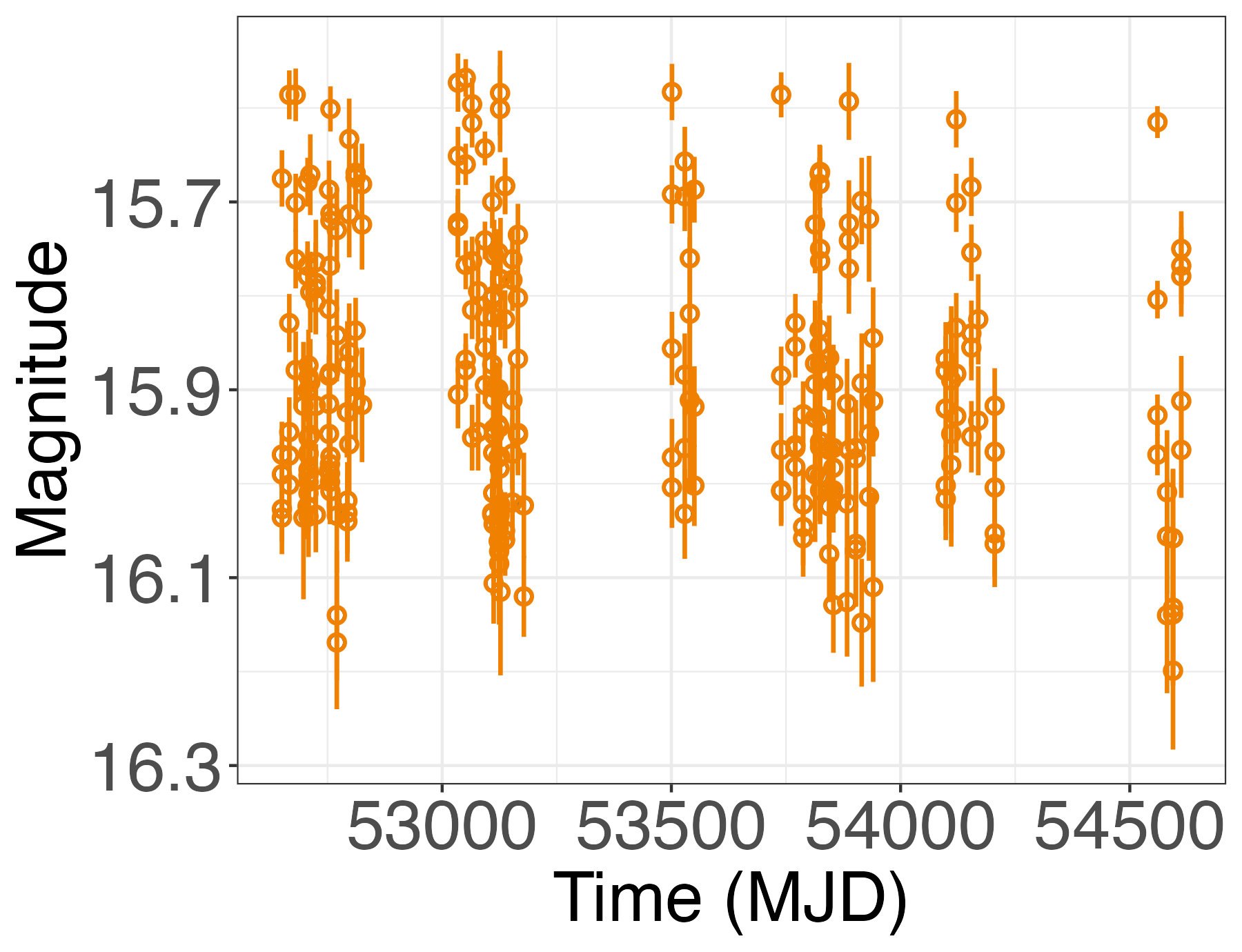}
         \caption{Time series}
         \label{fig:linear-object}
     \end{subfigure}
     \hfill
     \begin{subfigure}[b]{0.32\textwidth}
        \centering
         \includegraphics[width=\textwidth]{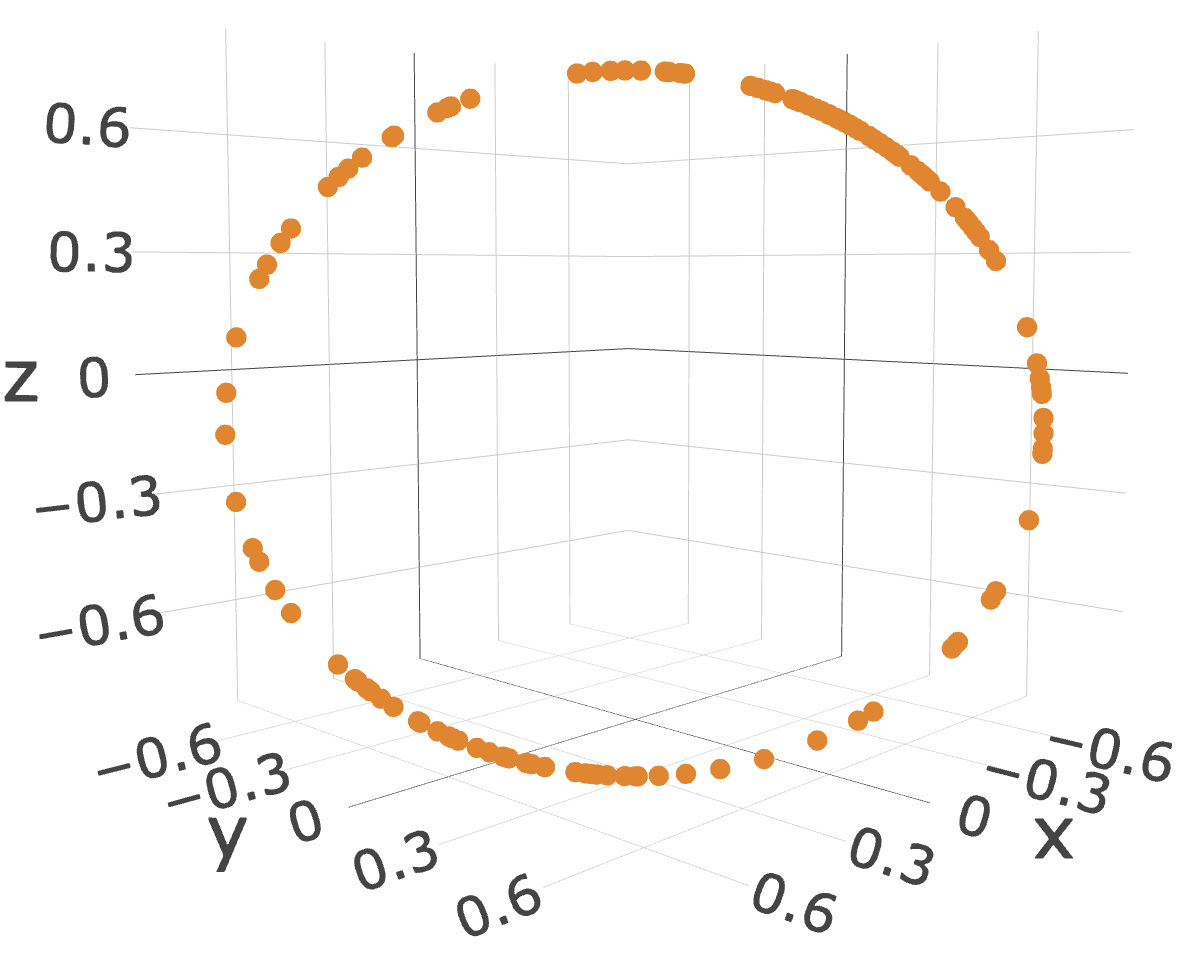}
         \caption{SSE}
         \label{fig:embedding-linear-object}
     \end{subfigure}
     \hfill
     \begin{subfigure}[b]{0.32\textwidth}
         \centering
         \includegraphics[width=\textwidth]{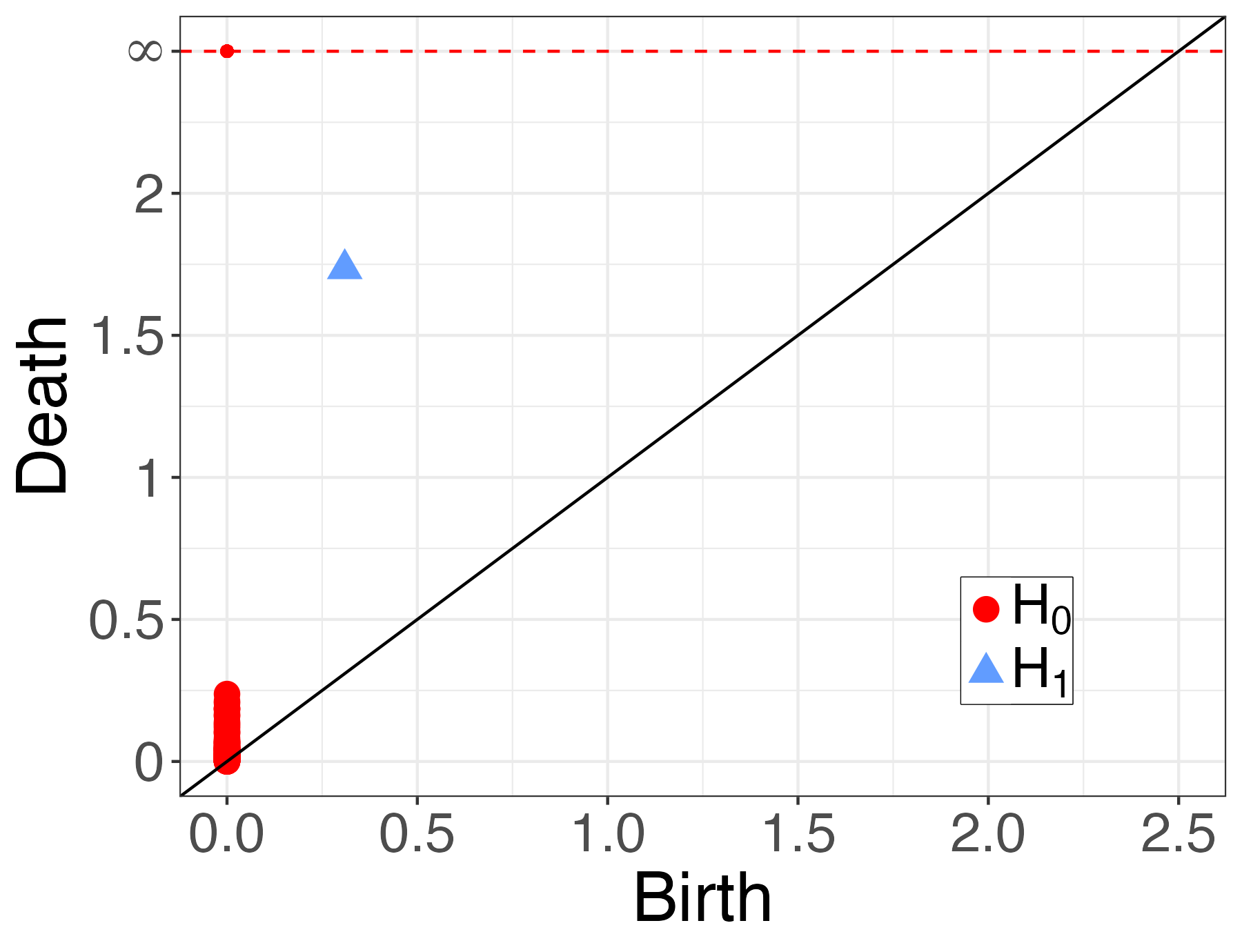}
         \caption{Persistence diagram}
         \label{fig:persistence-diagram-linear-object}
     \end{subfigure}
    \caption{LINEAR object ID 11375941. (a) The time series of the measured magnitudes (orange circles) with error bars (vertical bars).  (b) The SSE of the time series.   (c) The persistence diagram for the SSE with a single highly persistent $H_1$ feature as expected.}
    \label{fig:embedding-persistence-diagram}
\end{figure}

Figure \ref{fig:linear-object} shows the observed magnitude over time, revealing no visible periodic pattern due to irregular sampling. The TDE method is unsuitable for such data, but the proposed SSE method can construct a geometric representation. Using $M = 2$, and rescaling the time points to integers and taking $\tau = 1$, the SSE in Figure~\ref{fig:embedding-linear-object} reveals a circular geometric object, indicating high periodicity. The $H_1$ feature on the persistence diagram (Figure~\ref{fig:persistence-diagram-linear-object}) is at the point $(b, d) = (0.31, 1.74)$.
The periodicity score obtained using Equation \eqref{eqn:periodicity-score} is $0.82$, indicating high periodicity in he observed magnitude of LINEAR object ID 11375941. Using the Lomb-Scargle method, the optimal period was found to be $2.58$ with a p-value of $0.00$. These results confirm the SSE method's periodicity findings and highlight its utility in quantifying and visualizing periodicity. Additional astronomy applications are discussed in the supplementary material Section~\ref{sec:exoplanet}.


\section{Conclusion} \label{sec:conclusion}

The fusion of TDE with TDA holds significant promise for discerning system dynamics and quantifying properties like periodicity in uniformly-spaced time series. This work introduces a novel subsequence embedding method for irregularly-spaced time-series data. Irregular spacing can obscure patterns and introduce noise (e.g., Figure \ref{fig:linear-object}). While data imputation can create uniformly-spaced series, it often fails to accurately represent the true state space (e.g., Figure \ref{fig:henon-map-reconstructions}).
The proposed SSE method addresses these challenges, preserving the topology of the reconstructed state space and mitigating spurious homological features introduced by irregular spacing.

In Section~\ref{sec:period}, we note the need for statistical inference on periodicity scores to determine if the most persistent $H_1$ feature is due to a real periodic signal or chance. Existing methods for significance testing of homology generators, such as those using kernel density estimators, allow constructing confidence sets for homology generators \citep{fasy2014confidence,xu2019finding}. Extending this to homology generators based on direct filtration on the point-cloud space requires bounding the bottleneck-distance with the Hausdorff distance. Initial investigations produced wide confidence sets, indicating the need for a more tailored method, which remains a future research direction.

Finally, Algorithm \ref{alg:subsequence} constructs subsequences with a fixed regularity score $r$. Extending this to $r \pm \epsilon$ for small $\epsilon$ would increase the length of each constructed subsequence, and the number of points in the reconstructed space. Future work will explore the robustness of this method to different $\epsilon$ values and the trade-off between sample size and reconstruction accuracy.




\bibliographystyle{apalike}
\bibliography{main.article.arxiv}

\renewcommand{\thesection}{S\arabic{section}}
\setcounter{section}{0}

\section*{\centerline{Supplementary Material}} \label{sec:supp}
\setcounter{figure}{0}
\renewcommand{\thefigure}{S\arabic{figure}}
\setcounter{table}{0}
\renewcommand{\thetable}{S\arabic{table}}

\section{Stability and Convergence Results} \label{sec:supp_proofs}

\subsection{Proof of Proposition 4.1}

\begin{proof}
It suffices to bound the Hausdorff distance between $\mathbf{F}$ and $\mathbf{F}^\prime$, then using  the stability theory in persistence homology \citep{chazal2021introduction}, the bound on their persistence diagrams with respect to the Bottleneck distance can be established. The proof proceeds as follows.

There exists a subset $\mathbf{x}_i^* \subset \mathbf{x}^*$ that exactly equals $F^*(\mathbf{s}(t_{p, i}))$ (the $i$-th row of the $p$-th subsequence of the embedding matrix $\mathbf{F}^*$).  This fact stems from the construction of $\mathbf{F}^*$, whose rows are uniformly-spaced samples of $\mathbf{x}^*$. The same guarantee holds for the pairs $(\mathbf{F}, \mathbf{x})$, and $(\mathbf{F}^\prime, \mathbf{x}^\prime)$.
The distance between the projection $\Pi_{\mathbf{\boldsymbol{\Phi}}}(\mathbf{x}_i^*)$ and $F(\mathbf{s}(t_{p, i}))$ is given by
\begin{equation}
    \lVert \Pi_{\mathbf{\boldsymbol{\Phi}}}(\mathbf{x}_i^*) -  F(\mathbf{s}(t_{p, i})) \rVert_2 = \lVert \Pi_{\mathbf{\boldsymbol{\Phi}}} \left( \mathbf{x}_i^* - F(\mathbf{s}(t_{p, i})) \right) \rVert_2,
    \label{eqn:init-dist}
\end{equation}
where $\lVert \cdot \rVert_2$ denotes the $l^2$-norm. Equation~\eqref{eqn:init-dist} is under the assumption that the choice of frequency threshold does not smooth out the peaks in the true signal $\mathbf{x}$.
Observe that each $F^\prime(\mathbf{s}(t_{p, i}))$ is isometric to $\Pi_{\boldsymbol{\Phi}}(\mathbf{x}^*_{i})$,  where $\mathbf{x}^*_{i}$ is a subset of length $M+1$ of the original noisy scalar time series.
Then the Gromov-Hausdorff distance between $\mathbf{F}^\prime$ and $\mathbf{F}$ can be expressed as
\begin{equation}
    \text{d}_{\text{GH}}\left( \mathbf{F}^\prime, \mathbf{F} \right)  =  \text{d}_{\text{GH}}\left( \widehat{\Pi}_{\boldsymbol{\Phi}}(\mathbf{x}^*), \mathbf{F} \right), \label{eq:gh_expression}
\end{equation}
where $\widehat{\Pi}_{\boldsymbol{\Phi}}(\mathbf{x}^*)$ denotes embedding of the vector ${\Pi}_{\boldsymbol{\Phi}}(\mathbf{x}^*)$. Using the same isometric property, the Hausdorff distance between $\mathbf{F}^\prime$ and $\mathbf{F}$ can be expressed in terms of Equation~$\eqref{eqn:init-dist}$. This follows from the fact that
\begin{align}
\lVert \Pi_{\mathbf{\boldsymbol{\Phi}}} \left( \mathbf{x}_i^* - F(\mathbf{s}(t_{p, i})) \right) \rVert_2 
    \begin{aligned}[t]
    & = \lVert \Pi_{\mathbf{\boldsymbol{\Phi}}} \left( F^*(\mathbf{s}(t_{p, i}))  -F(\mathbf{s}(t_{p, i})) \right) \rVert_2 \\
    & \le 
    \lVert F^*(\mathbf{s}(t_{p, i}))  -F(\mathbf{s}(t_{p, i})) \rVert_2  \lVert (\boldsymbol{\Phi}^H\boldsymbol{\Phi})^\dagger \boldsymbol{\Phi}^H \rVert_2.
    \end{aligned}
    \label{eqn:denoising-bound}
\end{align}
The matrix $\boldsymbol{\Phi}^H\boldsymbol{\Phi}$ has the form:
\begin{equation}
    \boldsymbol{\Phi}^H\boldsymbol{\Phi} = 
    \begin{bmatrix}
        n & \sum_{k = 1}^n e^{j 2\pi (w_2 - w_1)f_k} & \cdots & \sum_{k = 1}^n e^{j 2\pi (w_n - w_1)f_k} \vspace{0.5cm} \\ 
        \sum_{k = 1}^n e^{-j 2\pi (w_2 - w_1)f_k} & n & \cdots & \sum_{k = 1}^n e^{j 2\pi (w_n - w_2)f_k} \vspace{0.5cm} \\
        \vdots & \vdots & \cdots & \vdots \vspace{0.5cm} \\
        \sum_{k = 1}^n e^{-j 2\pi (w_n - w_1)f_k} &  \sum_{k = 1}^n e^{-j 2\pi (w_n - w_2)f_k} & \cdots & n
    \end{bmatrix}.
\end{equation}
The modulus of off-diagonal elements of $\boldsymbol{\Phi}^H\boldsymbol{\Phi}$ is bounded as
\begin{equation}
    \left|\sum_{k = 1}^n e^{j 2\pi (w_{l_1} - w_{l_2})f_k}\right|
    \le \sum_{k = 1}^n \left| e^{j 2\pi (w_{l_1} - w_{l_2})f_k}\right| = n.
    \label{eqn:modulus-bound}
\end{equation}
Observe that $\lVert (\boldsymbol{\Phi}^H\boldsymbol{\Phi})^\dagger \boldsymbol{\Phi}^H \rVert_2 \le  \lVert (\boldsymbol{\Phi}^H\boldsymbol{\Phi})^\dagger \rVert_2  \lVert\boldsymbol{\Phi}^H \rVert_2$. First we bound $\lVert\boldsymbol{\Phi}^H \rVert_2$, by directly using the definition:
\begin{equation}
    \lVert\boldsymbol{\Phi}^H \rVert_2 = \sqrt{\lambda_{\text{max}}(\boldsymbol{\Phi}\boldsymbol{\Phi}^H)},
\end{equation}
where $\lambda_{\text{max}}(\boldsymbol{\Phi}\boldsymbol{\Phi}^H)$ is the maximum eigenvalue of $\boldsymbol{\Phi}\boldsymbol{\Phi}^H$.  Since $\boldsymbol{\Phi}\boldsymbol{\Phi}^H$ is Toeplitz, a bound on the maximum eigenvalue can be established as follows \citep{hertz1992simple}.
Let $\boldsymbol{\psi} = \left[ \psi_1, \psi_2, \cdots, \psi_n \right]^\top$ be a vector such that $\psi_1 = 1$ and 
\begin{equation}
    \psi_k = 2* \cos\left( \frac{\pi}{ \lfloor (n-1)/(k-1) \rfloor + 2 }  \right), \quad k = 2, \cdots, n.
\end{equation}
Also, let $\boldsymbol{\zeta} = \left[|(\boldsymbol{\Phi} \boldsymbol{\Phi}^H)_{1,1}|, | (\boldsymbol{\Phi} \boldsymbol{\Phi}^H)_{1,2}|, \cdots, |(\boldsymbol{\Phi} \boldsymbol{\Phi}^H)_{1,n}|\right]^\top$, that is, the modulus of the terms in first row of $\boldsymbol{\Phi} \boldsymbol{\Phi}^H$.
Let $\lambda_k$ be the $k$-th eigenvalue of $\boldsymbol{\Phi} \boldsymbol{\Phi}^H$. Then it follows that \citep{hertz1992simple}:
\begin{equation}
    \max_{1 \le k \le n}(\lambda_k) \le \boldsymbol{\zeta}^T \boldsymbol{\psi}.
\end{equation}
Further observe that $\max_{1 \le k \le n} (\psi_k) = 2$, hence, together with the bound on the values of $\boldsymbol{\zeta}$ from Equation~\eqref{eqn:modulus-bound} it follows that
\begin{align}
    \boldsymbol{\zeta}^T \boldsymbol{\psi}
    \begin{aligned}[t]
        & = n + \sum_{k = 2}^n |(\boldsymbol{\Phi} \boldsymbol{\Phi}^H)_{1,k}| \psi_k 
        & \le \left(n + n \sum_{k = 2}^n 2\right) = 2n^2 - n.
    \end{aligned}
    \label{eqn:bound-eig-val}
\end{align}
Hence the norm $\lVert {{\boldsymbol{\Phi}}}^H \rVert_2 \le 2n^2 - n$. It now remains to bound the quantity $\lVert (\boldsymbol{\Phi}^H\boldsymbol{\Phi})^\dagger \rVert_2$. By computing the singular value decomposition of $\lVert (\boldsymbol{\Phi}^H\boldsymbol{\Phi})^\dagger \rVert_2$, it follows directly that 
\begin{equation}
    \lVert (\boldsymbol{\Phi}^H\boldsymbol{\Phi})^\dagger \rVert_2 \le \frac{1}{\sigma^2_{\text{min}}(\boldsymbol{\Phi})},
    \label{eqn:svd-bound}
\end{equation}
where $\sigma^2_{\text{min}}(\boldsymbol{\Phi}) > 0$ is the smallest non-zero singular value of $\boldsymbol{\Phi}$. Using the fact that 
\begin{equation}
    \sum_{k = 1}^n  \lambda_k(\boldsymbol{\Phi}^H\boldsymbol{\Phi}) = \text{Tr}(\boldsymbol{\Phi}^H\boldsymbol{\Phi}) = n^2,
\end{equation}
where $\text{Tr}(\boldsymbol{\Phi}^H\boldsymbol{\Phi})$ is the matrix trace, it follows that the smallest non-zero eigenvalue is bounded as $0 < \lambda_{\text{min}}(\boldsymbol{\Phi}^H\boldsymbol{\Phi}) \le n$ which implies that $\sigma^2_{\text{min}}(\boldsymbol{\Phi}^H\boldsymbol{\Phi}) \le n$. For any such that $\sigma^2_{\text{min}}(\boldsymbol{\Phi}^H\boldsymbol{\Phi})$, we can always find a $0< \gamma \le 1$ such that $\sigma^2_{\text{min}}(\boldsymbol{\Phi}^H\boldsymbol{\Phi}) > {\gamma n}$. Hence the bound in Equation~\eqref{eqn:svd-bound} can be extended to 
\begin{equation}
    \lVert (\boldsymbol{\Phi}^H\boldsymbol{\Phi})^\dagger \rVert_2 \le \frac{1}{\sigma^2_{\text{min}}(\boldsymbol{\Phi})} \le \frac{1}{\gamma n}.
    \label{eqn:svd-bound-2}
\end{equation}
Now using the bound in Equations~\eqref{eqn:bound-eig-val} and ~\eqref{eqn:svd-bound-2}, the bound in Equation~\eqref{eqn:denoising-bound} has the form
\begin{equation}
    \lVert \Pi_{\mathbf{\boldsymbol{\Phi}}} \left( \mathbf{x}_i^* - F(\mathbf{s}(t_{p, i})) \right) \rVert_2 \le 
    \frac{2n - 1}{\gamma}\lVert F^*(\mathbf{s}(t_{p, i}))  -F(\mathbf{s}(t_{p, i})) \rVert_2
    \label{eqn:denoising-bound-2}
\end{equation}
The bound on the Hausdorff distance between $\mathbf{F}^\prime$ and $\mathbf{F}$ is then expressed as
\begin{align}
\text{d}_\text{H}(\widehat{\Pi}_{\boldsymbol{\Phi}}(\mathbf{x}^*), \mathbf{F})
    \begin{aligned}[t]
    & \le \sup\limits_{i, p}\lVert \Pi_{\mathbf{\boldsymbol{\Phi}}} \left( \mathbf{x}_i^* - F(\mathbf{s}(t_{p, i})) \right) \rVert_2 \\
    & \le \frac{2n - 1}{\gamma} \sup\limits_{i, p}\lVert F^*(\mathbf{s}(t_{p, i})) - F(\mathbf{s}(t_{p, i})) \rVert_2, 1 \le i \le n_p - M\tau_p, 1 \le p \le P.
    \end{aligned}
    \label{eqn:hausdorff-bound}
\end{align}
From the equality in Equation~\eqref{eq:gh_expression}, it holds that
\begin{equation}
    \text{d}_\text{B}(\text{Dgm}({\mathbf{F}}), \text{Dgm}({\mathbf{F}^\prime})) \le 2\text{d}_{\textbf{GH}}\left( \mathbf{F}, \mathbf{F}^\prime \right)  =  2\text{d}_{\textbf{GH}}\left( \widehat{\Pi}_{\boldsymbol{\Phi}}(\mathbf{x}^*), \mathbf{F} \right). 
\end{equation}
The Gromov-Hausdorff distance is further bounded above by the Hausdorff distance. From Equations~\eqref{eqn:hausdorff-bound}, the bound on the bottleneck distance between $\text{Dgm}({\mathbf{F}})$ and $\text{Dgm}({\mathbf{F}^\prime})$ is established as
\begin{equation}
    \text{d}_\text{B}(\text{Dgm}({\mathbf{F}}), \text{Dgm}({\mathbf{F}^\prime})) \le \frac{4n - 2}{\gamma} \left(\sup\limits_{i, p}||F^*(\mathbf{s}(t_{p, i})) - F(\mathbf{s}(t_{p, i}))||_2\right),
\end{equation}
where $1 \le i \le n_p - M\tau_p, \quad 1 \le p \le P$.
If the observed time series is `noise-free' such that $\mathbf{x}^* = \mathbf{x}$, then $\sup\limits_{i, p}||F(\mathbf{s}(t_{p, i})) - F^*(\mathbf{s}(t_{p, i}))||_2=0, \forall i, p$, and $\text{d}_\text{B}(\text{Dgm}({\mathbf{F}}), \text{Dgm}({\mathbf{F}^\prime})) = 0$.
\end{proof}

\subsection{Proof of Lemma 4.2}

\begin{proof}
    By construction, $\mathbf{F}^2 \subset \widehat{\mathbf{F}}^2$, thus for any $F^2(s(t_{p, i_1})) \in \mathbf{F}^2$, $\exists \widehat{F}^2(s(t_{p, i_2})) \in \widehat{\mathbf{F}}^2$ such that $F^2(s(t_{p, i_1})) = \widehat{F}^2(s(t_{p, i_2}))$. Further observe that $|\mathbf{F}^2|_u = |\hat{\mathbf{F}}^2|_u$, where $|.|_u$ is a measure of the cardinality of unique observations. Hence it follows that $\text{Dgm}(\mathbf{F}^2) \equiv \text{Dgm}(\widehat{\mathbf{F}}^2)$, and the conclusion is a direct consequence of this equivalence.
\end{proof}

\subsection{Proof of Theorem 4.5}

\begin{proof}
By construction of the subsequence and assumption \textbf{A2}, the Hausdorff distance between $\mathbf{F}_m$ and $\mathbf{F}_\vartheta$ has the form
\begin{equation}
    \text{d}_\text{H}(\mathbf{F}_m, \mathbf{F}_\vartheta) = \sup_{F_\vartheta \in \mathbf{F}_\vartheta} \min_{1 \le i \le m} \lVert F(s(t_i)) - F_\vartheta \rVert_2.
    \label{eqn:subset-hausdorff}
\end{equation}
Let $\mathbf{F}_0 \subset \mathbf{F}_\vartheta $ be a set of ball centers such that
\begin{equation}
    \mathbf{F}_\vartheta \subset \bigcup_{F_0 \in \mathbf{F}_0} B(F_0, \delta),
    \label{eqn:covering}
\end{equation}
i.e., the minimal covering of $\mathbf{F}_\vartheta$ consisting of balls of radius $\delta$ around $F_0 \in \mathbf{F}_0$.
For any $F_\vartheta \in \mathbf{F}_\vartheta$ and $F_0 \in \mathbf{F}_0$, the following inequality holds:
\begin{align}
    \min_{1 \le i \le m} \lVert F(s(t_i))-F_\vartheta  \rVert_2
    \begin{aligned}[t]
     & \le \lVert F(s(t_j)) - F_\vartheta \rVert_2\\
     & = \lVert F(s(t_j)) - F_0  + F_0  - F_\vartheta \rVert_2\\
     & \le \lVert F(s(t_j)) - F_0 \rVert_2 + \lVert F_0 - F_\vartheta \rVert_2,\quad j = 1, \cdots, m.
    \end{aligned}
    \label{eqn:min-bound}
\end{align}
Observe that $\lVert F_0 - F_\vartheta \rVert_2$ is bounded by the radius $\delta$, hence using Equation \eqref{eqn:covering}, it follows that 
\begin{equation}
    \min_{1 \le i \le m} \lVert F(s(t_i)) - F_\vartheta \rVert_2 \le \delta + \max_{F_0 \in \mathbf{F}_0} \min_{1 \le i \le m} \lVert F(s(t_i)) - F_0 \rVert_2 \le \varepsilon,
\end{equation}
for some $\epsilon > 0$. Further, taking the supremum over all $F_\vartheta$, the relation still holds:
\begin{equation}
    \sup_{F_\vartheta \in \mathbf{F}_\vartheta}\min_{1 \le i \le m} \lVert F(s(t_i)) - F_\vartheta \rVert_2 \le \delta + \max_{F_0 \in \mathbf{F}_0} \min_{1 \le i \le m} \lVert F(s(t_i)) - F_0 \rVert_2 \le \varepsilon,
\end{equation}
Then the probability that $\sup_{F_\vartheta \in \mathbf{F}_\vartheta}\min_{1 \le i \le m} \lVert F(s(t_i)) - F_\vartheta \rVert_2$ exceeds $\varepsilon$ is bounded as
\begin{align}
    \text{Pr}\left(\sup_{F_\vartheta \in \mathbf{F}_\vartheta}\min_{1 \le i \le m} \lVert F(s(t_i)) - F_\vartheta \rVert_2 > \varepsilon\right)
    \begin{aligned}[t]
        & \le \text{Pr}\left(\delta + \max_{F_0 \in \mathbf{F}_0} \min_{1 \le i \le m} \lVert F(s(t_i)) - F_0 \rVert_2 > \varepsilon\right)\\
        & = \text{Pr}\left(\max_{F_0 \in \mathbf{F}_0} \min_{1 \le i \le m} \lVert F(s(t_i)) - F_0 \rVert_2 > \varepsilon - \delta \right)
    \end{aligned}
\end{align}
From~Equation \eqref{eqn:subset-hausdorff}, $\sup_{F_\vartheta \in \mathbf{F}_\vartheta} \min_{1 \le i \le m} \lVert F(s(t_i)) - F_\vartheta \rVert_2 = \text{d}_\text{H}(\mathbf{F}_m, \mathbf{F}_\vartheta)$, hence a bound on the probability of $\text{d}_\text{H}(\mathbf{F}_m, \mathbf{F}_\vartheta)$ exceeding $\varepsilon$ can be obtained as 
\begin{equation}
    \text{Pr}\left(\text{d}_\text{H}(\mathbf{F}_m, \mathbf{F}_\vartheta) > \varepsilon\right) \le \text{Pr}\left(\max_{F_0 \in \mathbf{F}_0} \min_{1 \le i \le m} \lVert F(s(t_i)) - F_0  \rVert_2 > \varepsilon - \delta\right).
    \label{eqn:prob-bound-hd}
\end{equation}
Observe that $\mathbf{F}_\vartheta$ endowed with the Hausdorff metric is complete and separable \citep{attouch1991topology}. Then for $\varepsilon$ small enough, the following bound holds \citep{cuevas2004boundary}:
\begin{equation}
    \text{Pr}\left(\max_{F_0 \in \mathbf{F}_0} \min_{1 \le i \le m} \lVert F(s(t_i)) - F_0 \rVert_2 > \varepsilon - \delta\right) \le C \left(1- \kappa\omega(\varepsilon - \delta)^{M+1}\right)^m.
    \label{eqn:prob-bound}
\end{equation}
The constant $C$ is the number of points in the covering of $\mathbf{F}_\vartheta$, i.e., $|\mathbf{F}_0|$, and $\omega$ is the Lebesgue measure of the unit ball in $\mathbb{R}^{M+1}$.
Note  that since $0 \le \kappa \omega(\varepsilon - \delta)^{M+1} \le 1$, it follows that $\left(1 - \kappa \omega(\varepsilon - \delta)^{M+1}\right)^m \le e^{-m\kappa \omega(\varepsilon - \delta)^{M+1}}$.  This allows for Equations \eqref{eqn:prob-bound-hd} and \eqref{eqn:prob-bound} to be rewritten as 
\begin{equation}
     \text{Pr}\left(\text{d}_\text{H}(\mathbf{F}_m, \mathbf{F}_\vartheta) > \varepsilon\right) \le \text{Pr}\left(\max_{F_0 \in \mathbf{F}_0} \min_{1 \le i \le m} \lVert F(s(t_i)) - F_0 \rVert_2 > \varepsilon - \delta\right) \le C e^{-m\kappa \omega(\varepsilon - \delta)^{M+1}}.
    \label{eqn:prob-bound-2}
\end{equation}
Choose some $K > \left( \frac{2}{\kappa \omega} \right)^\frac{1}{M+1}$ and let $\varepsilon_m(r) = \left(r - \frac{m-l}{m}\right)$, where $l$ is the number of missing observations in the initial time series and $m \gg l$, then for $m$ large enough, it follows that 
\begin{equation}
     \text{Pr}\left(  \left( \varepsilon_m(r) \right)^{-\frac{1}{M+1}} \text{d}_\text{H}(\mathbf{F}_m, \mathbf{F}_\vartheta) > K \right) \le C e^{-m\kappa \omega\left( (\varepsilon_m(r)\frac{2}{\kappa\omega})^{\frac{1}{M+1}} - \delta\right)^{M+1}}.
    \label{eqn:prob-bound-3}
\end{equation}
The above bound can be obtained by simply substituting $\left(\frac{2\varepsilon_m(r)}{\kappa\omega}\right)^\frac{1}{M+1}$ for $\varepsilon$ in Equation \eqref{eqn:prob-bound-2}.
Now consider the sum 
\begin{equation}
    \sum_{m}  e^{-m\kappa \omega\left( (\varepsilon_m(r)\frac{2}{\kappa\omega})^{\frac{1}{M+1}} - \delta\right)^{M+1}},
\end{equation}
and observe that it is convergent if $\varepsilon_m(r) \ge \left( \frac{\delta}{K} \right)^{M+1}$. This condition can always be satisfied given the restriction $K > \left( \frac{2}{\kappa \omega} \right)^\frac{1}{M+1}$ and for an appropriate choice of $\kappa$ and $\delta$. Then by the Borel-Cantelli lemma \citep{emile1909probabilites,cantelli1917sulla,chung1952application}, since 
\begin{equation}
    \sum_m \text{Pr}\left(  \left( \varepsilon_m(r) \right)^{-\frac{1}{M+1}} \text{d}_\text{H}(\mathbf{F}_m, \mathbf{F}_\vartheta) > K \right) < \infty,
\end{equation}
it follows that
\begin{equation}
    \lim_{m \xrightarrow{} \infty} \sup  \left( \varepsilon_m(r) \right)^{-\frac{1}{M+1}} \text{d}_\text{H}(\mathbf{F}_m, \mathbf{F}_\vartheta) \le K.
    \label{eqn:final-proof}
\end{equation}
\end{proof}


\section{Discussion} \label{sec:discussion}
The proposed SSE methodology generalizes TDEs to allow for more realistic time-series data that are not uniformly spaced.  The SSE method and TDA analysis of the reconstructed state space can be used to detect periodic signals, but there is potential for this framework to provide more useful information. This discussion highlights this claim by demonstrating that different time-series signals with common frequency domain and periodicity properties can exhibit different structural shapes in their reconstructed state spaces. The reconstruction provides qualitative and quantitative information that may be used for downstream analysis.  We begin with artificial time-series data, and then provide a motivating application related to the detection and characterization of exoplanets.

\subsection{Denoising and Stability}
\label{sec:denoising-simulation}
\begin{figure}[ht!]
     \centering
     \begin{subfigure}[b]{0.45\textwidth}
         \centering
         \includegraphics[width=\textwidth, height=0.53\textwidth]{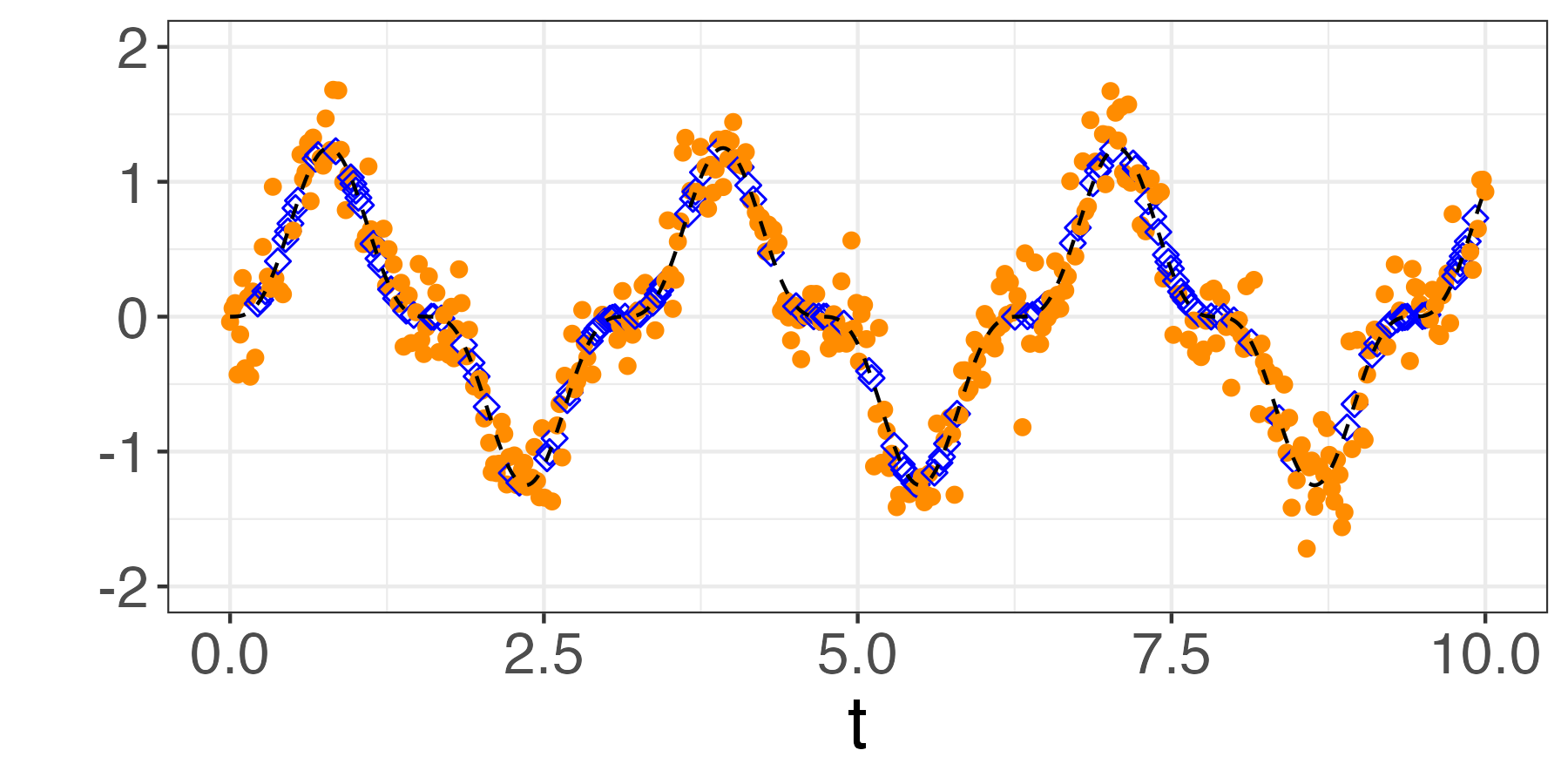}
         \caption{Noisy original time series}
         \label{fig:noisy-signal}
     \end{subfigure}
     \begin{subfigure}[b]{0.45\textwidth}
        \centering
         \includegraphics[width=\textwidth, height=0.53\textwidth]{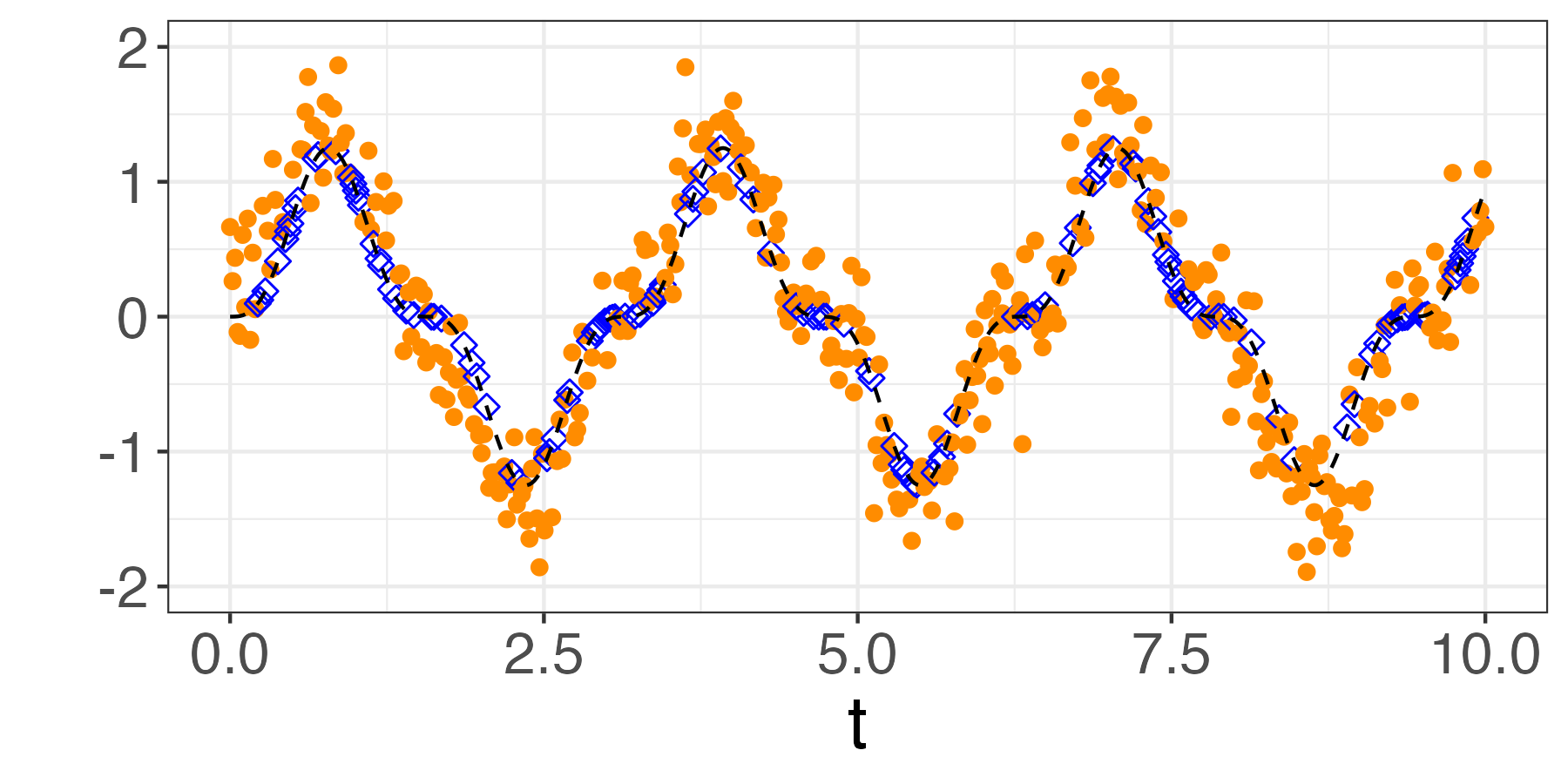}
         \caption{Smoothed (threshold=5)}
         \label{fig:smoothed-signal-5}
     \end{subfigure}
     \begin{subfigure}[b]{0.45\textwidth}
         \centering
         \includegraphics[width=\textwidth, height=0.53\textwidth]{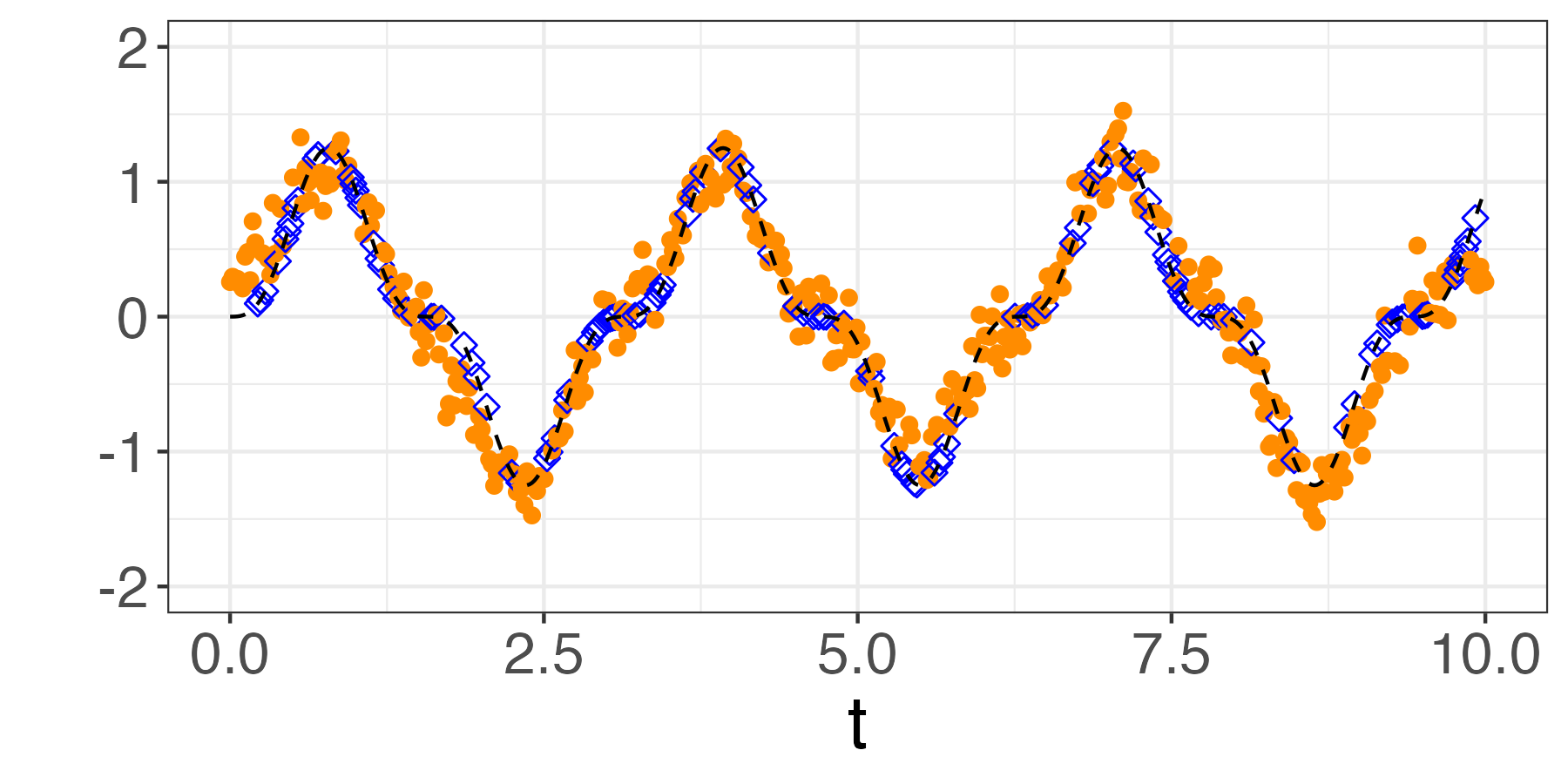}
         \caption{Smoothed (threshold=15)}
         \label{fig:smoothed-signal-15}
     \end{subfigure}
     \begin{subfigure}[b]{0.45\textwidth}
         \centering
         \includegraphics[width=\textwidth, height=0.53\textwidth]{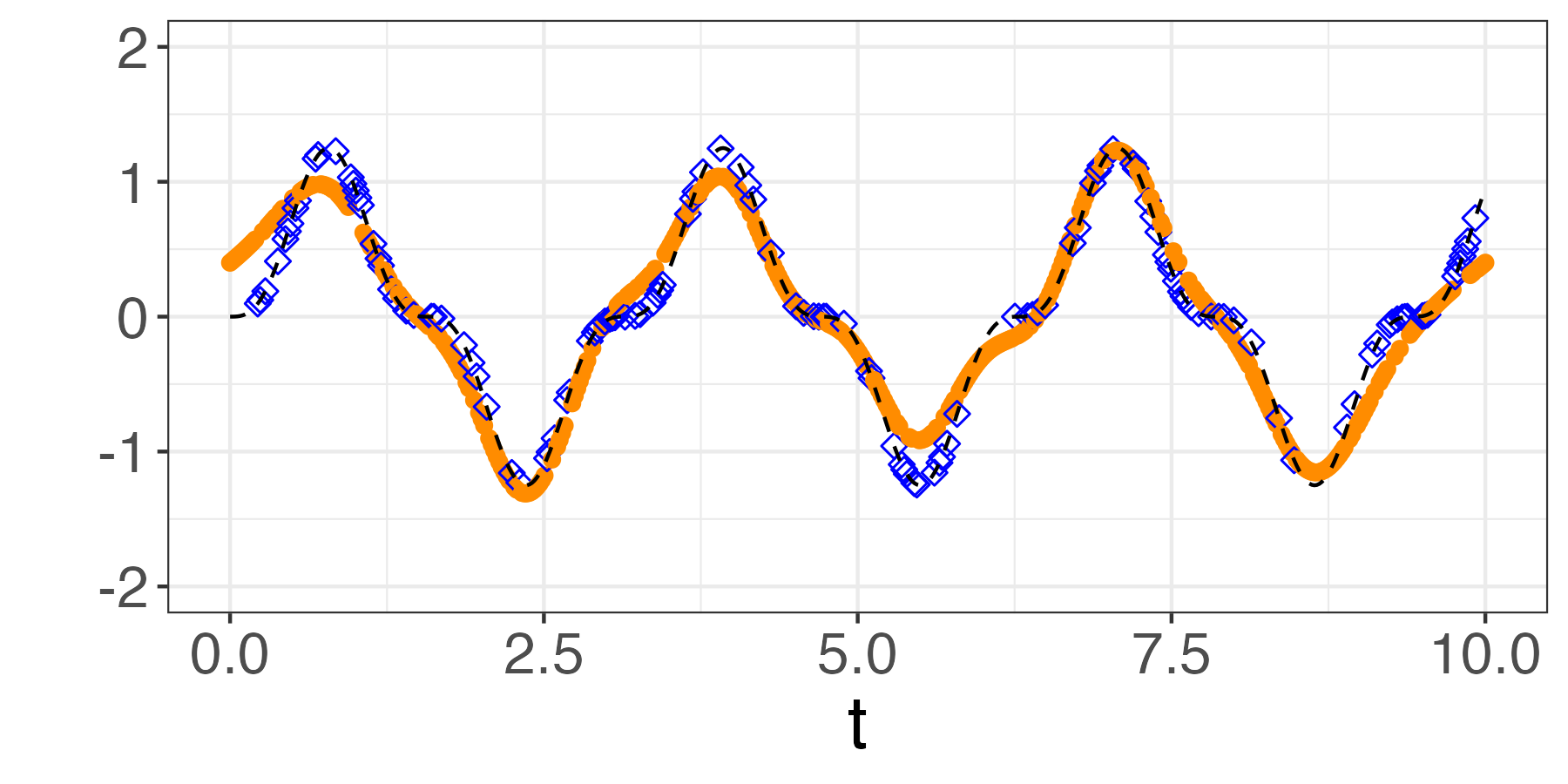}
         \caption{Smoothed (threshold=25)}
         \label{fig:smoothed-signal-25}
     \end{subfigure}
    \caption{Illustration of the denoising method of Section $4.2$ of the main article. The time series was perturbed with noise drawn from a $N(0,0.25)$, and the probability of a missing observation is $0.25$. (a) The original 500 time series measurements. The round orange points are observed values, while the blue diamonds are the missing values displayed at the true signal value without noise. The other sub-figures display the time series after denoising with a frequency threshold of 5 (b), 15 (c), and 25 (d).}
    \label{fig:smoothing-illustration}
\end{figure}

To evaluate the performance of denoising procedure presented in in this work, and Proposition $4.1$, the time series in Figure \textcolor{red}{3a} of the main article was replicated at varying noise levels and sample sizes. The probability that any value is unobserved at a given time point is fixed at $0.25$. Four noise levels $\{0, 0.25, 0.5, 2\}$ and five samples sizes $\{50, 100, 500, 1000, 5000\}$ were used in the simulations. For each noise level and sample size combination, the denoising method outlined in Section $4.2$ of the main article was performed and the bottleneck distance between the corresponding persistence diagrams and the theoretical upper-bound are computed. The upper bound computed does not include the multiplicative factor $\frac{2n - 1}{\gamma}$. Figure \ref{fig:smoothing-illustration} shows a  noisy time series and the outcome after denoising at various frequency thresholds.
%
For an appropriate choice of frequency threshold, which is generally application dependent, the true underlying signal can be satisfactorily recovered.

\begin{figure}[ht]
 \centering
 \includegraphics[width=1\linewidth,clip=true]{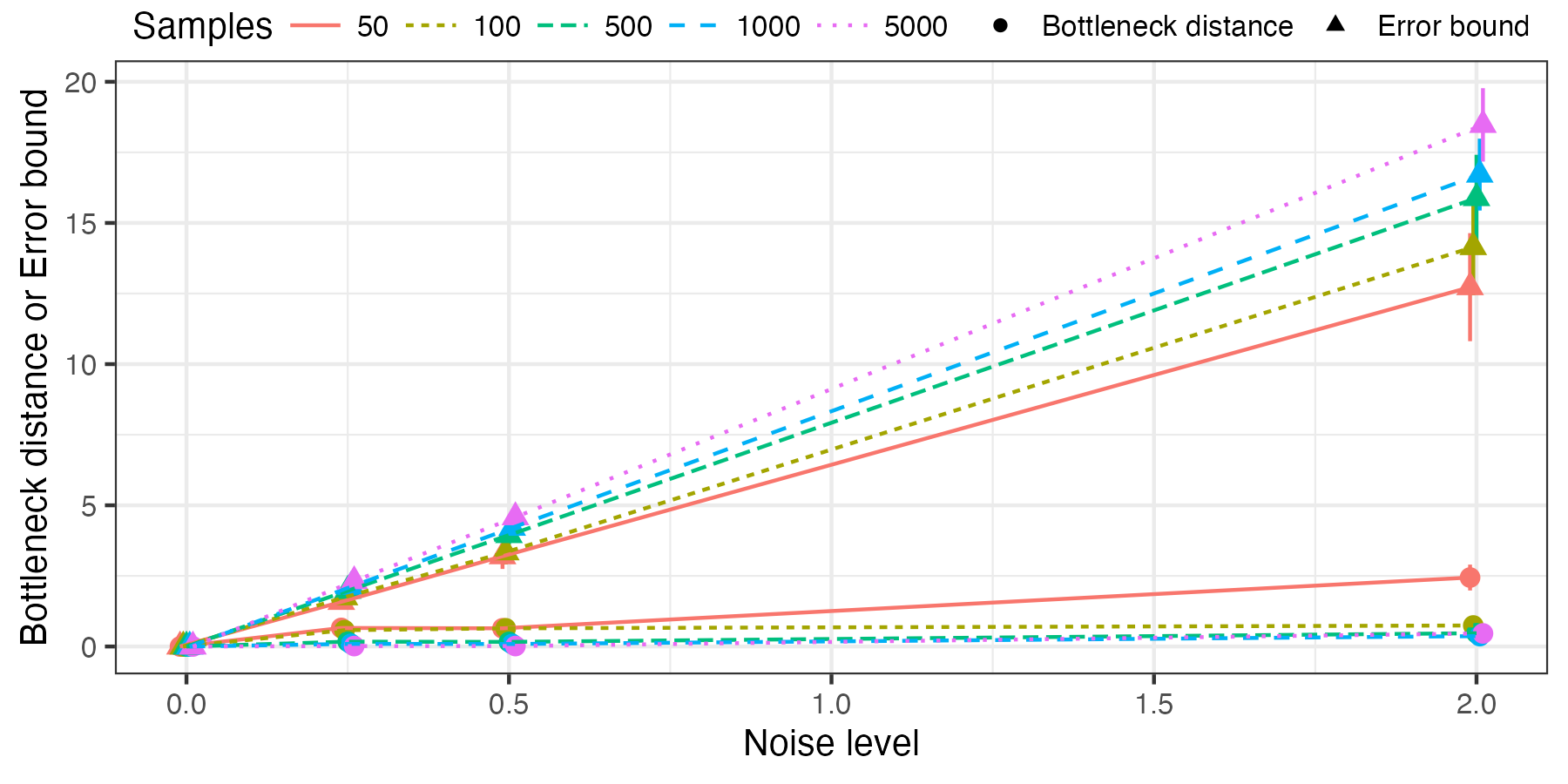}
 \caption{Stability results of the denoising procedure (see Proposition $4.1$ in the main text). The solid points represent the mean values from $100$ repetitions, the vertical lines on these points indicate the error bars (which are too small to see in many cases), and the colors and line types indicate the sample size.  The vertical axis represents the bottleneck distance for the circle points and the error bound (without the multiplicative factor $\frac{2n - 1}{\gamma}$) for the triangle points.}
 \label{fig:bottleneck-error-bound}
\end{figure}

For each combination of the noise-level and sample size, the process is repeated 100 times and standard errors are obtained. 
The results are presented in Figure \ref{fig:bottleneck-error-bound}.
The bottleneck distance is bounded above by the error bound for all noise levels and sample sizes as expected. At the same noise level, smaller sample sizes tend to have larger bottleneck distances. This can partly be explained by the fact that the reconstructed space is more sparse (i.e., points in the space are more spread out since there are fewer points). The $H_0$ features are more likely to persist longer in such sparse settings. The reverse is true for the error bound in Proposition $4.1$, which for the same noise level, is higher for larger sample sizes. This follows from the fact that larger sample sizes increases the chance of observing highly noisy terms. These results demonstrate the denoising procedure's efficacy in controlling noise effects on the SSE's topological features.

Proposition 4.1 establishes a conservative bound on the bottleneck distance between persistence diagrams of a noise-free and a denoised time series using Fourier methods. The factor $(2n -1)/\gamma$ reflects the poor-conditioning of the Fourier matrix in non-uniform domains. Empirical evidence suggests this bound could be improved in more restricted settings, a topic for future research.

\subsection{Topology-Informed Features}
\begin{figure}[ht!]
    \centering
    \includegraphics[width=1\textwidth,clip=true]{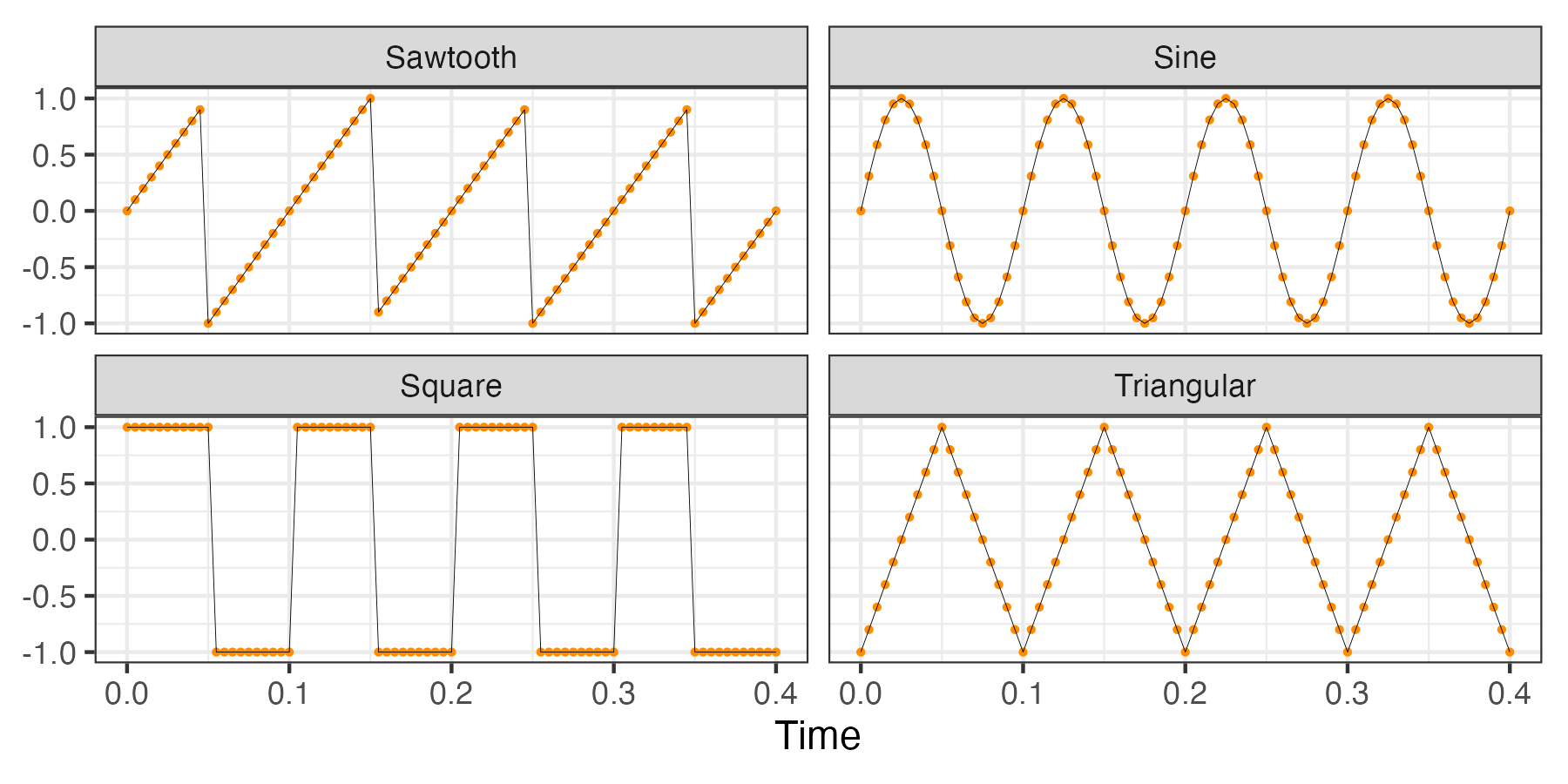}
    \caption{Four time-series signals. Top-left: sawtooth wave; Top-right: sine wave; Bottom-left: square wave; Bottom-right: triangular wave. All four signals were generated to have the same period and the observations were sampled at the same frequency.}
    \label{fig:four-waves}
\end{figure}

\begin{figure}[ht!]
    \centering
    \includegraphics[width=1\textwidth,clip=true]{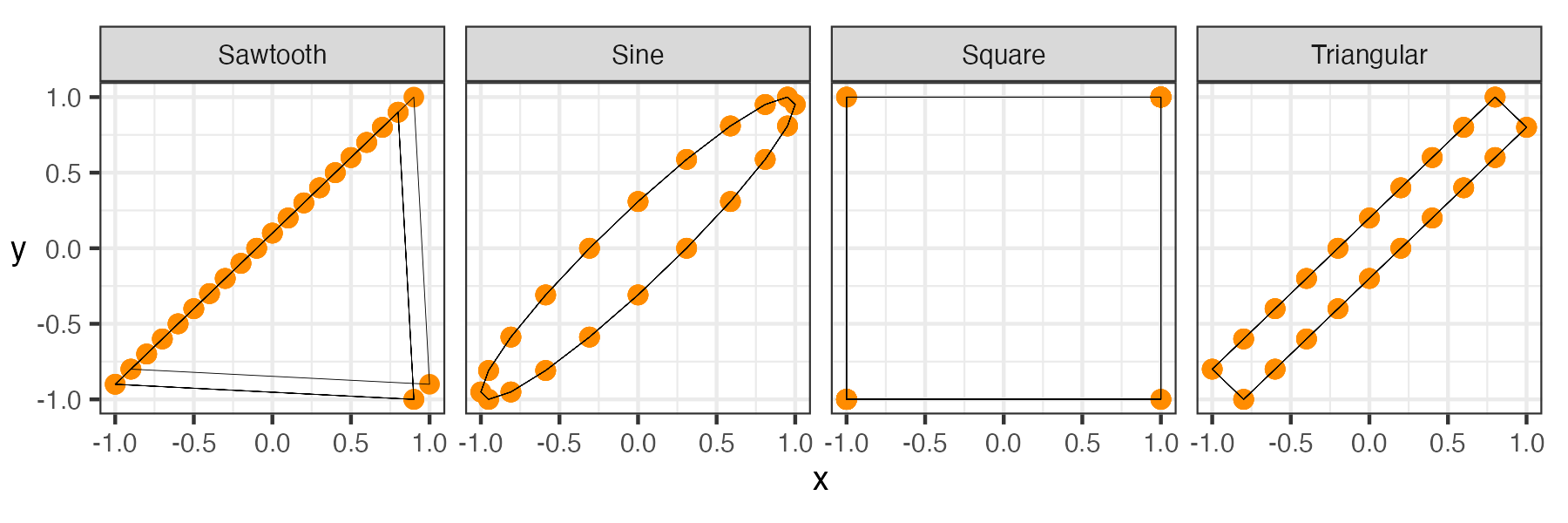}
    \caption{The embedding in $\mathbb{R}^2$ of the four time series in Figure \ref{fig:four-waves}. Each embedding highlights the unique shape of the original time series.}
    \label{fig:four-waves-embedding}
\end{figure}

\begin{figure}[ht!]
    \centering
    \includegraphics[width=1\textwidth,height=0.55\textwidth,clip=true]{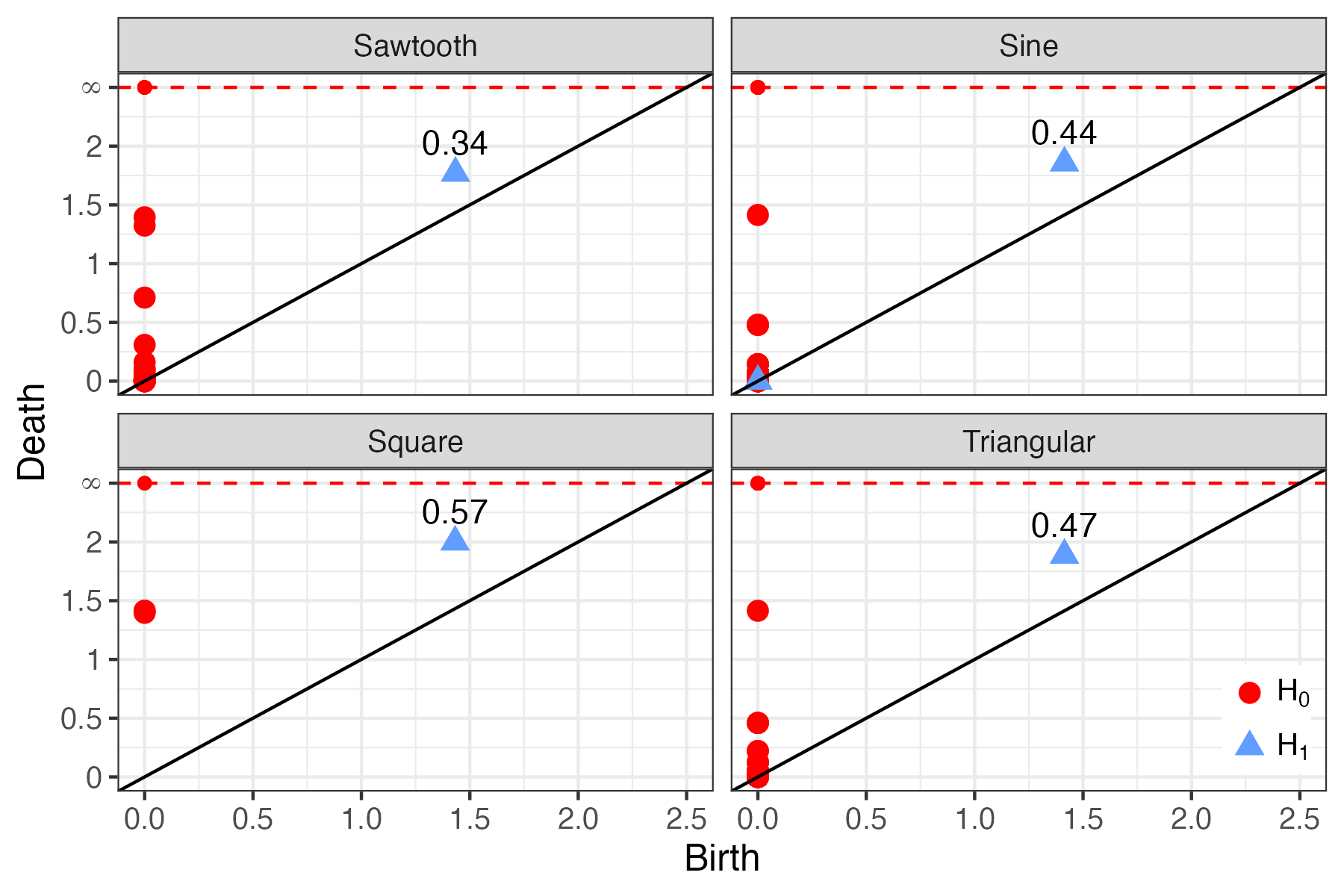}
    \caption{The persistence diagrams for the four time series in Figure \ref{fig:four-waves}. The persistence of the most significant $H_1$ feature in each diagram is noted above the corresponding feature. The square wave has a single highly persistent $H_1$ feature consistent with its shape.}
    \label{fig:our-waves-pds-non-periodic}
\end{figure}

Traditional time series methods (e.g., \citealt{lomb1976least}), such as Fourier analysis and autocorrelation, focus primarily on frequency components and periodicity. The SSE method coupled with TDA through techniques like persistent homology allow for further exploration of the time series beyond the observed scalar values. For instance, consider the { square wave} in Figure \ref{fig:four-waves}-bottom-left, where its abrupt transitions challenges frequency-based analyses \citep{lomb1976least}. Its sharp edges and discontinuities result in significant spectral leakage when analyzed using Fourier methods \citep{lomb1976least}. However, its SSE captures these sharp transitions (Figure \ref{fig:four-waves-embedding})-bottom-left, while its associated persistence diagram in Figure \ref{fig:our-waves-pds-non-periodic}-bottom-left effectively distinguishes between high and low states. Specifically, the high and low states of the square wave result in exactly a single cluster for each homology dimension. The persistence diagrams were computed on point-wise centered and scaled embedding maps.
The {sawtooth wave} (Figure \ref{fig:four-waves}-top-left), {sine wave} (Figure \ref{fig:four-waves}-top-right), and the triangular wave (Figure \ref{fig:four-waves}-bottom-right) display distinct geometric structures in their embeddings (Figure \ref{fig:four-waves-embedding}). Note that their persistence diagrams (Figure \ref{fig:our-waves-pds-non-periodic}) do not exhibit more significant $H_1$ features compared to the square wave.

\subsection{Shape-Agnostic Analysis}

\begin{figure}[ht!]
    \centering
    \includegraphics[width=1\textwidth,clip=true]{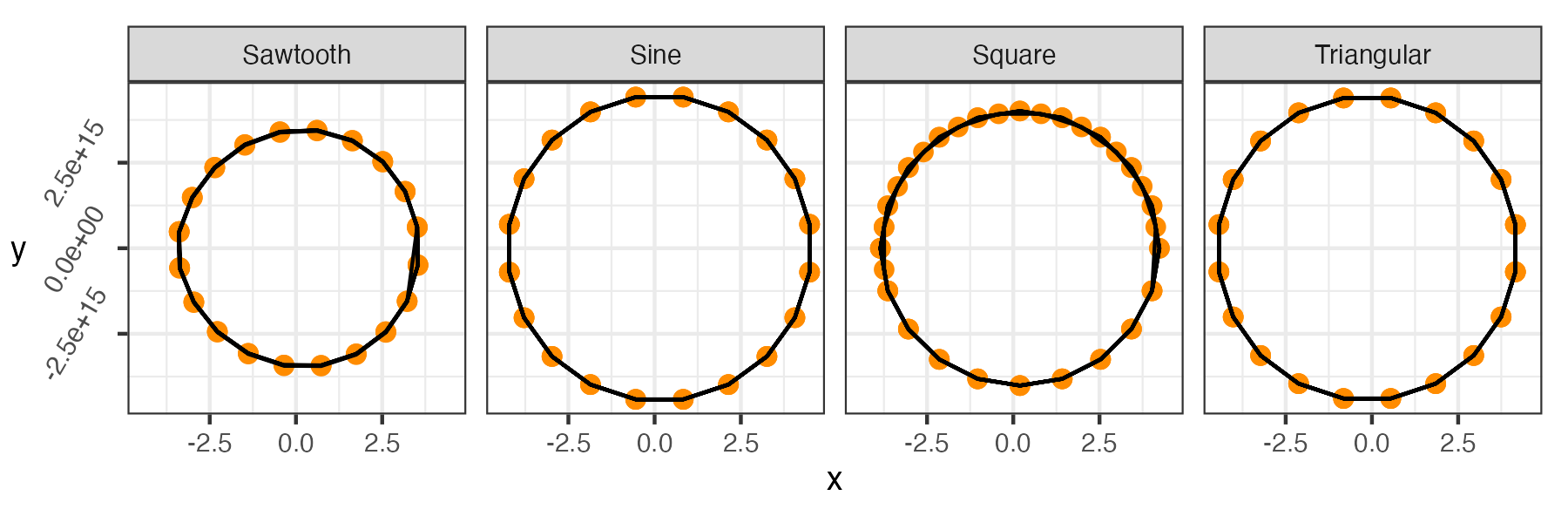}
    \caption{The embedding in $\mathbb{R}^{M+1}$ of the four time series in Figure \ref{fig:four-waves}. $M=19$ was chosen such that $(M+1)\tau$ for $\tau=1/200$ approximate the period of the signals.}
    \label{fig:four-waves-embedding-periodic}
\end{figure}

\begin{figure}[ht]
    \centering
    \includegraphics[width=1\textwidth,height=0.55\textwidth,clip=true]{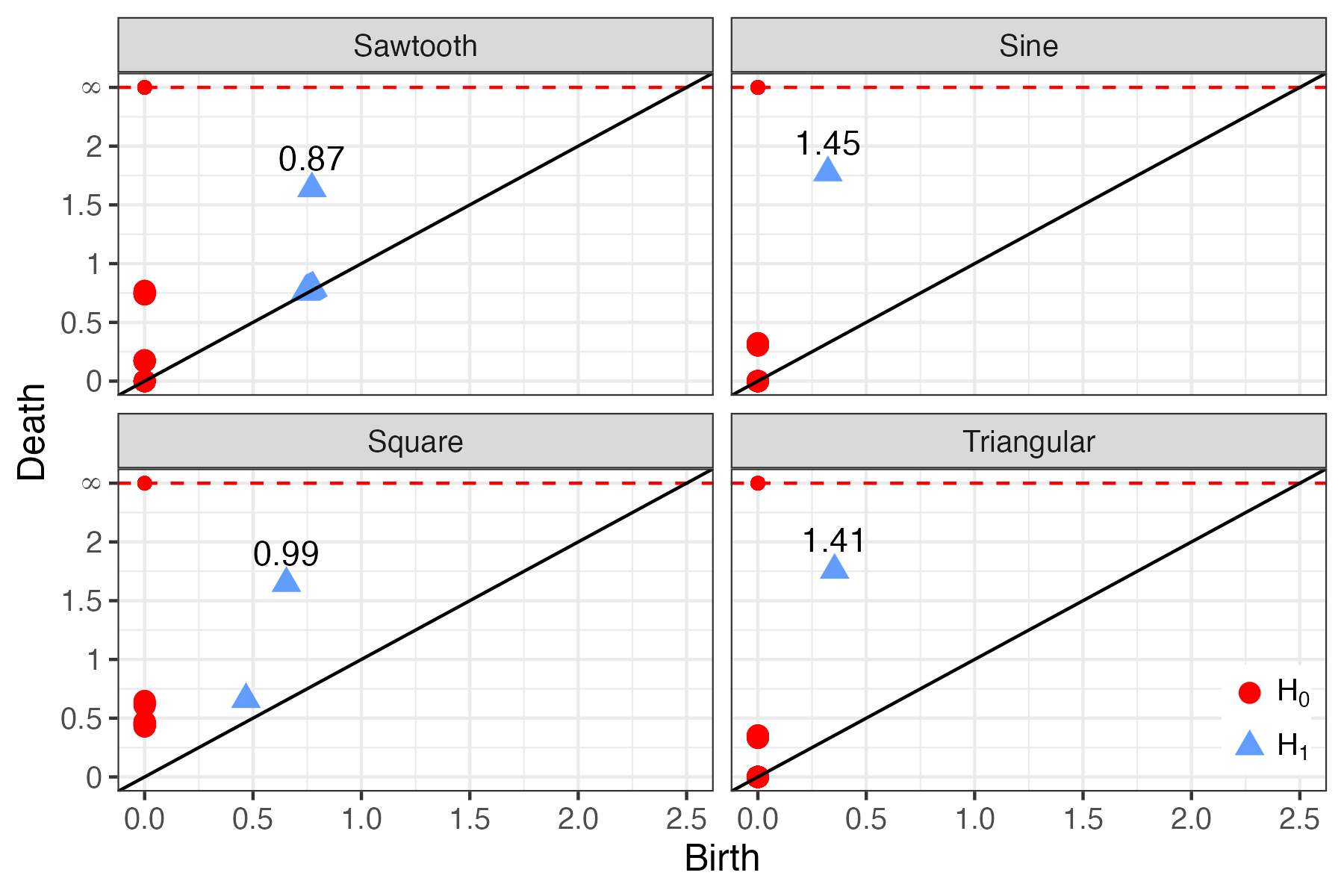}
    \caption{The persistence diagrams associated with the embeddings Figure \ref{fig:four-waves-embedding-periodic}. All four signals have at least one-significant $H_1$ feature, which captures the cyclical pattern in the original time series.}
    \label{fig:four-waves-pds-periodic}
\end{figure}

Different time-series signals with the same periodicity and frequency information can exhibit unique and additional characteristics when examined through SSE and TDA. In particular, the SSE and TDA analysis pipeline further reveals the presence and longevity of topological features, emphasizing the stability of any cyclic structure in a way that complements traditional time-series methods.
For example, the time-series signals in Figure \ref{fig:four-waves} can be embedded in a space that emphasizes the stability of their cyclical structure. If one chooses an embedding dimension $M+1$ and step-size $\tau$ such that $(M+1)\tau$ approximates the period of the time series, the cyclical structure can be well approximated. The four signals were each embedded in $\mathbb{R}^{20}$, and a scatterplot of their first two principal components are shown in Figure \ref{fig:four-waves-embedding-periodic}. The circular (loops) shapes are indicative of the cyclical structures in the time series. Their corresponding persistence diagrams, computed on the pointwise-centered and scaled SSE matrices are shown in Figure \ref{fig:four-waves-pds-periodic} where the persistence of the $H_1$ feature can be viewed as quantifying the stability and longevity of these cyclical structures.

\subsection{Exoplanet Time-Series Data} \label{sec:exoplanet}
\begin{figure}[ht!]
    \centering
    \includegraphics[width=1\textwidth,clip=true]{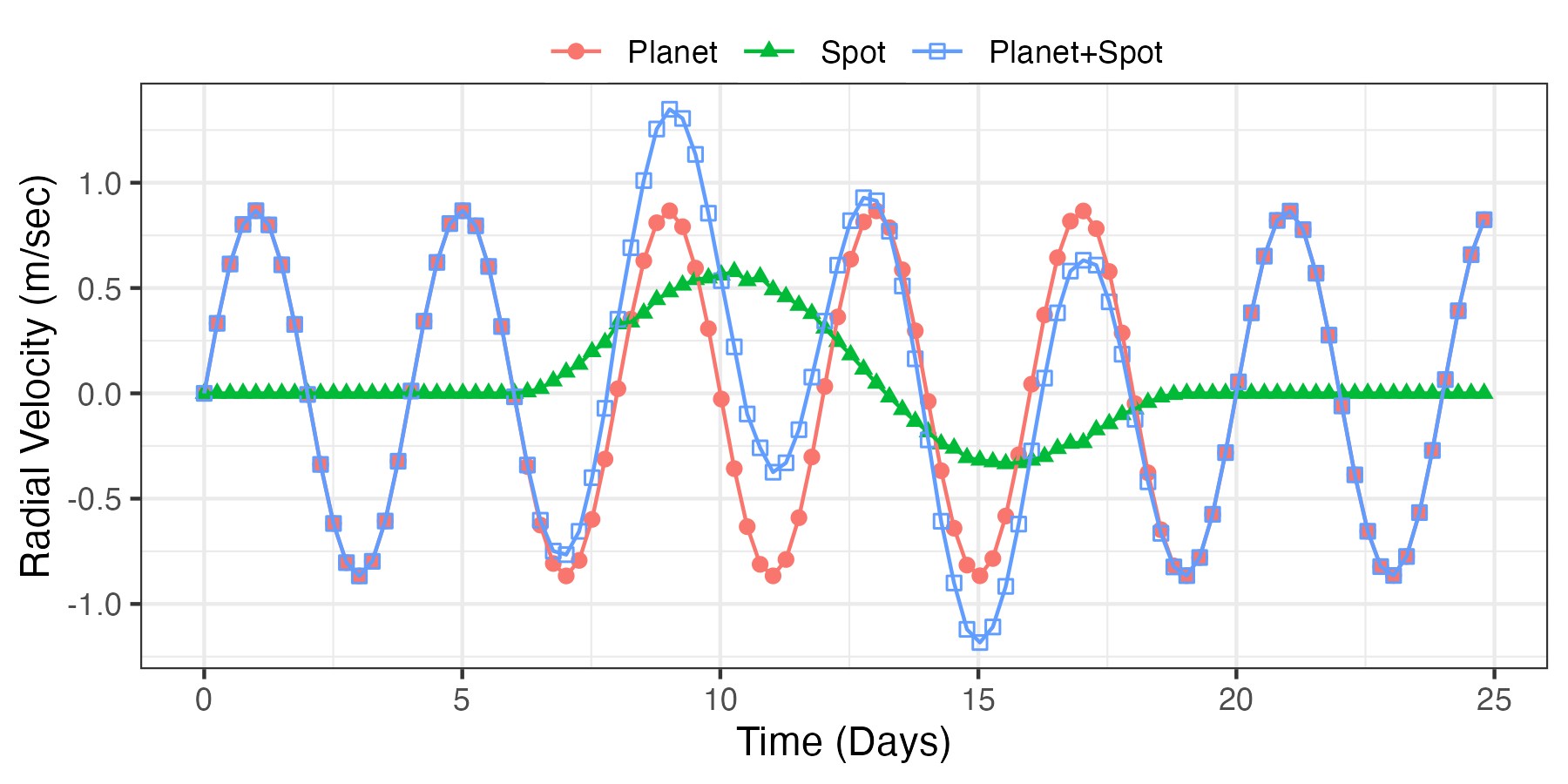}
    \caption{Exoplanet time-series data.  Simulated RV data of an exoplanet (red circles), a 0.05\% spot (blue squares), and the planet and spot combined (green triangles).}
    \label{fig:exo-planet-signals}
\end{figure}

Exoplanets are planets that orbit stars other than our sun.  The radial velocity (RV) method is one of the most prolific methods for the detection of exoplanets, and the method used to discover the first exoplanet orbiting a sun-like star \citep{mayor1995jupiter}.  The RV method is sometimes referred to as the ``wobble method" because an exoplanet is detected through the motion, or the wobble, of its host star.  More specifically, an exoplanet is detected by estimating the forward and backward velocity (i.e., the RV) of the star across time, and observing a particular periodic signature in the time series suggesting the RV of the star is induced by an exoplanet.  A simulated example of the Keplerian model of the RV of star across time when an exoplanet is orbiting in a circular orbit (i.e., with an eccentricy of zero) is displayed as the red point and line in Figure~\ref{fig:exo-planet-signals}.  More background on statistical aspects of the RV method can be found in \citep{hara2023statistical}.
Though the RV method has been successful at detecting exoplanets, challenges exist in detecting low-mass exoplanets, such as earth analogs, which, all else equal, produce smaller amplitude signals compared to more massive exoplanets.\footnote{The exoplanet detected in \cite{mayor1995jupiter} had an RV time series semi-amplitude around 60 m/s, and the RV signal for an earth-like exoplanet orbiting a sun-like star is around 10 cm/s.}  

Until the recent development of extreme precision radial velocity (EPRV) spectrographs, the detection of low-mass exoplanet RV signals was not generally feasible \citep{fischer2016state}.  
Estimation of planetary RV signals is especially challenging in the low-mass regime because activity and variability in the atmosphere of a star (e.g., the formation and evolution of star spots) can produce periodic variation imprinted on the spectra that can hide or mimic exoplanet signals \citep{huelamo2008tw,dumusque2016radial, davis2017insights}.  
For example, the green triangle line in Figure~\ref{fig:exo-planet-signals} displays the RV times series induced by a single simulated star spot orbiting a star (discussed in more detail below).  While the spot is behind the star, the RV is zero.  However, when the spot rotates in view, it produces a sinusoid-like signal that is challenging to distinguish from an exoplanet signal in real RV observations.
New statistical methods have been developed to address the challenge of detecting low-mass exoplanets in the presence of stellar activity (e.g., \citealt{rajpaul2015gaussian, dumusque2018measuring, holzer2021hermite, holzer2021stellar, jones2022improving}), but none of these methods fully or generally mitigate the issues \citep{zhao2022expres}.  

While we do not solve the issue with detecting low-mass exoplanets in the presence of stellar activity, ideas from TDE and TDA may prove to be fruitful.  In particular, RV measurements of stars other than our sun are highly non-uniform,\footnote{Figure 3 of \cite{zhao2022expres} displays examples of the non-uniformity of RV time-series observations.}
the proposed SSE method is a necessary foundation to use TDEs for exoplanet detection using the RV method.  

To illustrate how TDEs and TDA may be used in this setting, we consider a simple simulated example of a planet RV time series, a star spot RV time series, and the combination of a planet and spot RV times series in Figure~\ref{fig:exo-planet-signals}.  
 The host star is set to have a stellar rotation period of 25.05 days, the planet RV time series has an orbital period of 4 days and is on a circular orbit resulting in a semi-amplitude of 0.87 m/s, and the single spot covers 0.05\% of the observed surface area of the star and orbits at 30 degrees latitude generated using the Spot Oscillation And Planet (SOAP) 2.0 code \citep{dumusque2014soap}, which induces a semi-amplitude of 0.58 m/s.  Realistic RV data has noise, irregular and sparse sampling, more complicated stellar variability, and many other complexities; our purpose here is to illustrate how extensions of SSE and TDA may be helpful in this setting.

\begin{figure}[ht!]
    \centering
    \includegraphics[width=1\textwidth,clip=true]{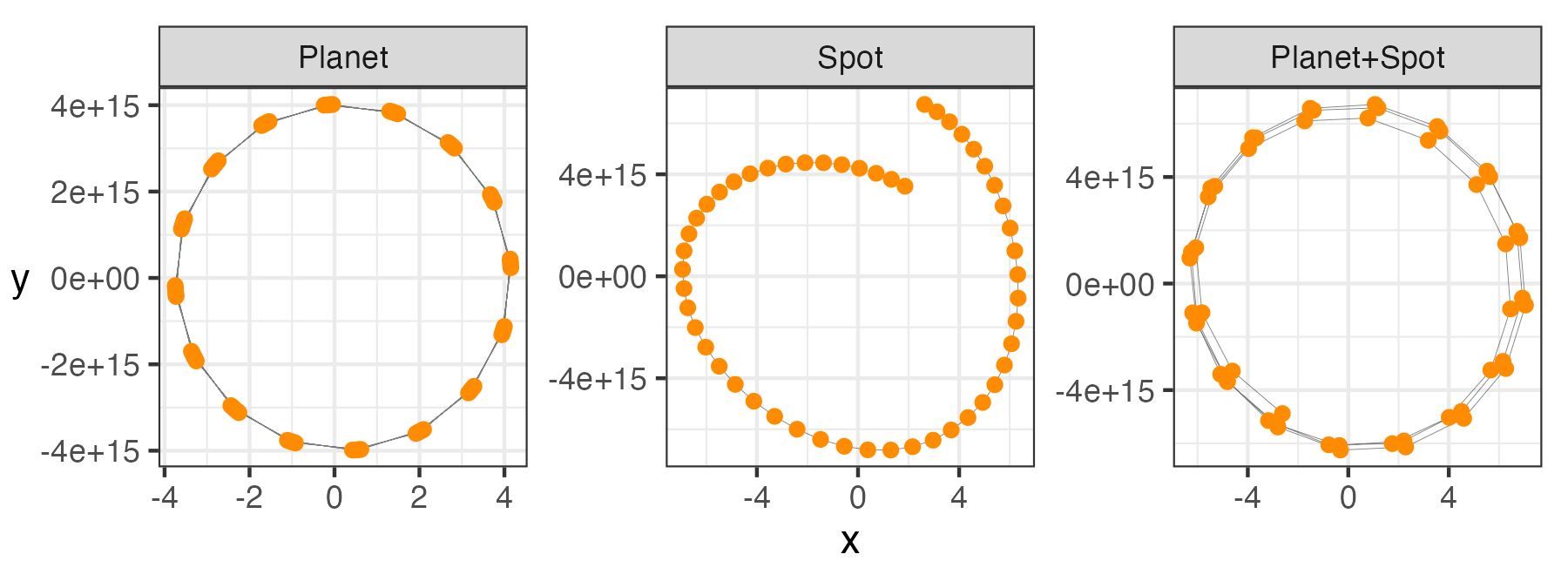}
    \caption{The embedding of the three signals. Planet was embedded in $\mathbb{R}^{16}$ whiles Spot and Spot+Planet were each embedded in $\mathbb{R}^{50}$. The planet signal is prominent in the Planet+Spot embedded space.}
    \label{fig:exo-planet-embedding}
\end{figure}

\begin{figure}[ht!]
    \centering
    \includegraphics[width=1\textwidth,clip=true]{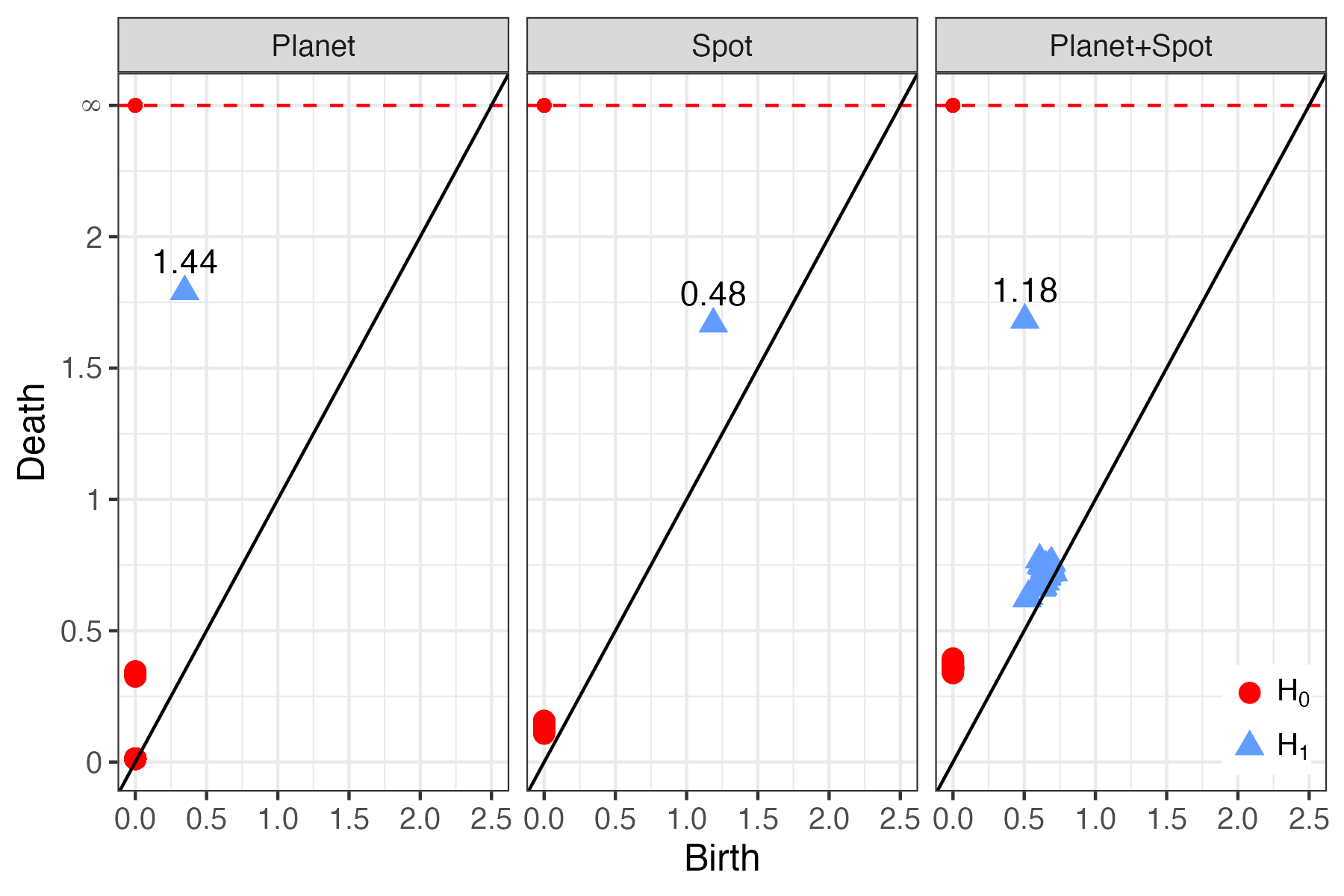}
    \caption{The persistence diagrams associated with the three signals. The text above the most significant $H_1$ feature in each diagram represents the persistence of that feature. The Planet+Spot, as expected has a few noisy $H_1$ features.}
    \label{fig:exo-planet-pds-periodic}
\end{figure}

The SSE matrices for the three time series are equivalent to the TDE matrices since the observations are uniformly-spaced (i.e., the time difference between observations in each of the three time series is $0.2505$ days). To quantify the longevity and stability of any cyclical structures, the embedding window $((M+1)\tau)$ was chosen to approximate the period of the time series.  For the planet RV time series, choosing $M = 15$ and taking $\tau = 0.2505$ approximates the period of $4$ days. Similarly, for the star spot RV time series, taking a $M = 49$ with the same $\tau=0.2505$ approximates the period of $12.75$ days. For the combination of planet and spot RV times series, the embedding window was chosen to approximate the dominant period in the two time series, which is choosing $M = 49$ and $\tau=0.2505$. We note that the embedding window can also be chosen to approximate the less dominant period. This can sometimes result in negligible decrease in the persistence of the most persistent $H_1$ feature. The embedding matrix denoted $\mathbf{F} \subset \mathbb{R}^{M+1}$ for each time series is pointwise-centered and scaled. Figure \ref{fig:exo-planet-embedding} shows the plot of the two principal components of the embedding matrices.  The planet's TDE displays a clear circular pattern reflecting the periodic nature of the planetary signal in its RV time series.  In contrast, the spot's TDE, while still appearing somewhat circular, does not form a closed loop.  The combined planet and spot RV time series has a TDE that appears to have several closely matched loops.  These differences in structure in the TDEs are apparent in the corresponding persistence diagram, which is shown in Figure \ref{fig:exo-planet-pds-periodic}.  The planet's persistence diagram has the most persistent $H_1$ feature, while the spot's persistence diagram has a less persistent, though still apparent, $H_1$ feature.  The persistence diagram of the combined planet and spot has a persistent $H_1$ feature, but also has a cluster of low persistence $H_1$ features.  

The long-term goal of this sort of RV analysis is to detect and characterize the planetary signal in the presence of stellar activity and other sources of (possibility periodic) variability.  Unlike the simplified simulated data used here to illustrate concepts, real RV times series are highly non-uniform and noisy, and contain many complicated and periodic stellar variability signals, along with instrumental biases and other sources of uncertainty.  Dominant periodic signals in RV time series can vary widely among planetary, stellar variability, instrumental, and other sources.

\subsection{Summary}
Our methodology has implications for real-world applications, such as the detection of exoplanets in the presence of stellar variability. The proposed SSE method addresses challenges posed by non-uniformly sampled time-series data for TDEs. The SSE method preserves the topology of the state space, mitigating the artificial noisy structures introduced by irregular time points, or structural shifts introduced by imputation methods. Coupling the SSE with TDA further enhances its robustness, since TDA is robust to noise and local distortions in the geometry of the state spaces. This advancement ensures that the reconstructed state space accurately reflects the true dynamics of the system, even with non-uniform and noisy sampling. This is particularly relevant in real-world applications where data is often sampled at irregular intervals with an inherent noisy structure.

Future research will further explore the full potential of this methodology, integrating it with conventional statistical techniques, developing more efficient algorithms, and applying these methods to a broader range of real-world data to uncover new insights and structures not readily apparent on scalar time-series data.

\end{document}